\documentclass[iop, twocolumn]{aastex63}
\usepackage{verbatim,subfigure}
\usepackage{multirow}
\hypersetup{citecolor=blue,urlcolor=red}
\usepackage{lineno}
\shorttitle{Pulsar interpulses}
\shortauthors{Sun et al.}

\begin{document}

\title{The emission geometry of pulsars with interpulses}

\correspondingauthor{N. Wang}
\email{na.wang@xao.ac.cn}

\author{S. N. Sun}
\affiliation{Xinjiang Astronomical Observatory, Chinese Academy of Sciences, Urumqi, Xinjiang 830011, P. R. China}
\affiliation{Xinjiang Key Laboratory of Radio Astrophysics, 150 Science1-Street, Urumqi 830011, P. R. China}
\affiliation{Key Laboratory of Radio Astronomy and Technology, Chinese Academy of Sciences, A20 Datun Road, Chaoyang District, Beijing, 100101, P. R. China}

\author{N. Wang}
\affiliation{Xinjiang Astronomical Observatory, Chinese Academy of Sciences, Urumqi, Xinjiang 830011, P. R. China}
\affiliation{Xinjiang Key Laboratory of Radio Astrophysics, 150 Science1-Street, Urumqi 830011, P. R. China}
\affiliation{Key Laboratory of Radio Astronomy and Technology, Chinese Academy of Sciences, A20 Datun Road, Chaoyang District, Beijing, 100101, P. R. China}

\author{W. M. Yan\href{https://orcid.org/0000-0002-7662-3875}}
\affiliation{Xinjiang Astronomical Observatory, Chinese Academy of Sciences, Urumqi, Xinjiang 830011, P. R. China}
\affiliation{Xinjiang Key Laboratory of Radio Astrophysics, 150 Science1-Street, Urumqi 830011, P. R. China}
\affiliation{Key Laboratory of Radio Astronomy and Technology, Chinese Academy of Sciences, A20 Datun Road, Chaoyang District, Beijing, 100101, P. R. China}

\author{S. Q. Wang \href{https://orcid.org/0000-0003-4498-6070}}
\affiliation{Xinjiang Astronomical Observatory, Chinese Academy of Sciences, Urumqi, Xinjiang 830011, P. R. China}
\affiliation{CSIRO Astronomy and Space Science, PO Box 76, Epping, NSW 1710, Australia}
\affiliation{Xinjiang Key Laboratory of Radio Astrophysics, 150 Science1-Street, Urumqi 830011, P. R. China}
\affiliation{Key Laboratory of Radio Astronomy and Technology, Chinese Academy of Sciences, A20 Datun Road, Chaoyang District, Beijing, 100101, P. R. China}

\begin{abstract}

We present polarization profiles of 23 pulsars exhibiting interpulse (IP) emissions using the Five-hundred-meter Aperture Spherical Radio Telescope (FAST). By applying the rotating vector model (RVM), we measured the inclination and impact angles for 16 pulsars, allowing us to investigate their beam geometries. Our analysis indicates that for 5 pulsars, the IP emissions likely originate from the same magnetic pole as the main pulse (MP), whereas for the remaining 11 pulsars, the IP and MP emissions originate from opposite magnetic poles. For the 7 pulsars that do not conform to the RVM, we are unable to determine whether the IP emissions originate from the same or opposite magnetic pole as the MPs.
By analyzing the beam configurations of pulsars, we found that the emission within the beam is not fully active across both longitude and latitude. Filling factors ranging from 0.27 to 0.99 were obtained for pulsars with opposite pole IP emissions, suggesting an underestimation of emission height when applying the geometrical method.
The emissions for MPs and IPs occur at different heights in the pulsar magnetosphere, with the difference in emission height ranging from tens to thousands of kilometers. We also found that some pulsars have wide emission beams, indicating that radio emissions may occur in regions of high altitude within the pulsar magnetosphere.

\end{abstract}


\section{INTRODUCTION}

\iffalse
\begin{figure*}
\centering
\subfigure[J1932+1059]{
\includegraphics[width=1.0\columnwidth]{J1932+1059_corner.pdf}}
\subfigure[J1935+2025]{
\includegraphics[width=1.0\columnwidth]{J1935+2025_corner.pdf}}
\caption{Posterior distributions of the parameters from the RVM fits for PSRs J1932+1059 and J1935+2025. The dashed vertical lines in the top panel of each column indicate the median and the 16 and 84 percentiles of the distribution, measured in degrees. }
\label{fig:1mcmc}
\end{figure*}
\fi

Pulsars generally exhibit highly polarized radio emissions, with mean fractions of linear polarization at 20\% and circular polarization at 10\%~\citep{jk2018,pkj+2023}. For some pulsars, the fraction of linear polarization can reach up to 100\%. The polarization properties are stable for most pulsars. Polarimetric observations allow for the determination of both the geometry and emission physics of pulsars.
Based on the polarization properties of the Vela pulsar (PSR B0833$-$45), \citet{rc1969} proposed the rotating vector model (RVM) to explain the S-shaped swings of the linear polarization position angle (PPA). According to the RVM, the emission is polarized parallel to the plane of curvature of the magnetic field, and the PPA across the pulse phase ($\phi$) can be described as~\citep{rc1969}: \begin{equation}\label{eq60} \tan(\psi-\psi_{0})=\frac{\sin\alpha\sin(\phi-\phi_{0})}{\sin\zeta\cos\alpha-\cos\zeta\sin\alpha\cos(\phi-\phi_{0})}, \end{equation} where $\zeta=\alpha+\beta$ is the angle between the rotation axis and the observer’s line-of-sight (LOS). The angle between the rotation axis and the magnetic axis is denoted by $\alpha$, while $\beta$ represents the angle of closest approach of the magnetic axis to the LOS. $\psi_{0}$ and $\phi_{0}$ correspond to the position angle and the pulse phase, respectively, at the inflection point of the PPAs.

The polarization properties of pulsars provide a method to determine the emission height, as demonstrated by the delay-radius method. For pulsars, if radio emission occurs above the pulsar surface, the PPAs will lag behind the pulse profile due to aberration and retardation effects~\citep{bcw1991}. The height of the emission region can be estimated by measuring the delay ($\Delta\phi$) between the steepest gradient point of the PPAs and the profile center~\citep{bcw1991}: \begin{equation}\label{eq2} h_{\rm em}=\frac{\Delta \phi R_{\rm lc} }{4}, \end{equation} where $R_{\rm lc} = {c P}/{2\pi} $ is the light-cylinder radius, with $c$ denoting the speed of light and $P$ representing the pulsar's rotation period. This method has been used to estimate emission heights for many pulsars (e.g., \citealt{bcw1991,1997A&A...324..981V,2004A&A...421..215M,2008MNRAS.391.1210W}).

Another method for estimating emission height considers the geometry of the pulsar, known as the geometrical method. By assuming pulsars have dipolar magnetic field lines and that emission can extend to the last open field line, the emission height can be calculated by measuring the half-opening angle ($\rho$) of the emission at the last open field line~\citep{r1990}: \begin{equation}\label{eq3} h_{\rm em}=\frac{4}{9}R_{\rm lc}\rho, \end{equation} where $\rho$ can be calculated using the following expression~\citep{gk1993}: \begin{equation}\label{eq4} \cos \rho= \cos \alpha \cos \zeta + \sin \alpha \sin \zeta \cos (W/2), \end{equation} where $W$ is the observed pulse width.

The RVM has been widely applied to determine the geometries of pulsars with S-shaped PPAs (e.g., \citealt{jkk+2023}). However, many pulsars exhibit PPAs that deviate from the S-shape, making it difficult to determine their geometries.
Some pulsars exhibit orthogonal polarization modes (OPMs), in which the PPA suddenly jumps by approximately $90^{\circ}$ at specific longitudes within the pulse profile~\citep{jkk+2023}. The physics underlying OPMs is not well understood and may be related to propagation effects in pulsar plasma~\citep{scw+1984}.

For most pulsars, the profiles are narrow, typically less than 10\% of the phase~\citep{lm1988}. This narrowness results in a strong covariance between $\alpha$ and $\beta$ during RVM fitting, leading to significant uncertainties in geometry measurements~\citep{ew2001}. Further studies have found that pulsars with wide profiles enable more precise determinations of $\alpha$ and $\beta$, such as those with interpulse (IP) emissions. In these pulsars, the profile features an IP component that is separated from the main pulse (MP) by approximately half of the pulsar period. Usually, the intensity of the IP is only a few percent of the MP, but it can sometimes be as strong as the MP~\citep{jk2019}.

The simplest explanation is that the IP emits from the magnetic pole opposite that of the MP, with the magnetic axis nearly orthogonal to the rotational axis, i.e., the two-pole model. However, for some pulsars with IP emissions, such as PSR J0953+0755~\citep{hc1981}, a weak emission bridge connecting the MP and IP is observed, which is difficult to explain within the two-pole model. It has been proposed that the IPs of some pulsars may originate from the same pole as the MPs, i.e., the one-pole model~\citep{ml1977}. Phase-locked modulation has been reported in some pulsars, such as PSRs B1702$-$19, B1055$-$52, B1822$-$09, and B1929+10, where both MPs and IPs show periodic pulse intensity modulations with the same modulation period~\citep{wws2007,wws2012,ymw+2019, kyp+2021}. This phenomenon also supports the one-pole model. Polarization observations are crucial for determining whether the emission of the IP originates from the opposite pole as the MPs.

Due to its high sensitivity, the Five-hundred-meter Aperture Spherical radio Telescope (FAST) is an ideal instrument for studying pulsar polarizations (e.g., \citealt{2011IJMPD..20..989N,2022ApJ...934...57S,2023RAA....23j4002W,2024ApJ...964....6W}). High-sensitivity observations maybe detect weak emissions from pulsars, extending the range of pulse windows, which can enhance our understanding of pulsar emission geometry and location. Motivated by this, we conducted a polarization study of a sample of pulsars with IP emissions. In this paper, we observed 23 pulsars exhibiting IP emissions and applied the RVM to determine their geometries. The observations and data processing are presented in Section~\ref{sec:obs}. Section~\ref{sec:res} presents the results, and Section~\ref{sec:discussion} discusses and summarizes our results.

\section{OBSERVATIONS AND DATA PROCESSING}\label{sec:obs}

FAST was completed in September 2016, and the 19-beam receiver, which covers a frequency range of 1.05–1.45 GHz, was installed in 2018~\citep{lwq+2018,jyg+2019}. We used the central beam of the 19-beam receiver to observe 23 pulsars with IP emissions. The data were recorded with four polarizations, 8-bit samples, a 49.152\,$\mu$s sampling interval, and 4096 frequency channels. A polarization calibration noise signal was recorded either before or after the observation for calibration purposes. The observational details are shown in Table~\ref{obs}.

\begin{table}
\footnotesize
\centering
\caption{Observation parameters for 23 pulsars.}
\label{obs}
\begin{tabular}{ccccc}
\hline
Name & Period &  Obs. duration  & RM  & MJD 
\\
     & (s) &  (s) &  ${\rm (rad\,m^{-2}})$ &
\\
\hline
J0304+1932 & 1.38  & 1738 &  $-7.06 (2)$ & 59242.46
\\
J0627+0706 & 0.47 & 14258 & 210.31 (5)  & 60231.82  
\\
J0826+2637 & 0.53 & 3601 & 6.55 (3) &  58812.82
\\
J0953+0755 & 0.25 & 3421 & 1.57(2) & 59522.99  
\\
J1755$-$0903 & 0.19 & 1861 & 93.0 (2) & 60345.07
\\
J1816$-$0755 & 0.21 & 855 & 29.5 (2)  &  60277.27
\\
J1825$-$0935 & 0.76 & 1049 & 68.67 (3)  &  59441.54
\\
J1842+0358 & 0.23 & 1141 & 40.2 (3) & 60278.29
\\
J1843$-$0702 & 0.19 & 1741 & 195.8 (3)  &  60222.44
\\
J1849+0409 & 0.76 & 4020 &  28.92 (9) & 60218.51
\\
J1852$-$0118 & 0.45 & 2940 & 714.48 (8) & 60184.54  
\\
J1909+0749 & 0.23 & 1140 & $-238.1 (5)$  &  60284.27
\\
J1913+0832 & 0.13 & 615 & 495.4 (1)  &   59225.12
\\
J1918+1541 & 0.37 & 1667 & 0.4 (5) &  60301.23
\\
J1926+0737 & 0.31 & 1440 & 351 (2) &   60272.33
\\
J1932+1059 & 0.22 & 616&  $-5.21(2)$ & 59257.21 
\\
J1935+2025 & 0.08 & 1741 & 24.34 (7) &  60167.73
\\
J1946+1805 & 0.44 & 3301 & $-44.12 (2)$  & 59150.48
\\
J1952+3252 & 0.03 & 3501 & $-176.99 (7)$  &  59872.34
\\
J2023+5037 & 0.37 & 3479 & 46.8 (2)  & 59529.43
\\
J2032+4127 & 0.14 & 2941 & 215.7 (1) &  60184.64
\\
J2047+5029 & 0.44 & 2941& $-121.58 (7)$  & 60182.68
\\
J2208+4056 & 0.63 & 601 & $-40.8 (2)$  & 59140.59
\\ 
\hline
 \end{tabular}
\end{table}

For each pulsar, the observation data were folded using the {\tt DSPSR} software~\citep{vb2011}. Then, we used the {\tt PAZ} and {\tt PAZI} routines in the {\tt PSRCHIVE} software~\citep{hvm2004} to remove radio-frequency interference (RFI). The calibration observations were folded according to their period, and the pulsar observations were calibrated using {\tt PAC}. We measured the rotation measure (RM) of these pulsars using the {\tt RMFIT} tool, and the data were RM-corrected before summing over frequency. Our measured RM values for each pulsar are listed in the fourth column of Table~\ref{obs}.

Following the method of~\citet{jk2019}, we used a modified form of the original model of~\citet{rc1969} to fit the pulsars with IP emissions from the opposite magnetic poles as the MPs:
\begin{equation}
\label{eq50}
\psi = \arctan ( \frac{\sin\alpha\sin(\phi-\phi_{0}- \Delta )}{\sin\zeta\cos\alpha-\cos\zeta\sin\alpha\cos(\phi-\phi_{0} - \Delta)}) + \psi_{0}, 
\end{equation}
where the term $\Delta = 4 h_{\rm MI} /R_{\rm lc} $ represents the difference in emission height between the two poles ($h_{\rm MI}$). 
Note that PPAs are defined as increasing counter-clockwise on the sky. 
For the pulsars with IP emissions from the same magnetic poles as the MPs, we do not consider the difference in emission height between IP and MP, and Equation~\ref{eq60} is used.

For pulsars exhibiting OPMs, the OPMs do not always exhibit a 90$^\circ$ separation, which may be attributed to magnetospheric propagation effects. Therefore, we treated the degrees of OPM jumps as free parameters in the RVM fitting, while the pulse phases of OPM jumps were determined through visual inspection. 
We modified Equations~\ref{eq60} and \ref{eq50} as follows: 
\begin{equation}
\label{eq6}
\psi =\left\{
\begin{array}{ll}
\arctan ( \frac{\sin\alpha\sin(\phi-\phi_{0} )}{\sin\zeta\cos\alpha-\cos\zeta\sin\alpha\cos(\phi-\phi_{0})}) + \psi_{0}, & \\ \phi  < \phi_{\rm OPM} \\
\arctan ( \frac{\sin\alpha\sin(\phi-\phi_{0})}{\sin\zeta\cos\alpha-\cos\zeta\sin\alpha\cos(\phi-\phi_{0})}) + \psi_{0} + \psi_{\rm OPM}, & \\ \phi \geq \phi_{\rm OPM}  \\
\end{array} \right.
\end{equation}
and 
\begin{equation}
\label{eq5}
\psi =\left\{
\begin{array}{ll}
\arctan ( \frac{\sin\alpha\sin(\phi-\phi_{0}- \Delta )}{\sin\zeta\cos\alpha-\cos\zeta\sin\alpha\cos(\phi-\phi_{0}- \Delta)}) + \psi_{0}, & \\ \phi  < \phi_{\rm OPM} \\
\arctan ( \frac{\sin\alpha\sin(\phi-\phi_{0}- \Delta)}{\sin\zeta\cos\alpha-\cos\zeta\sin\alpha\cos(\phi-\phi_{0}- \Delta)}) + \psi_{0} + \psi_{\rm OPM}, & \\ \phi \geq \phi_{\rm OPM}  \\
\end{array} \right.
\end{equation}
where $\psi_{\rm OPM}$ and $\phi_{\rm OPM}$ represent the OPM jump degree and phase, respectively. Note that some pulsars exhibit arbitrary PA jumps that significantly deviate from $90^{\circ}$, and these jumps are not included in our RVM fitting.
Following the method of \citet{jk2019}, we performed Markov Chain Monte Carlo (MCMC) fitting using the Python package EMCEE\citep{fhl+2013} to determine the values of the unknown parameters.

\section{RESULTS}\label{sec:res}

\begin{figure*}
\begin{center}
\begin{tabular}{c}
\includegraphics[width=0.23\textwidth]{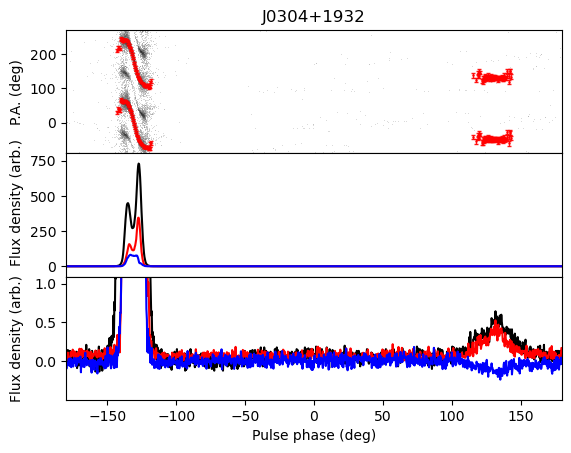}
\includegraphics[width=0.23\textwidth]{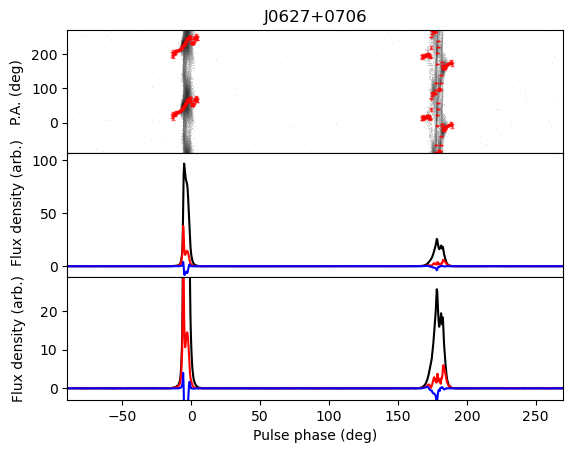}
\includegraphics[width=0.23\textwidth]{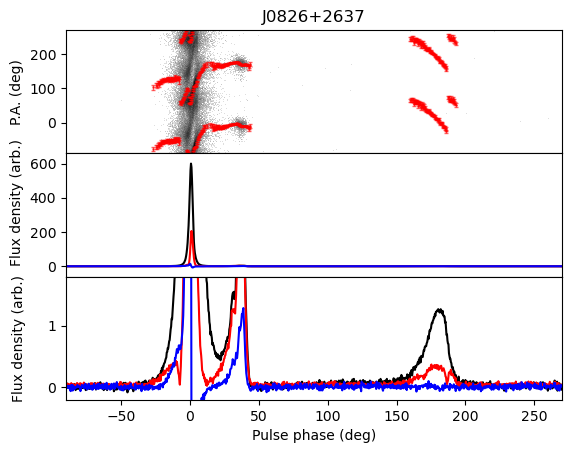}
\includegraphics[width=0.23\textwidth]{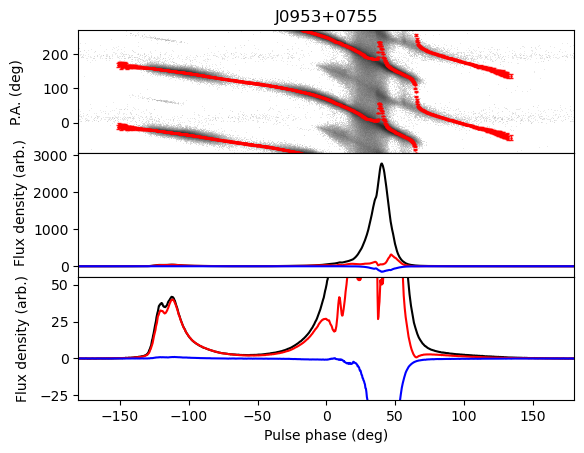} \\
\includegraphics[width=0.23\textwidth]{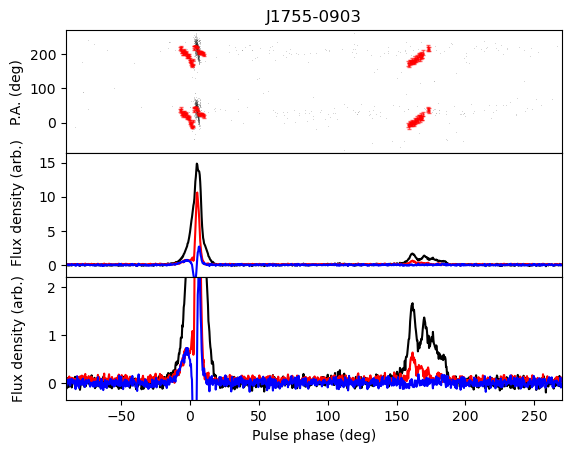} 
\includegraphics[width=0.23\textwidth]{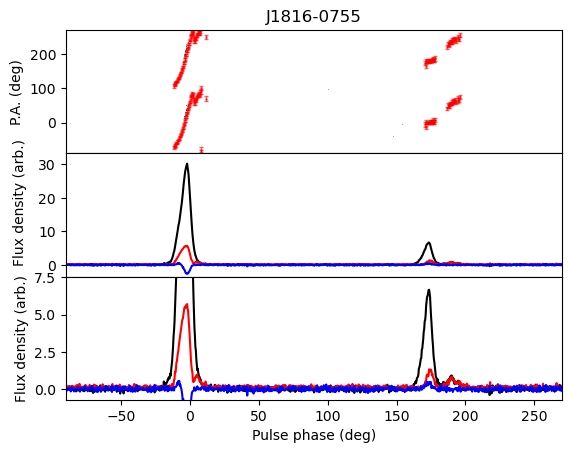}
\includegraphics[width=0.23\textwidth]{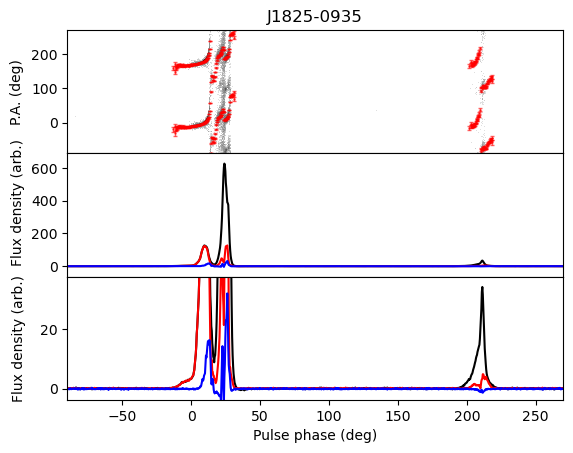}
\includegraphics[width=0.23\textwidth]{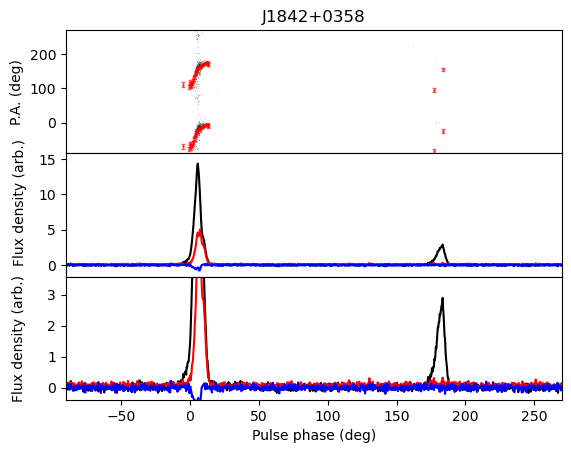}\\
\includegraphics[width=0.23\textwidth]{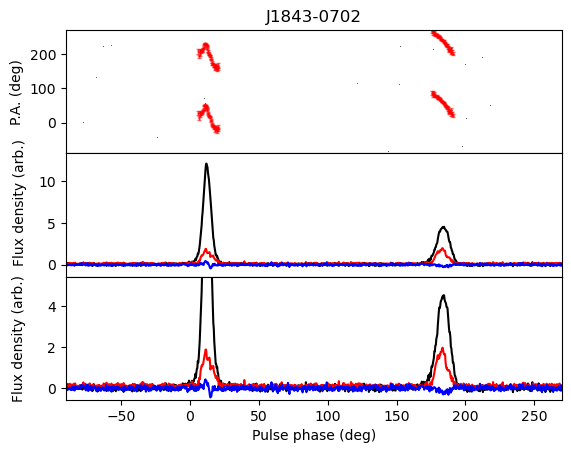}
\includegraphics[width=0.23\textwidth]{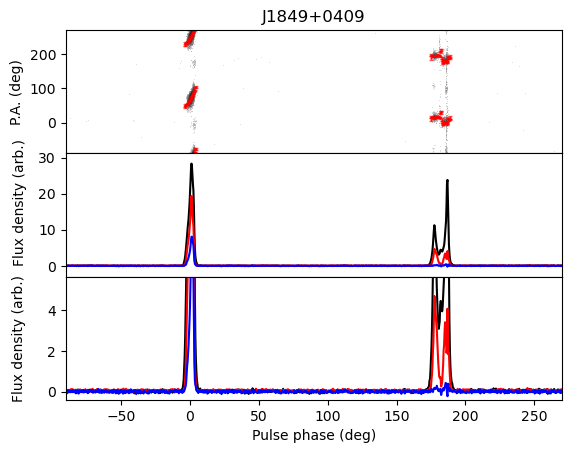}
\includegraphics[width=0.23\textwidth]{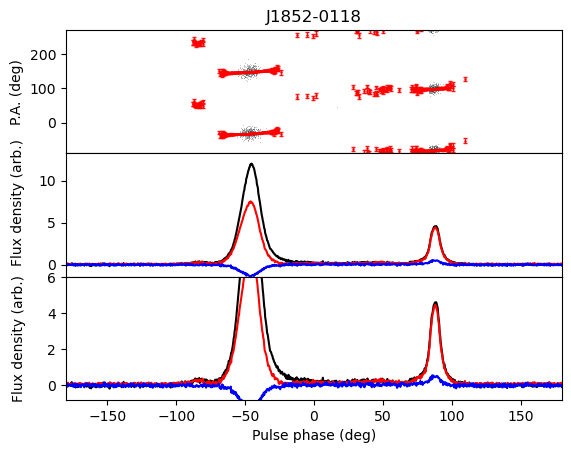}
\includegraphics[width=0.23\textwidth]{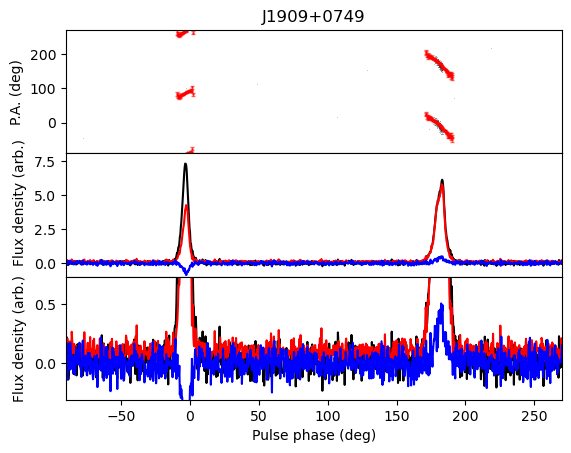}\\
\includegraphics[width=0.23\textwidth]{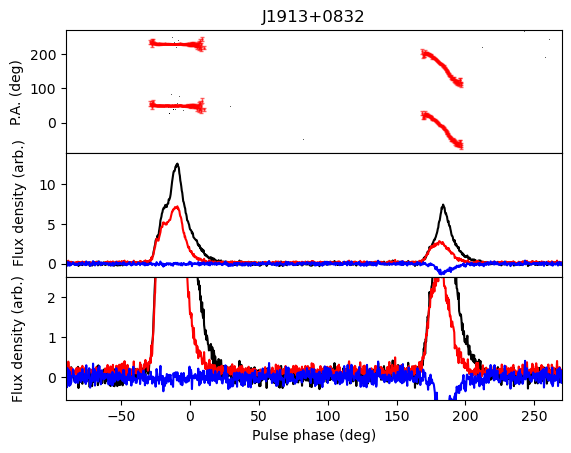}
\includegraphics[width=0.23\textwidth]{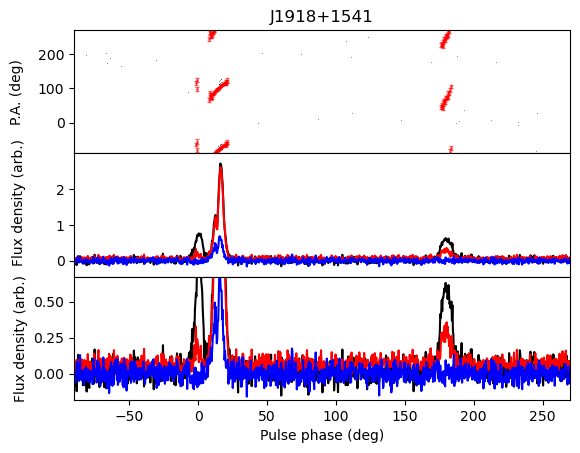}
\includegraphics[width=0.23\textwidth]{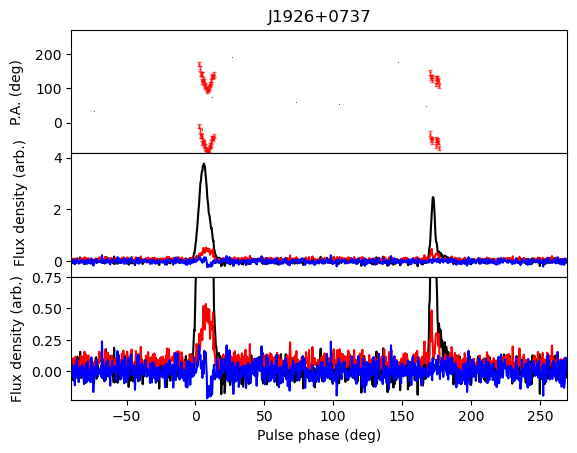}
\includegraphics[width=0.23\textwidth]{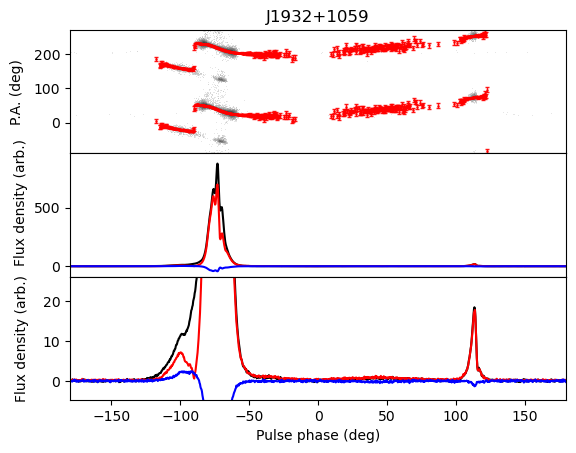}\\
\includegraphics[width=0.23\textwidth]{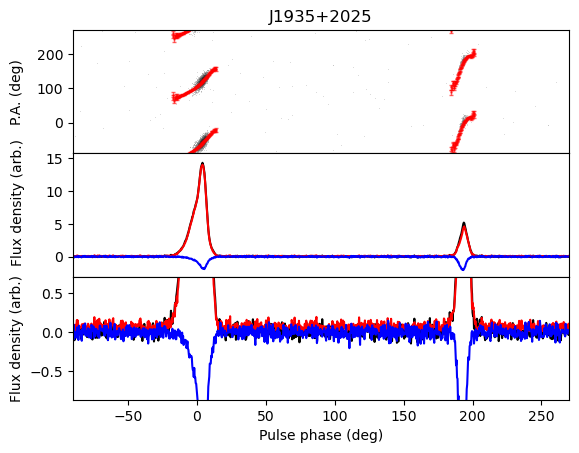}
\includegraphics[width=0.23\textwidth]{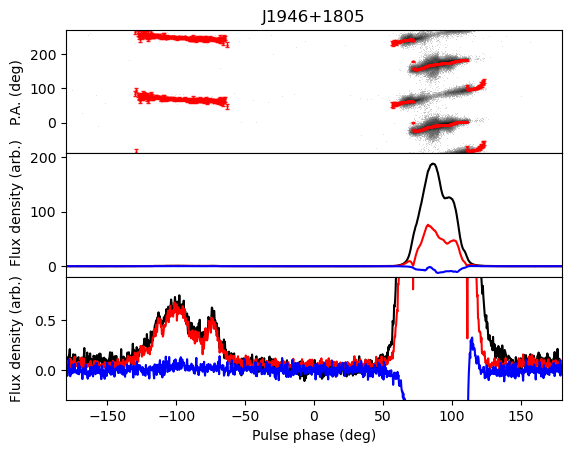}
\includegraphics[width=0.23\textwidth]{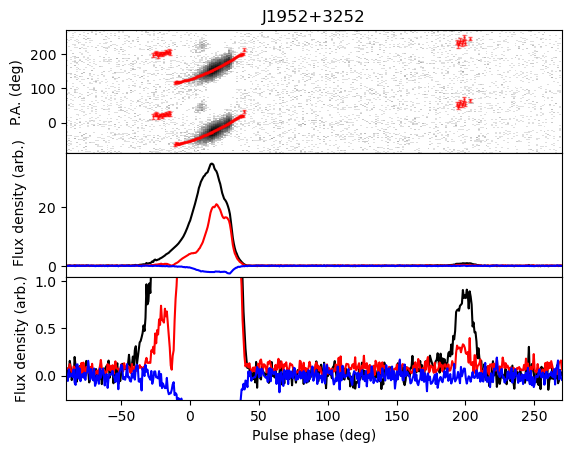}
\includegraphics[width=0.23\textwidth]{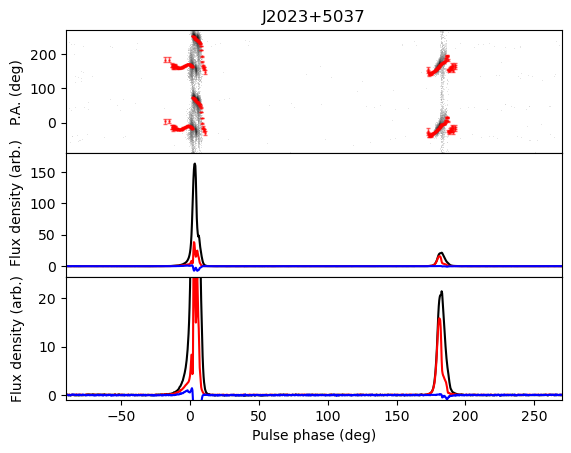}\\
\includegraphics[width=0.23\textwidth]{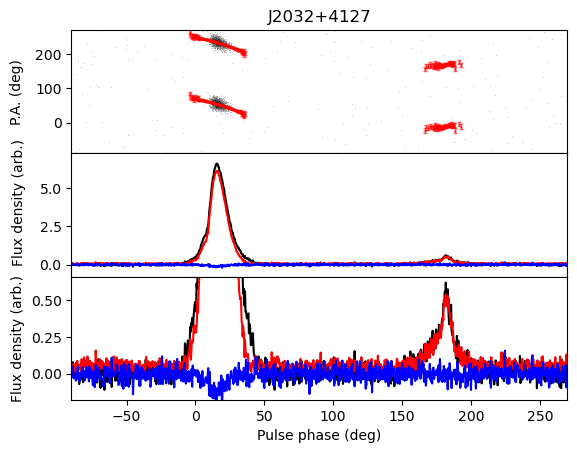}
\includegraphics[width=0.23\textwidth]{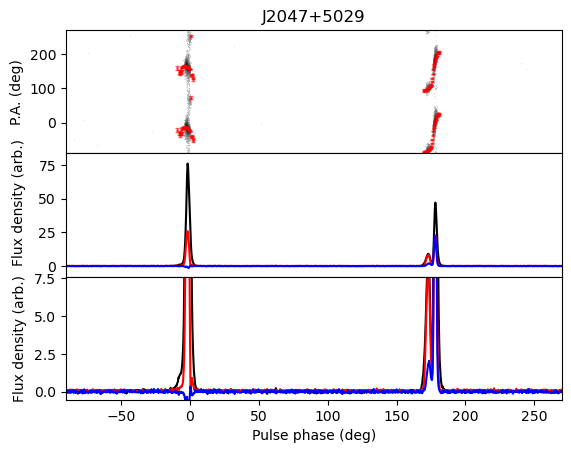}
\includegraphics[width=0.23\textwidth]{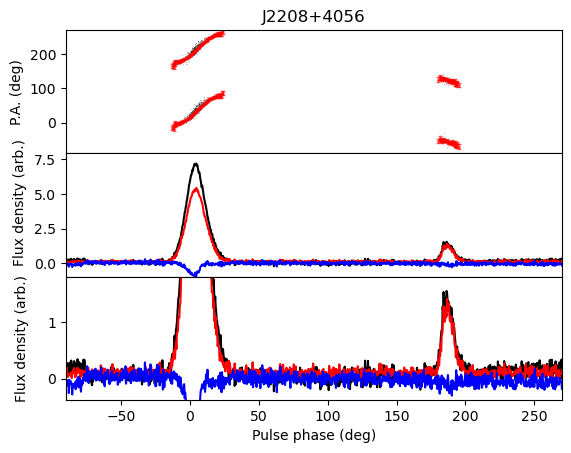}
\end{tabular}
\end{center}
\caption{Polarization profiles for 23 pulsars with IP emissions. The PPAs of the average profile (red dots), superimposed on the PPAs of single pulses (black dots), are shown in the upper panels. Note that only PPAs with uncertainties smaller than $7^{\circ}$ are included. The total intensities (black lines), linear polarizations (red lines), and circular polarizations (blue lines) are shown in the middle panels. The zoomed-in polarization profiles are presented in the bottom panels.}
\label{fig:prof}
\end{figure*}

\begin{figure*}
\begin{center}
\begin{tabular}{c}
\includegraphics[width=0.23\textwidth]{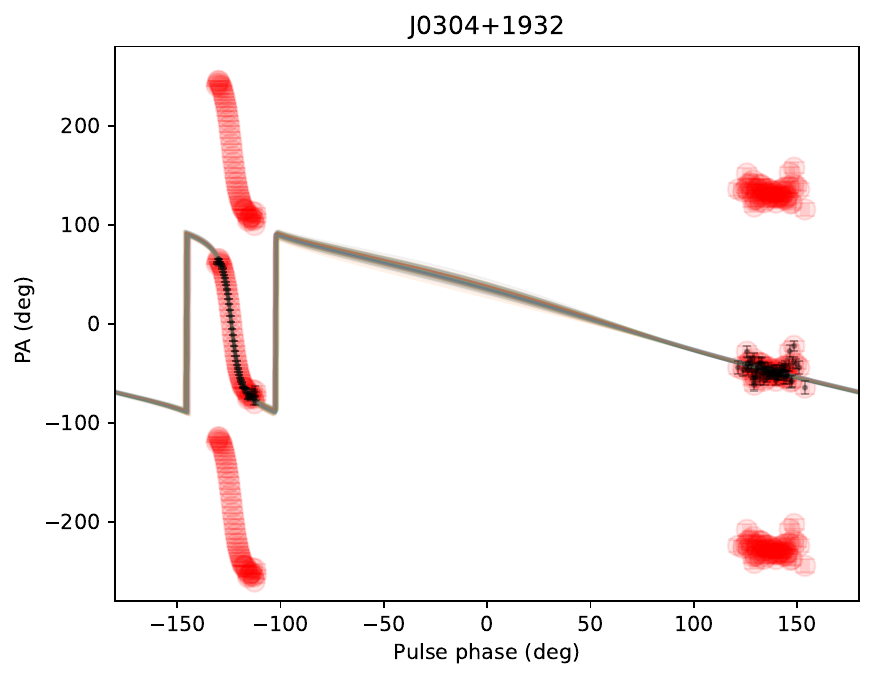}
\includegraphics[width=0.23\textwidth]{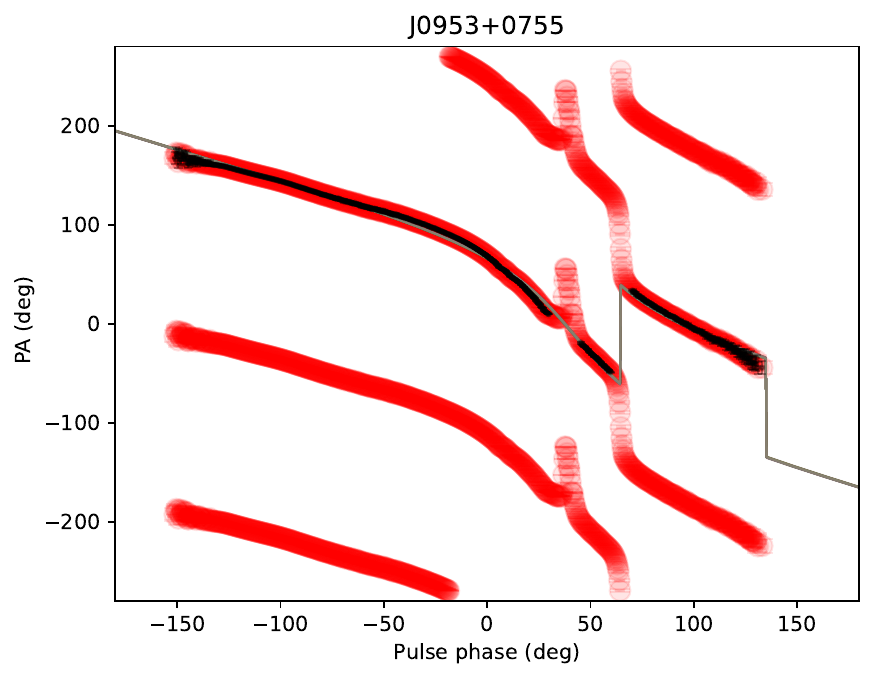}
\includegraphics[width=0.23\textwidth]{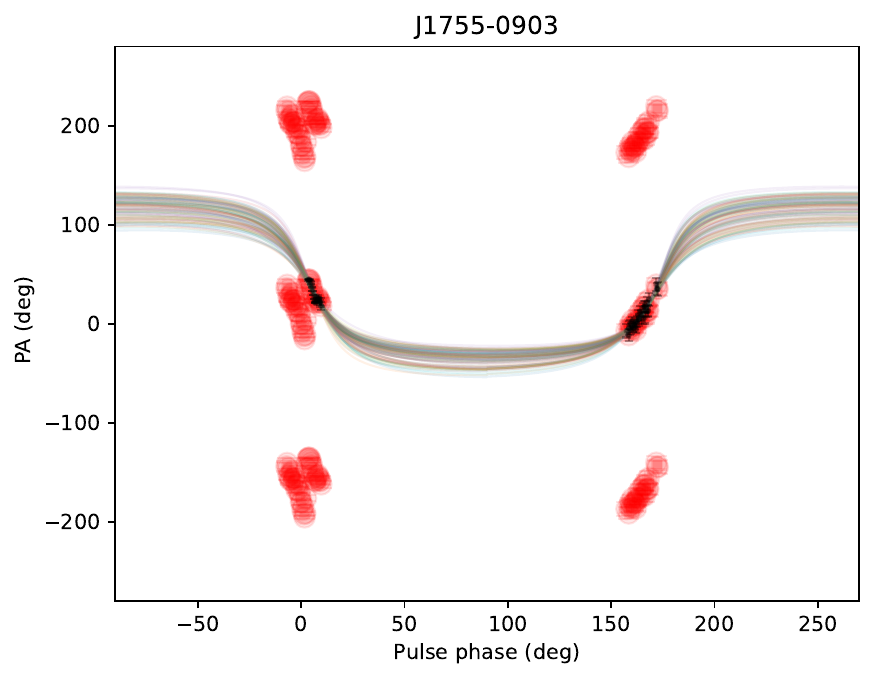}
\includegraphics[width=0.23\textwidth]{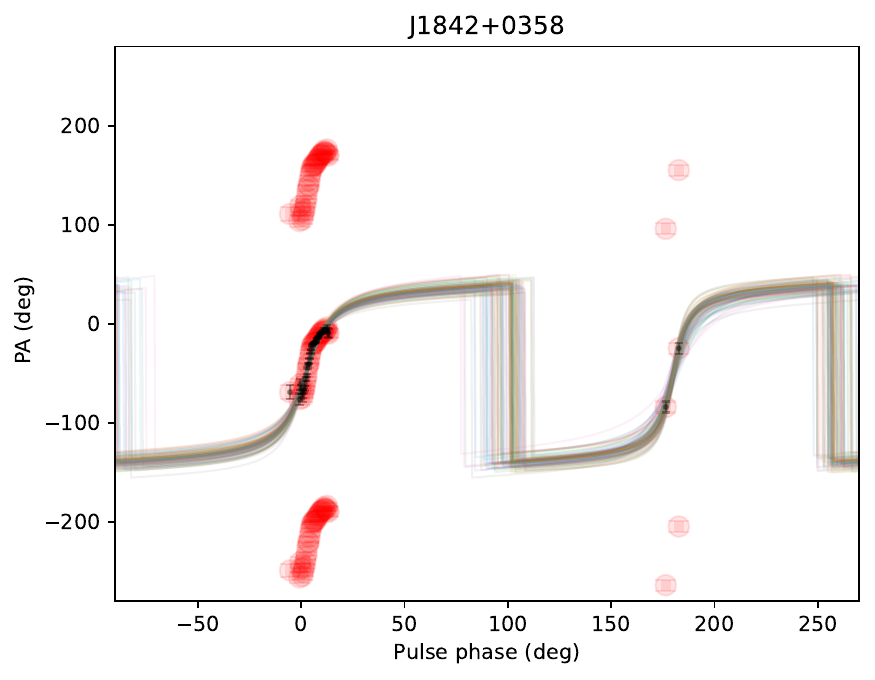}\\
\includegraphics[width=0.23\textwidth]{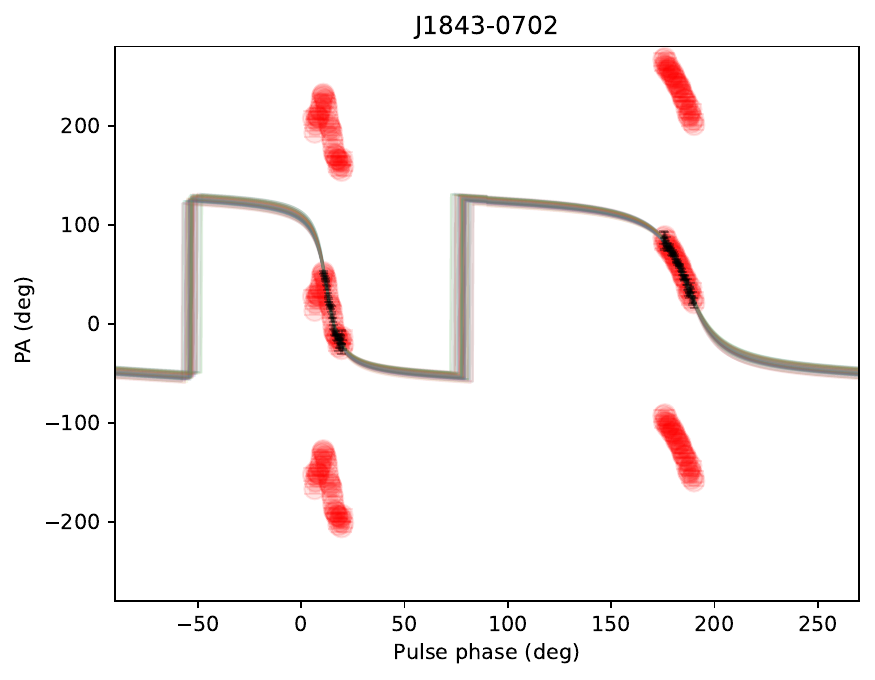}
\includegraphics[width=0.23\textwidth]{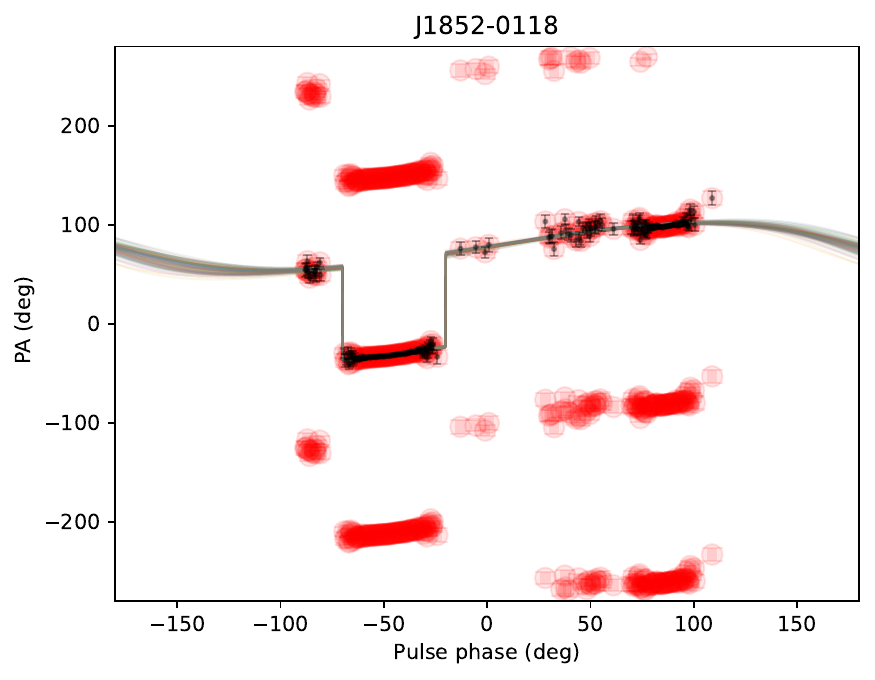}
\includegraphics[width=0.23\textwidth]{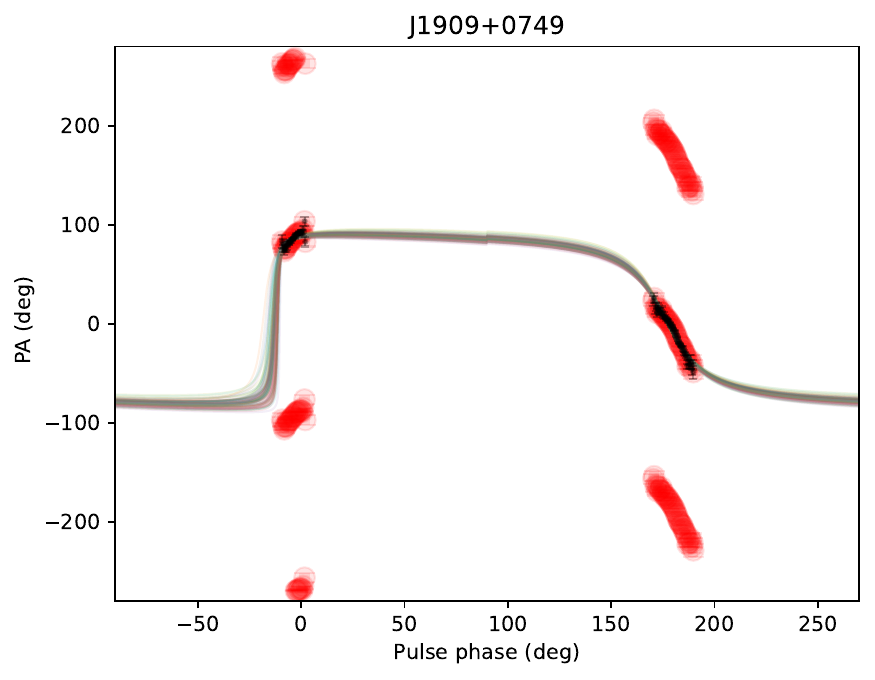}
\includegraphics[width=0.23\textwidth]{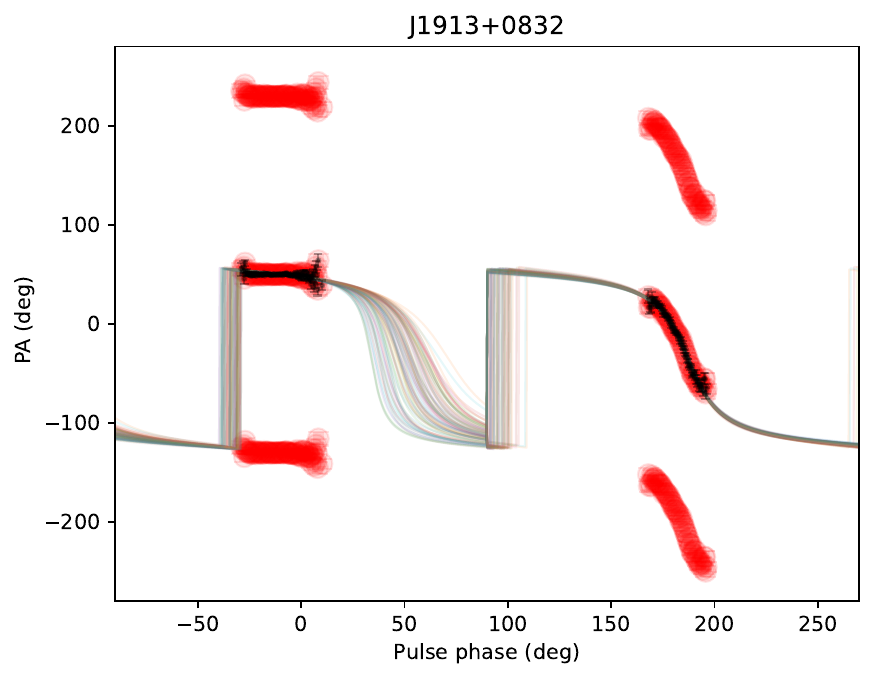}\\
\includegraphics[width=0.23\textwidth]{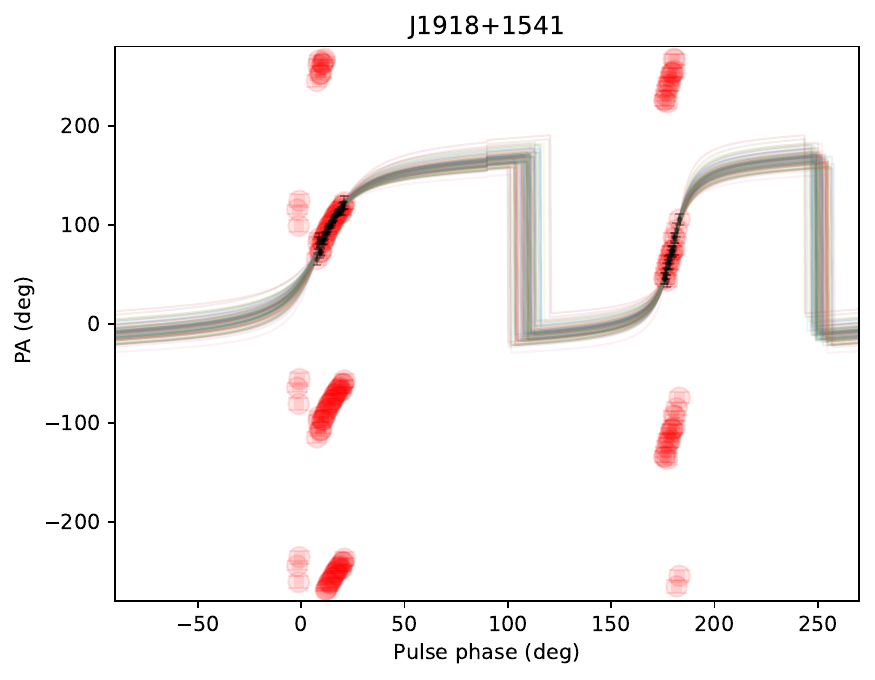}
\includegraphics[width=0.23\textwidth]{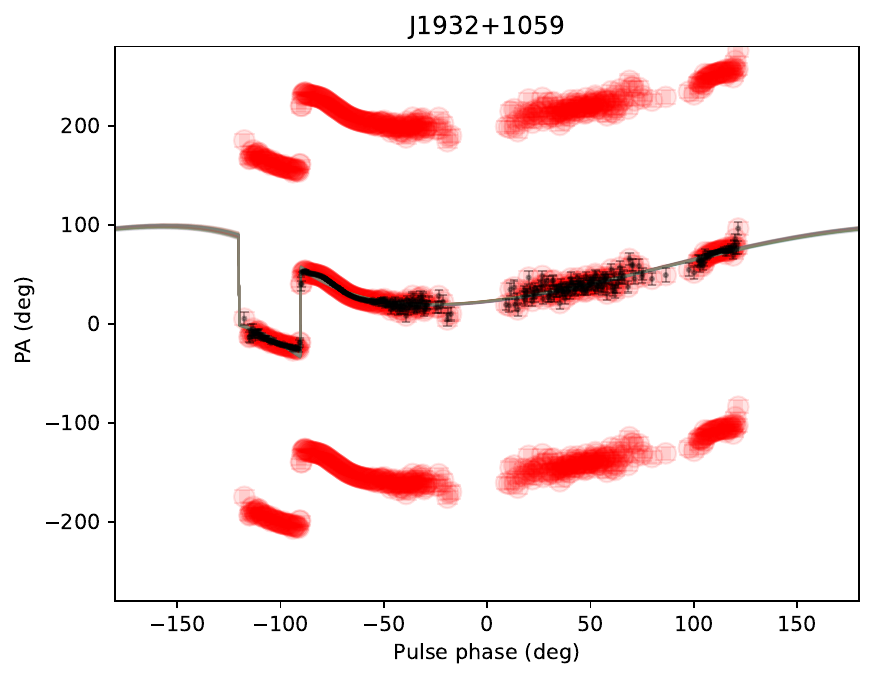}
\includegraphics[width=0.23\textwidth]{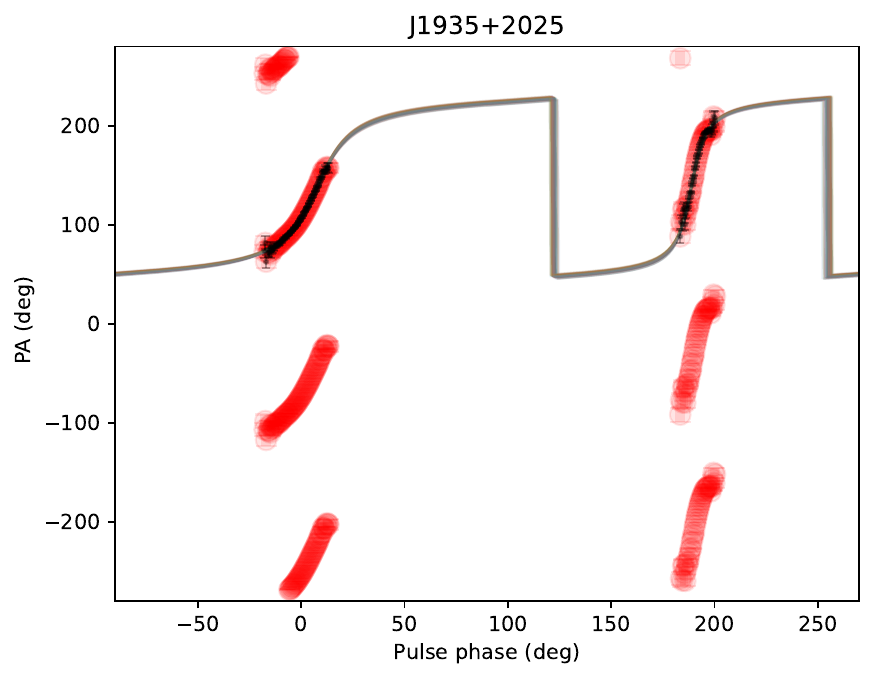}
\includegraphics[width=0.23\textwidth]{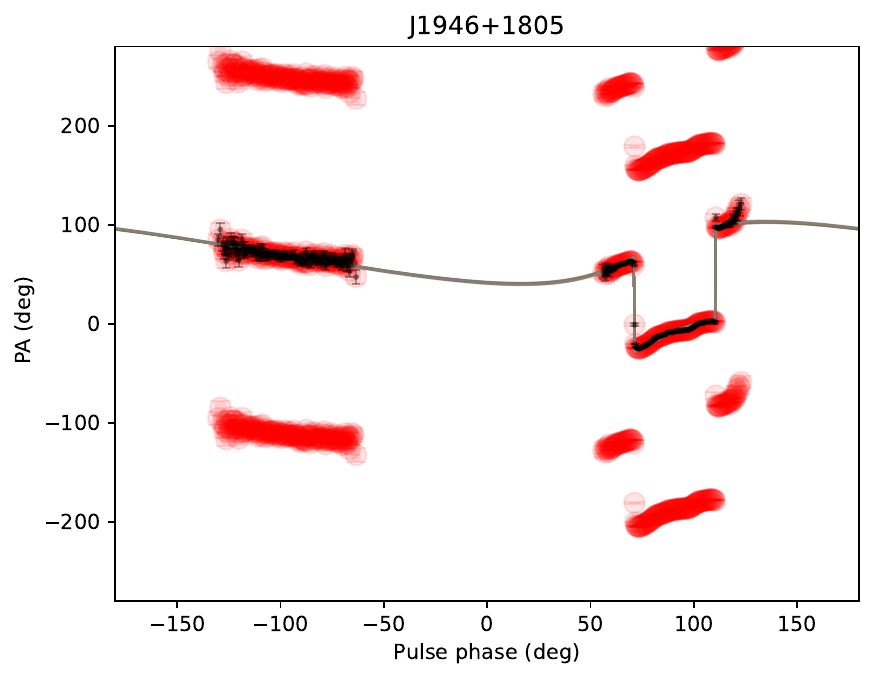}\\
\includegraphics[width=0.23\textwidth]{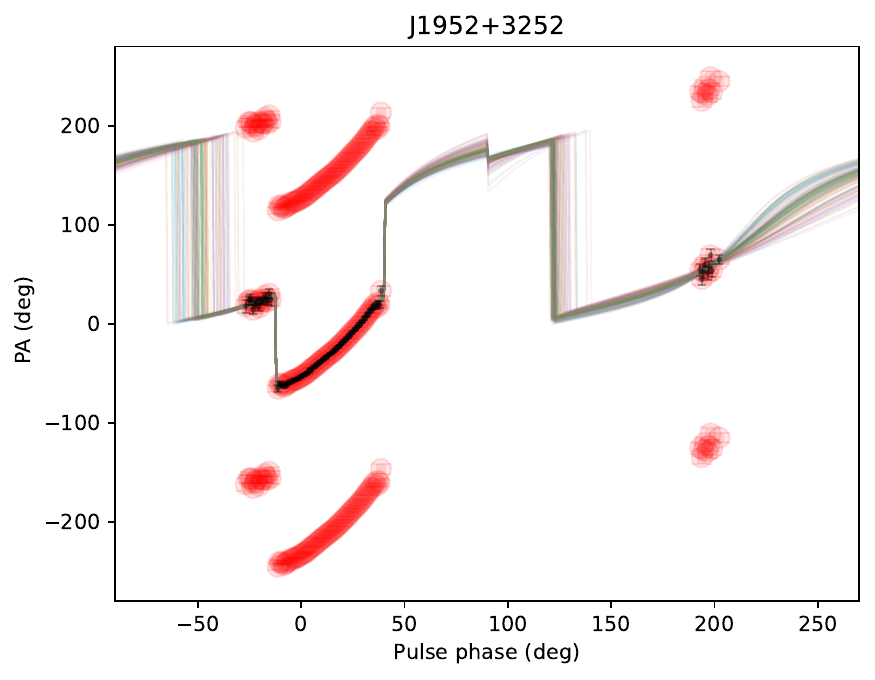}
\includegraphics[width=0.23\textwidth]{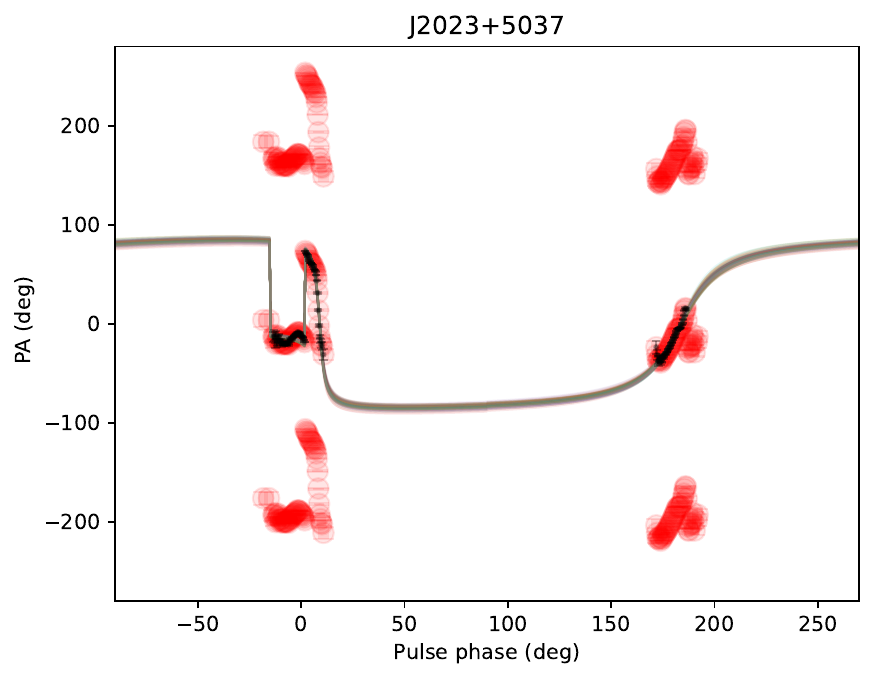}
\includegraphics[width=0.23\textwidth]{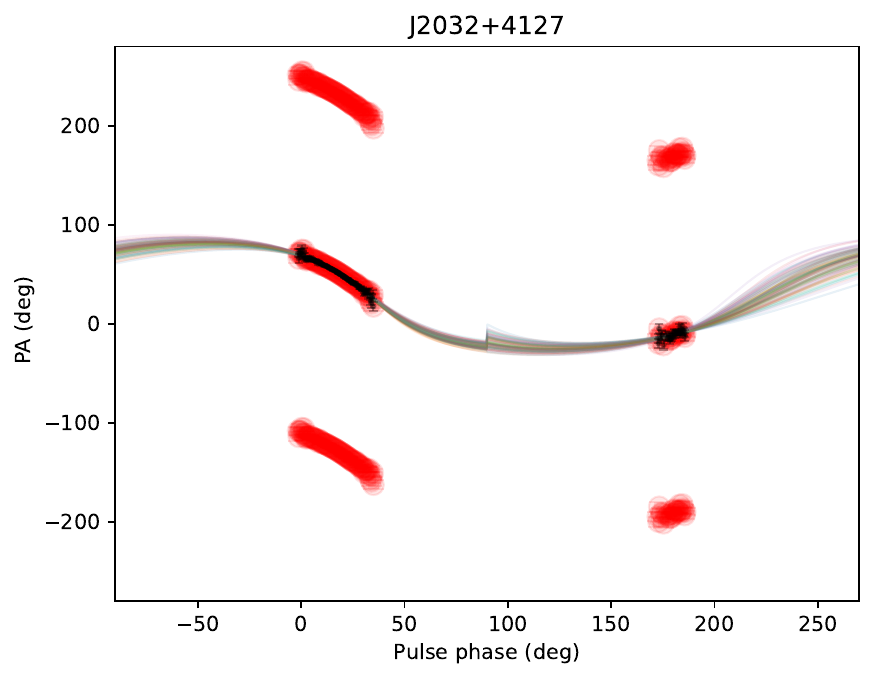}
\includegraphics[width=0.23\textwidth]{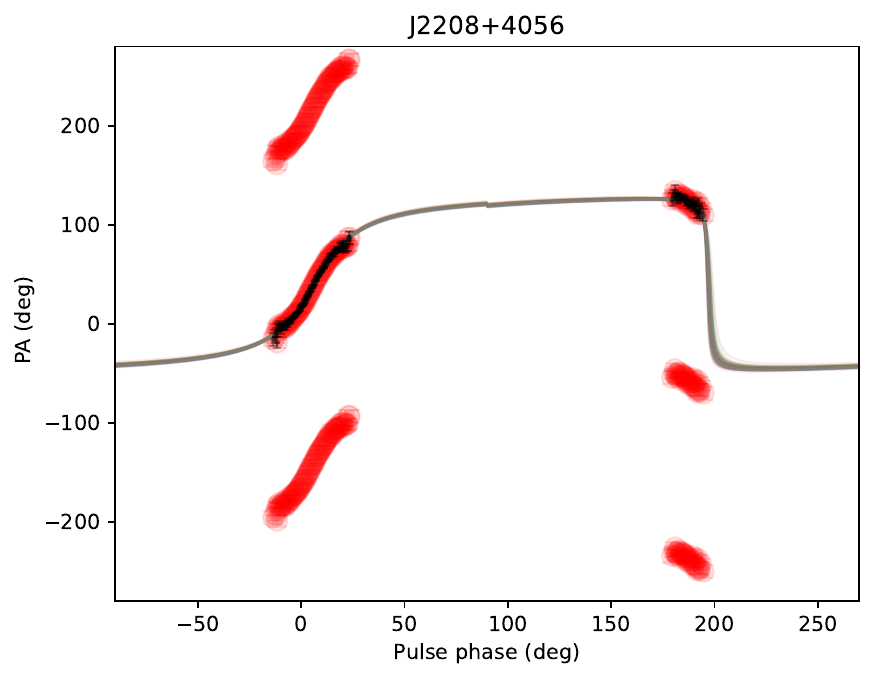}
\end{tabular}
\end{center}
\caption{PPA swings for 16 pulsars which conform the RVM. Each panel shows the observed PPAs (red dots) and the RVM fitting results (gray solid lines) across the entire pulse longitude with error bars determined by the fitting process. The black dots represent the PPA points included in the fits.}
\label{fig:pafit}
\end{figure*}

\begin{table*}
\setlength{\tabcolsep}{1pt}
\footnotesize
\renewcommand{\arraystretch}{1.5}
\centering
\caption{Results of RVM fitting for 16 pulsars which conform the RVM. The pulsar name is shown in the first column, the superscripts of BCW91, EW01, YZK23, JK19, SJK21 and WHX23 refer to \citet{bcw1991}, \citet{ew2001}, \citet{yzk+2023}, \citet{jk2019}, \citet{sjk+2021} and \citet{2023RAA....23j4002W}, respectively. The values of $\alpha_{M}$, $\beta_{M}$, and $\phi_{\rm M}$ for the main pulses are shown in the second, third, and forth columns, respectively. The $\zeta$ is shown in the fifth column. The values of $\alpha_{I}$, $\beta_{I}$, and $\phi_{\rm I}$ for the interpulses are shown in the sixth, seventh, and eighth columns, respectively. The values of $\Delta$ is shown in the ninth column, and the OPM jumps are shown in the tenth column. Note that for the pulsars with the same pole IP emissions, we present only the fitting results for MPs.}
\label{fit}
\begin{tabular}{ccccccccccc}
\hline
Name & $\alpha_{M}$ & $\beta_{M}$ & $\phi_{M}$ & $\zeta$ & $\alpha_{I}$ & $\beta_{I}$ & $\phi_{I}$ & $\Delta$ & jump 
\\
\hline
J0304+1932 & $141.02_{-4.06}^{+4.87}$  & $2.01_{-5.63}^{+6.75}$ & $-123.88_{-0.16}^{+0.17}$ & $143.03_{-3.90}^{+4.68}$  &  &  & &  & $-$
\\
$J0304+1932^{\rm YZK23}$ & $127\pm15$ & $2.8\pm0.4$
\\
J0953+0755 & $177.48_{-1.60}^{+1.27}$ & $1.01_{-1.86}^{+1.48}$ & $42.43_{-0.31}^{+0.30}$ & $178.49_{-0.96}^{+0.76}$ &  &  &  &   & $99.96_{-0.49}^{+0.48}$ 
\\
$J0953+0755^{\rm BCW91}$ & $174\pm30$ & $2.5\pm15$ &   &  & 
\\
J1755$-$0903 & $76.92_{-1.90}^{+1.99}$ & $12.97_{-2.29}^{+2.30}$ & $3.56_{-1.90}^{+2.50}$ & $89.89_{-1.27}^{+1.16}$ & $103.08_{-1.99}^{+1.90}$ & $-13.19_{-2.36}^{+2.23}$ & $173.66_{-5.14}^{+4.0}$  & $-9.90_{-4.78}^{+3.21}$   & $-$   
\\
$J1755$-$0903^{\rm SJK21}$ & $83 (1)$ & $5(2)$ & $-1(1)$ & $87(2)$ & $97(2)$ & $-10(2)$ & $174(2)$ & $-5(2)$
\\
J1842$+$0358 & $91.24_{-1.06}^{+0.76}$ & $-8.61_{-1.50}^{+1.07}$ & $2.05_{-0.81}^{+0.92}$ & $82.63_{-1.06}^{+0.75}$ & $88.76_{-0.76}^{+1.06}$ & $-6.13_{-1.30}^{+1.30}$ & $179.87_{-1.13}^{+1.22}$& $-2.18_{-0.79}^{+0.80}$ & $-$  
\\
J1843$-$0702 & $93.20_{-0.32}^{+0.32}$ & $4.46_{-0.37}^{+0.38}$ & $12.23_{-0.21}^{+0.22}$ & $97.66_{-0.19}^{+0.20}$ & $86.80_{-0.32}^{+0.32}$ & $10.86_{-0.37}^{+0.38}$ & $187.12_{-0.40}^{+0.39}$ & $-5.11_{-0.34}^{+0.32}$ & $-$ 
\\
$J1843$-$0702^{\rm SJK21}$ & $91(1)$ & $6(1)$ & $3(1)$ & $97(1)$ & $88(1)$ & $9 (1)$ & $180(2)$ & $-3(1)$
\\
J1852$-$0118 & $163.85_{-4.79}^{+6.90}$ & $25.02_{-20.13}^{+21.48}$
& $0.05_{-5.45}^{+4.64}$ & $41.17_{-19.55}^{+20.34}$ &  &  &  &   & $94.09_{-1.20}^{+1.44}$
\\
$J1852$-$0118^{\rm WHX23}$ & $27_{-18}^{+11}$ & $90_{-77}^{+0}$ &  & & 
\\
J1909+0749 & $97.67_{-0.27}^{+0.31}$ & $-1.19_{-0.52}^{+0.50}$ & $-12.73_{-1.46}^{+1.05}$ & $96.48_{-0.44}^{+0.39}$ & $82.33_{-0.31}^{+0.27}$ & $14.15_{-0.54}^{+0.47}$ & $176.26_{-1.69}^{+1.62}$  & $8.99_{-0.86}^{+1.23}$ & $-$ 
\\
$J1909+0749^{\rm SJK21 }$ & $112(6)$ & $-32(8)$ & $-4(4)$ & $80(7)$ & $68(6)$ & $12(9)$ & $185(6)$ & $-9(5)$
\\
$J1909+0749^{\rm WHX23}$ &  & & & & $82$ & $15.6$ & $180.1$
\\
J1913+0832 & $91.00_{-1.57}^{+1.29}$ & $9.48_{-2.08}^{+2.09}$ & $50.37_{-8.64}^{+8.56}$  & $100.48_{-1.37}^{+1.64}$  & $89.00_{-1.29}^{+1.57}$ & $11.48_{-1.88}^{+2.27}$  & $187.12_{-12.19}^{+12.22}$ & $-43.25_{-8.60}^{+8.72}$ & $-$ 
\\
$J1913+0832^{\rm WHX23}$ & $85.3_{-1.0}^{+0.5}$  & $9.3_{-1.2}^{+2.1}$ & $-161.2_{-0.6}^{+1.4}$ & & &
\\
J1918+1541 & $93.20_{-0.73}^{+0.71}$ & $-12.85_{-0.88}^{+0.91}$ & $9.37_{-1.32}^{+1.61}$ & $80.35_{-0.50}^{+0.57}$ &  $86.80_{-0.71}^{+0.73}$ & $-6.45_{-0.87}^{+0.93}$ & $179.99_{-1.54}^{+1.70}$ & $-9.38_{-0.80}^{+0.56}$ &  $-$ 
\\
$J1918+1541^{\rm SJK21}$ & $93(1)$ & $-12(1)$ & $-8(1)$ & $81(1)$ & $87(1)$ & $-6(1)$ & $164(2)$ & $-8(1)$
\\
$J1918+1541^{\rm WHX23}$ & $98_{-7}^{+34}$ & $-27_{-63}^{+12}$ & $-186.1_{-2.8}^{+4.1}$ & & & & 
\\
J1932+1059 & $31.23_{-1.07}^{+0.97}$ & $23.08_{-2.52}^{+2.41}$   & $-92.22_{-0.69}^{+0.67}$ & $54.31_{-2.28}^{+2.21}$ &  &  &  &   &  $89.29_{-1.59}^{+1.57}$
\\
$J1932+1059^{\rm EW01}$ & $35.97\pm0.95$ & $25.55 \pm 0.87$ & 
\\
J1935+2025 & $93.59_{-0.17}^{+0.17}$ & $-12.52_{-0.22}^{+0.22}$ & $7.58_{-0.25}^{+0.26}$ & $81.07_{-0.14}^{+0.14}$ & $86.41_{-0.17}^{+0.17}$ & $-5.34_{-0.22}^{+0.22}$  & $188.64_{-0.30}^{+0.31}$ & $1.06_{-0.17}^{+0.17}$ & $-$  
\\
$J1935+2025^{\rm JK19}$ & $93(2)$ & $-13$ & $-13(2)$ & $80(1)$ & $87(2)$ & $-7$ & $166(1)$ & $-1(1)$
\\
$J1935+2025^{\rm WHX23}$ & & & & & $87_{-1}^{+1}$ & $-6.2_{-0.6}^{+0.7}$ & $-174.3_{-0.7}^{+0.3}$
\\
J1946+1805 & $5.06_{-2.50}^{+2.95}$ & $-4.69_{-6.26}^{+5.66}$ & $-104.47_{-0.52}^{+0.52}$ &  $170.25_{-5.74}^{+4.83}$  &  &  &  &   &  $94.49_{-0.76}^{+0.76}$
\\ 
$J1946+1805^{\rm WHX23}$ & $158_{-17}^{+47}$ & $16_{-15}^{+12}$ &  &   & 
\\
J1952+3252 & $82.39_{-11.07}^{+5.20}$ & $-24.12_{-14.65}^{+7.16}$ & $28.94_{-1.24}^{+1.77}$ & $58.27_{-9.59}^{+4.92}$ & $97.61_{-5.20}^{+11.07}$ & $-39.34_{-10.91}^{+12.11}$ & $224.84_{-7.10}^{+15.00}$ & $15.90_{-6.99}^{+14.90}$ & $93.95_{-1.37}^{+1.23}$  
\\
$J1952+3252^{\rm WHX23}$ & $19_{-16}^{+114}$ & $-6.7_{-25.9}^{+5.9}$ & $27.4_{-16.5}^{+0.7}$
\\

J2023+5037 & $82.09_{-0.44}^{+0.40}$ & $1.88_{-0.64}^{+0.59}$  & $8.92_{-0.11}^{+0.10}$ & $83.97_{-0.47}^{+0.43}$ & $97.91_{-0.40}^{+0.44}$ & $-13.94_{-0.62}^{+0.62}$  & $183.59_{-0.66}^{+0.66}$ & $-5.33_{-0.65}^{+0.65}$  & $95.64_{-1.90}^{+1.93}$ & 
\\

J2032+4127 & $53.18_{-3.08}^{+2.37}$ & $28.78_{-5.85}^{+4.90}$ & $32.86_{-1.26}^{+1.14}$ & $81.96_{-4.97}^{+4.28}$ &  $126.82_{-2.37}^{+3.08}$ & $-44.86_{-5.51}^{+5.27}$ & $223.91_{-6.99}^{+9.36}$ & $11.05_{-6.88}^{+9.29}$ & $-$ 
\\
$J2032+4127^{\rm WHX23}$ & $54$ & $40.5$ & $17.7$
\\ 
J2208+4056 & $98.51_{-0.13}^{+0.14}$ &   $-15.89_{-0.20}^{+0.21}$  & $6.52_{-0.32}^{+0.32}$ &  $82.62_{-0.15}^{+0.15}$ & $ 81.49_{-0.14}^{+0.13}$ & $ 1.30_{-0.21}^{+0.20}$ &  $197.21_{-0.53}^{+0.62}$ & $10.69_{-0.42}^{+0.52}$ & $-$  
\\ 
$J2208+4056^{\rm WHX23}$ & $120_{-11}^{+46}$ & $-12.8_{-1.5}^{+9.3}$ & $0.8_{-0.3}^{+0.3}$
\\         
\hline
 \end{tabular}
\end{table*}

\begin{figure}
\centering
\includegraphics[width=0.45\textwidth]{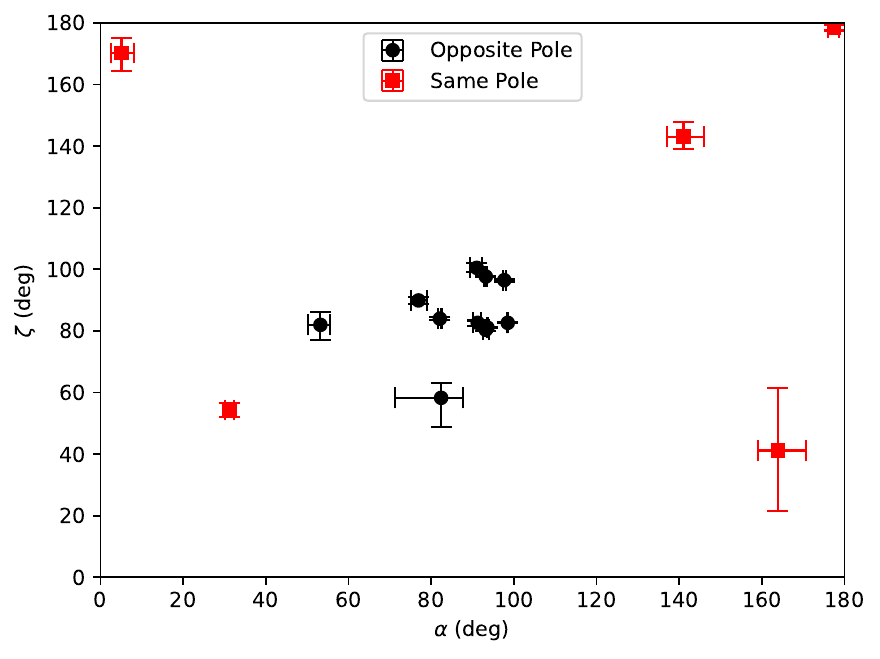}
\caption{The distribution of magnetospheric structural parameters $\alpha$ and $\zeta$ for 16 pulsars, in which the red squares and black circles indicate the pulsars with same pole and opposite pole IP emissions, respectively. }
\label{fig:dist}
\end{figure}

\begin{figure}
\centering
\includegraphics[width=0.45\textwidth]{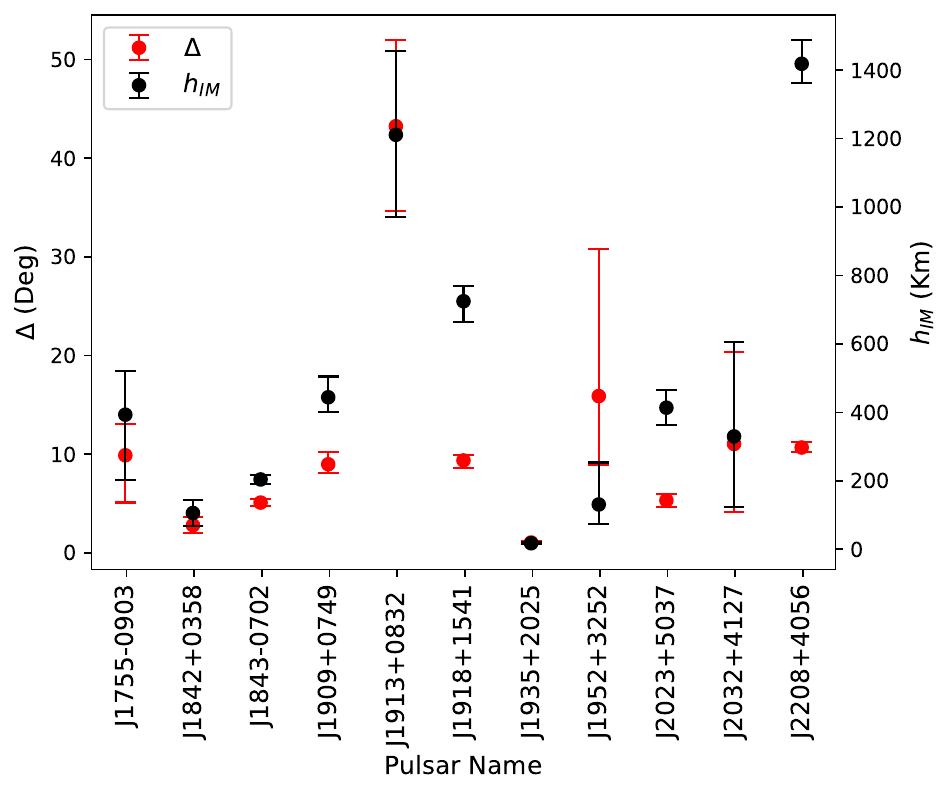}
\caption{The $\Delta$ (red circles) and $h_{\rm MI}$ (black circles) for 11 pulsars with opposite pole IP emissions. }
\label{fig:height_diff}
\end{figure}

\begin{figure*}
\flushleft
\subfigure{
\includegraphics[width=0.25\textwidth]{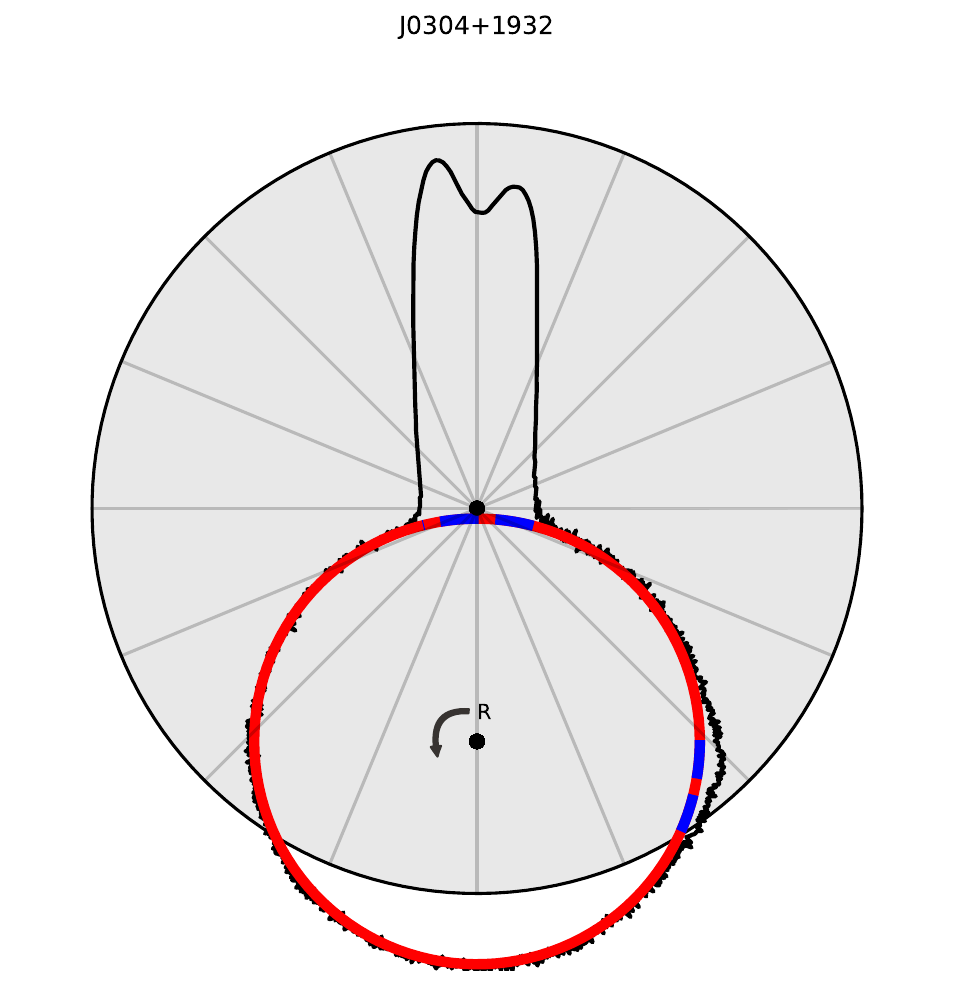}
\includegraphics[width=0.27\textwidth]{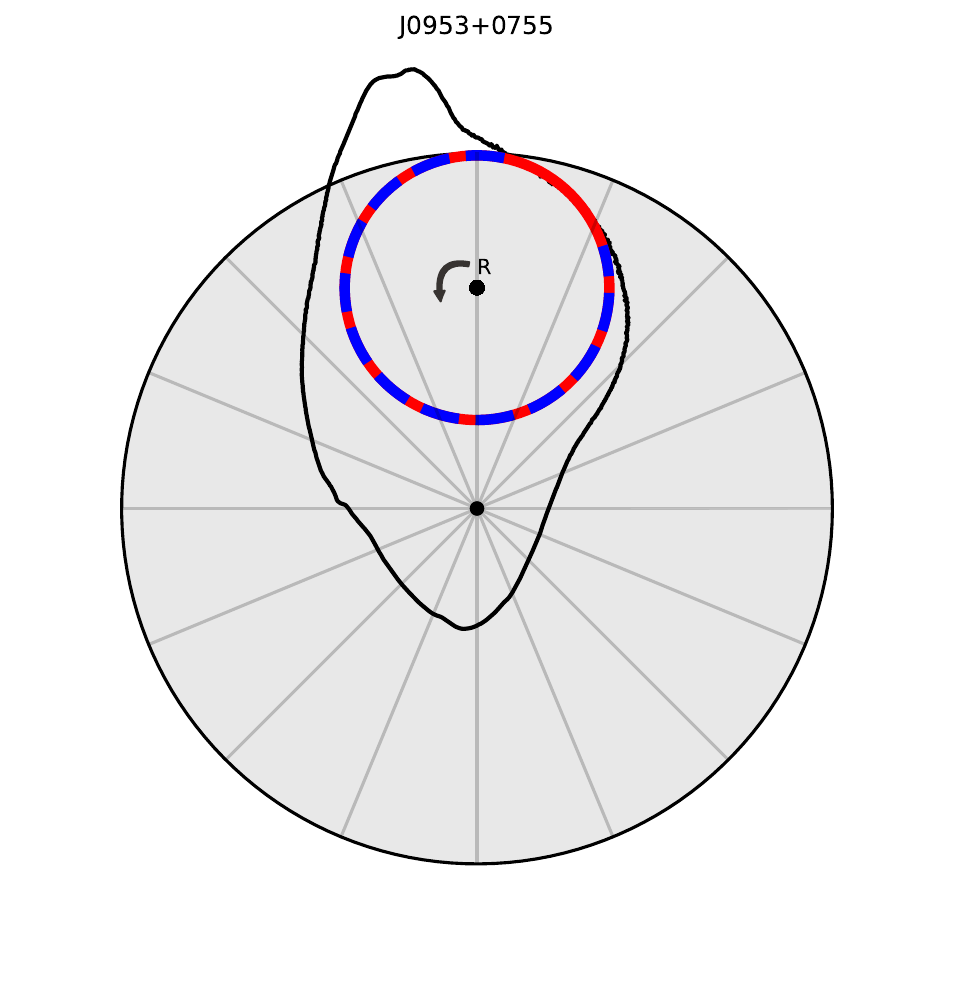}
\includegraphics[width=0.24\textwidth]{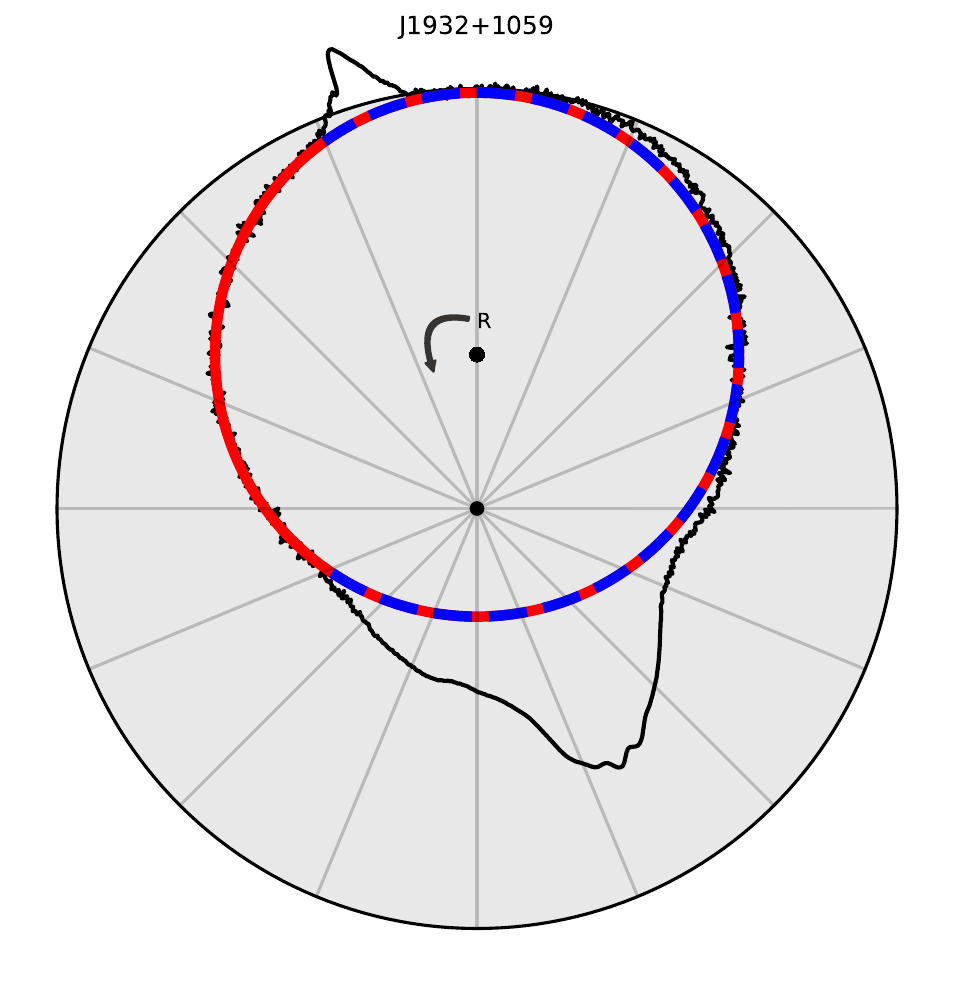}
\includegraphics[width=0.24\textwidth]{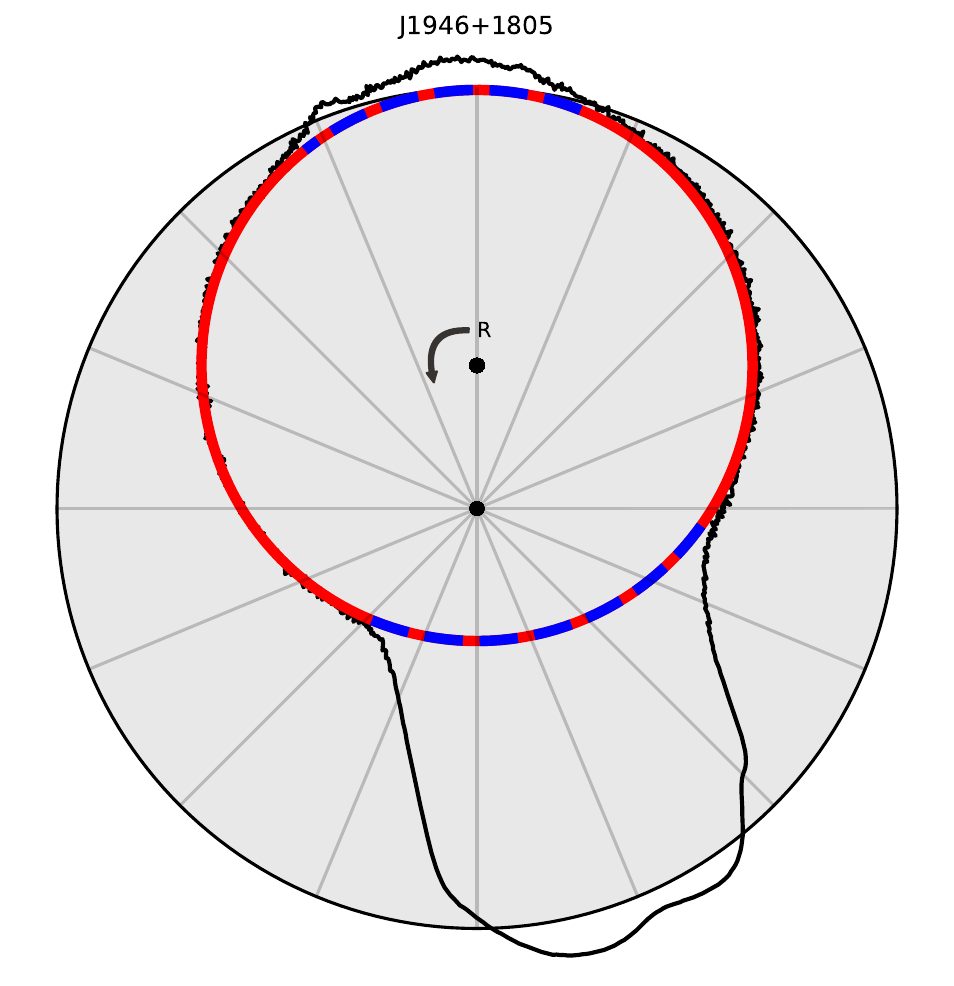}
}
\caption{The schematic configuration for the beams of 4 pulsars with same pole IP emissions. The red solid lines represent the LOS, while the blue dashed lines overlap on the red lines indicate the on pulse windows. The filled circles indicate the emission beams. The black solid lines within the beams represent the average profiles, with the baseline set to 1 and shown on a logarithmic scale.}
\label{fig:1beam}
\end{figure*}

\begin{figure*}
\flushleft
\subfigure{
\includegraphics[width=0.25\textwidth]{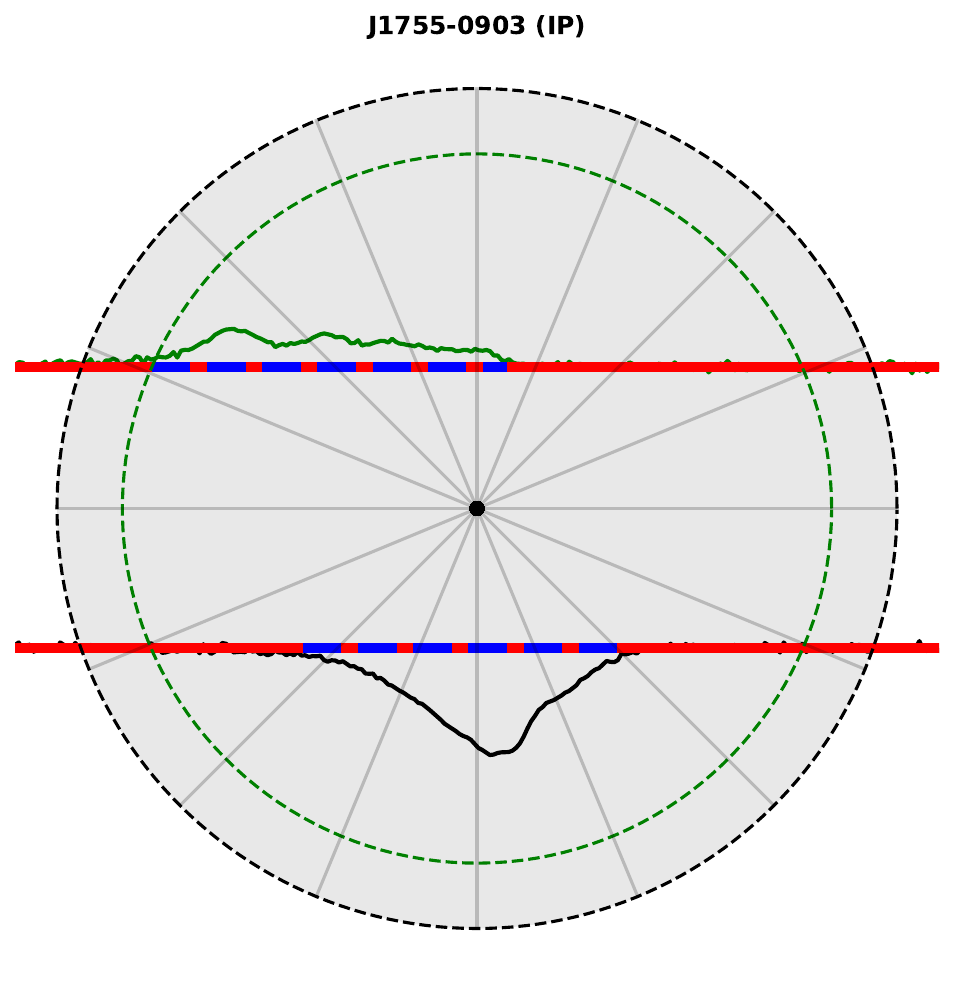}
\includegraphics[width=0.25\textwidth]{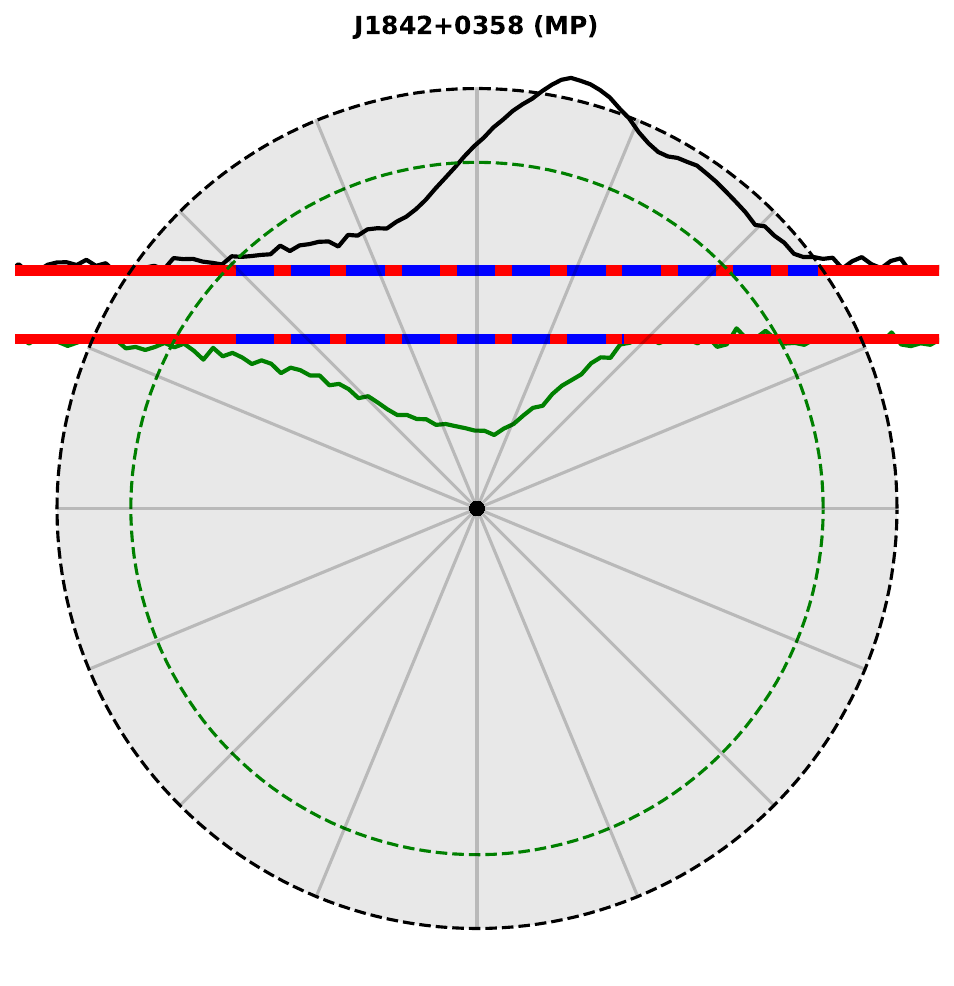}
\includegraphics[width=0.25\textwidth]{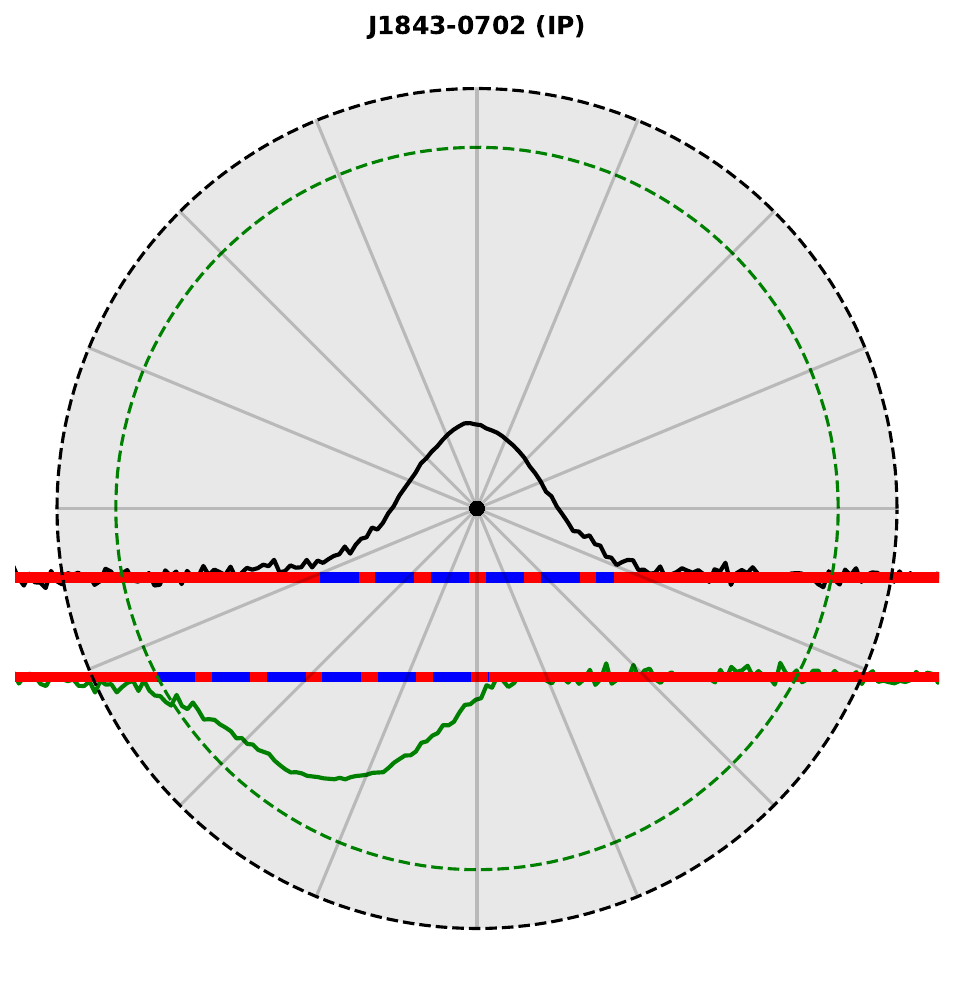}
\includegraphics[width=0.25\textwidth]{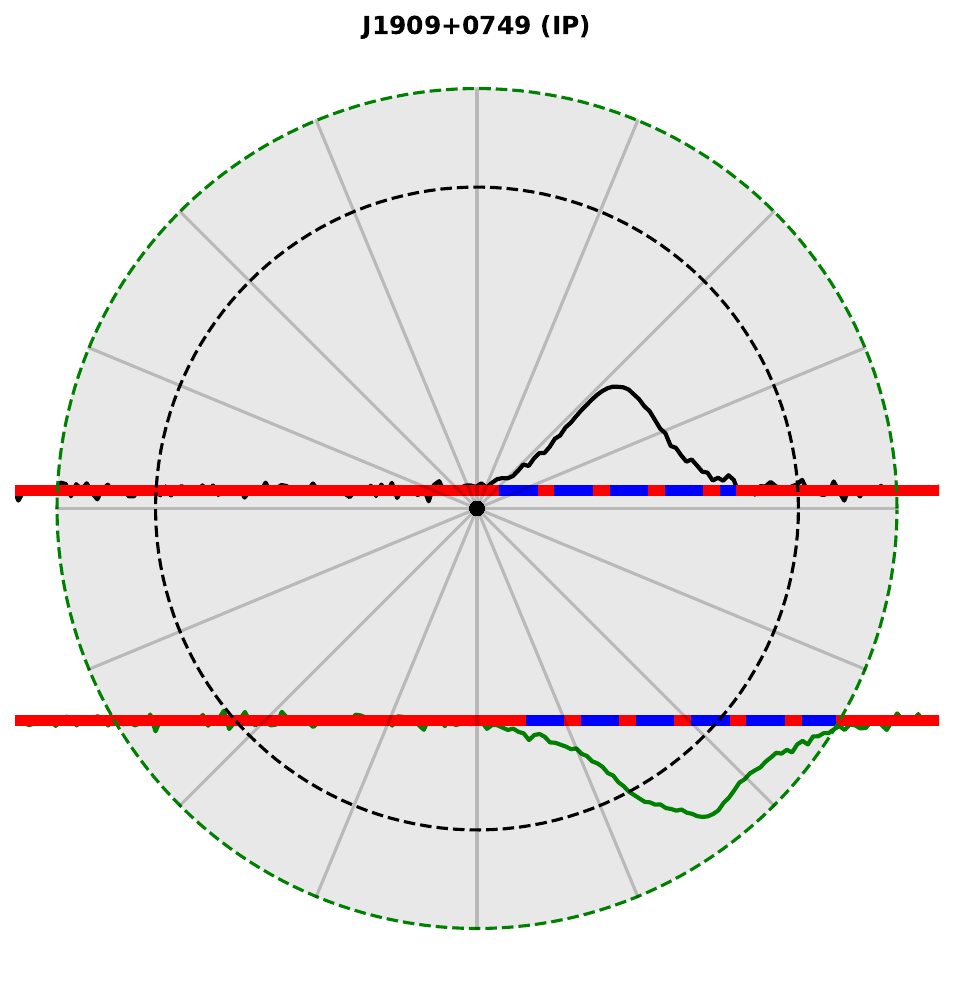}
}
\subfigure{
\includegraphics[width=0.25\textwidth]{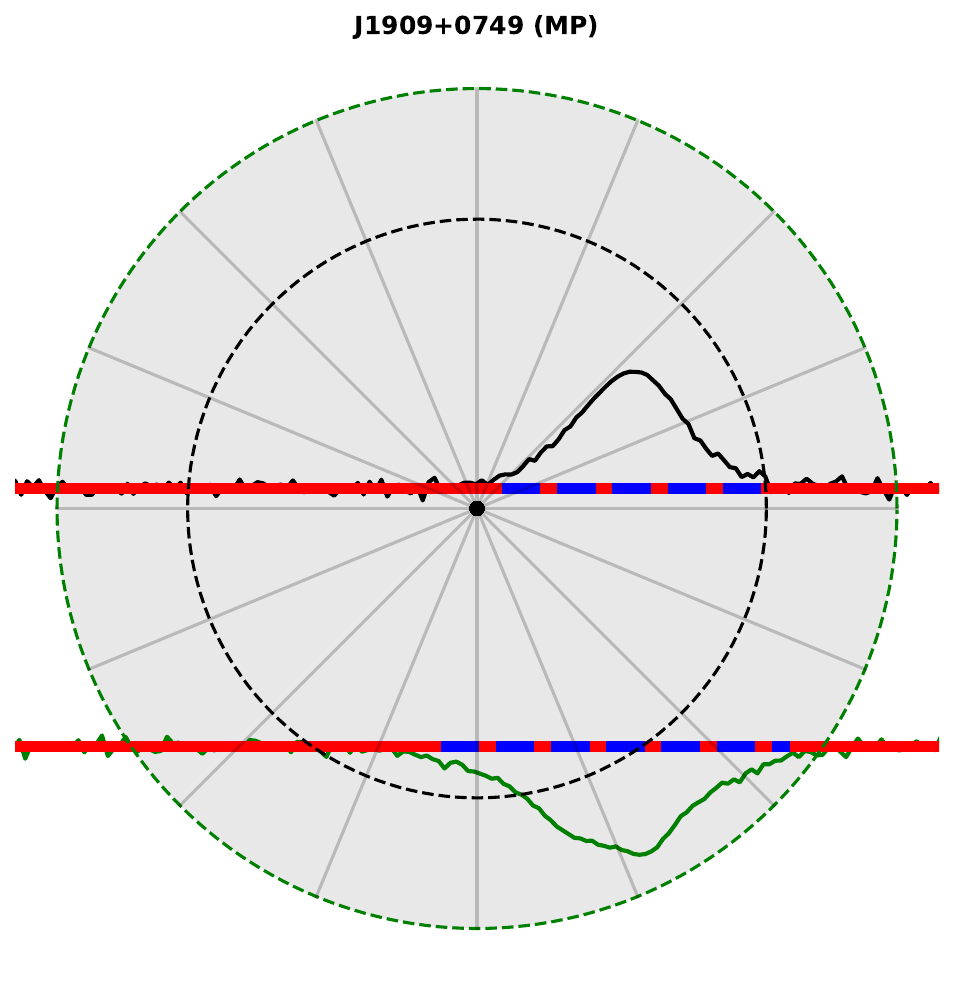}
\includegraphics[width=0.25\textwidth]{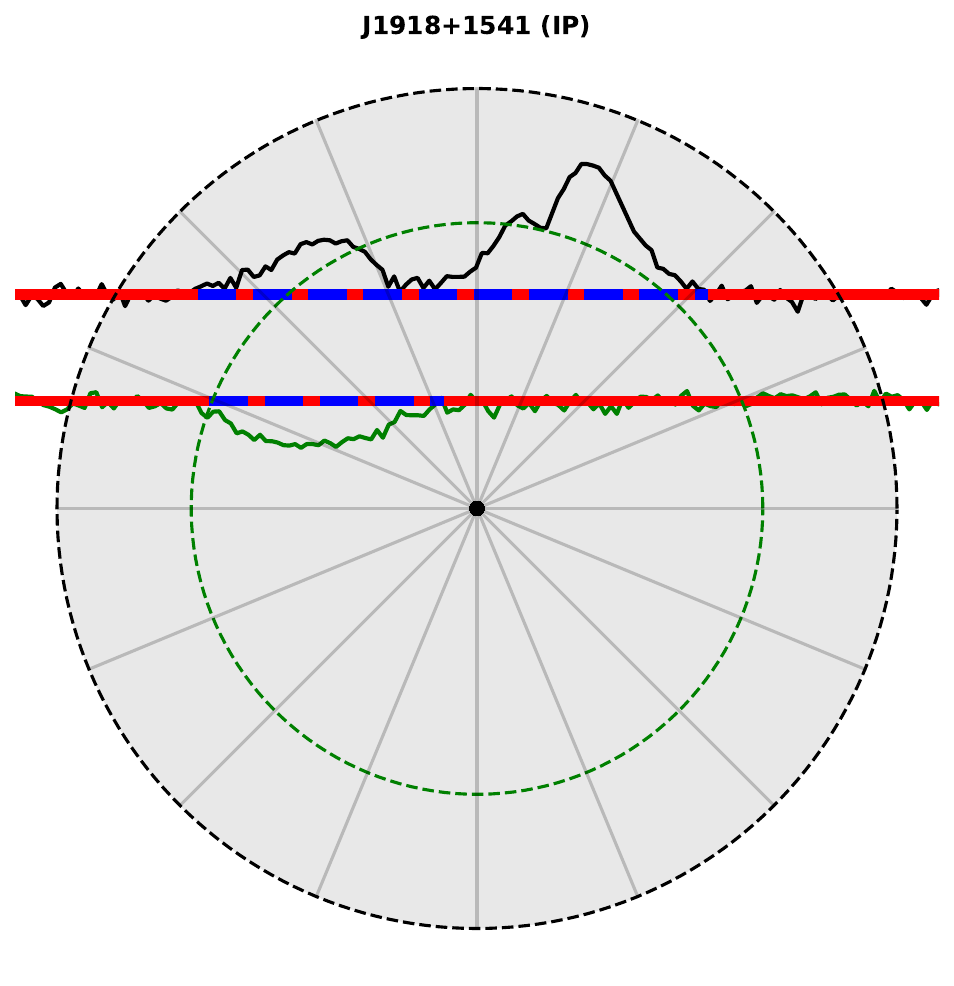}
\includegraphics[width=0.25\textwidth]{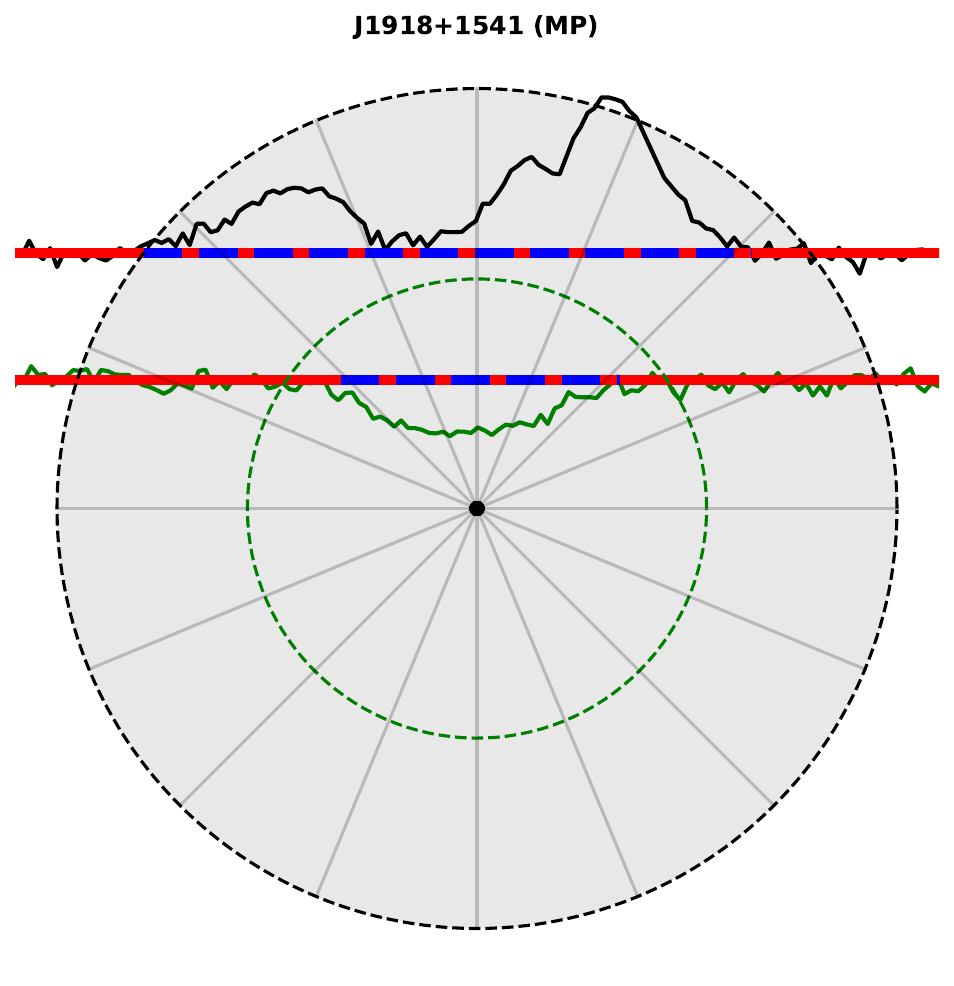}
\includegraphics[width=0.25\textwidth]{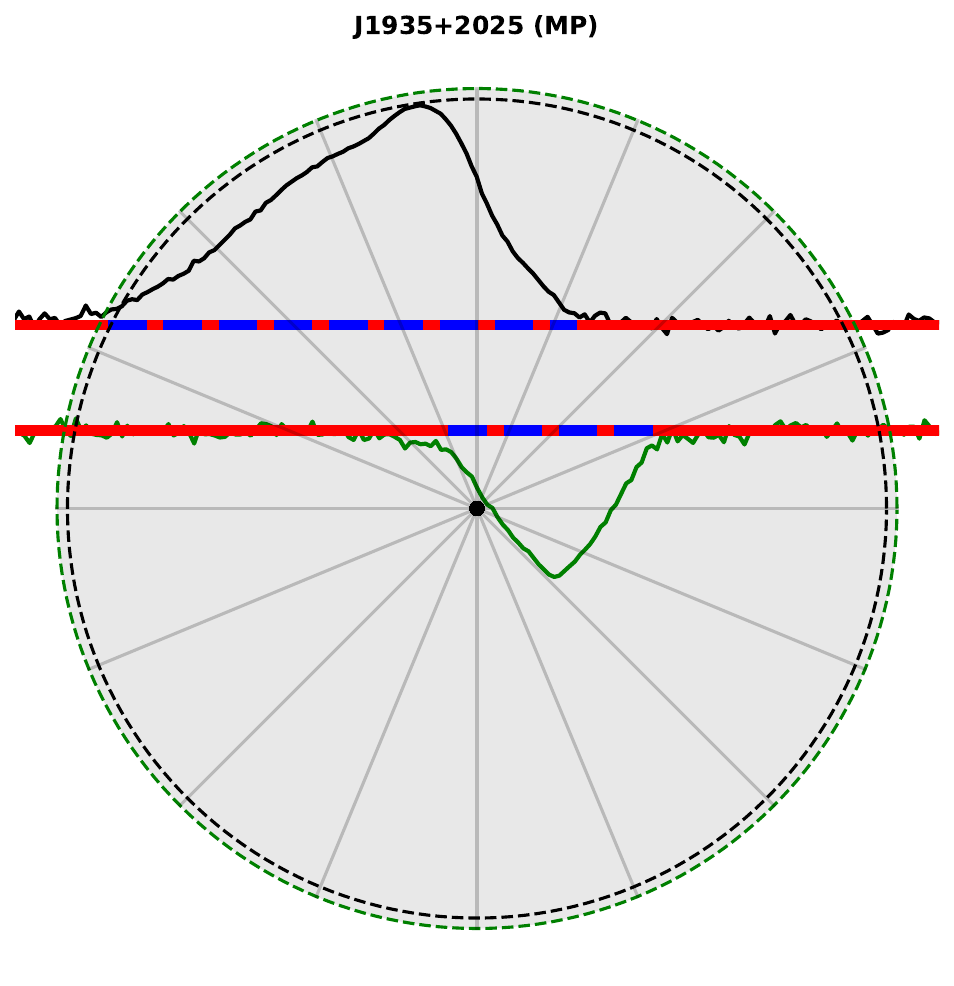}
}
\subfigure{
\includegraphics[width=0.25\textwidth]{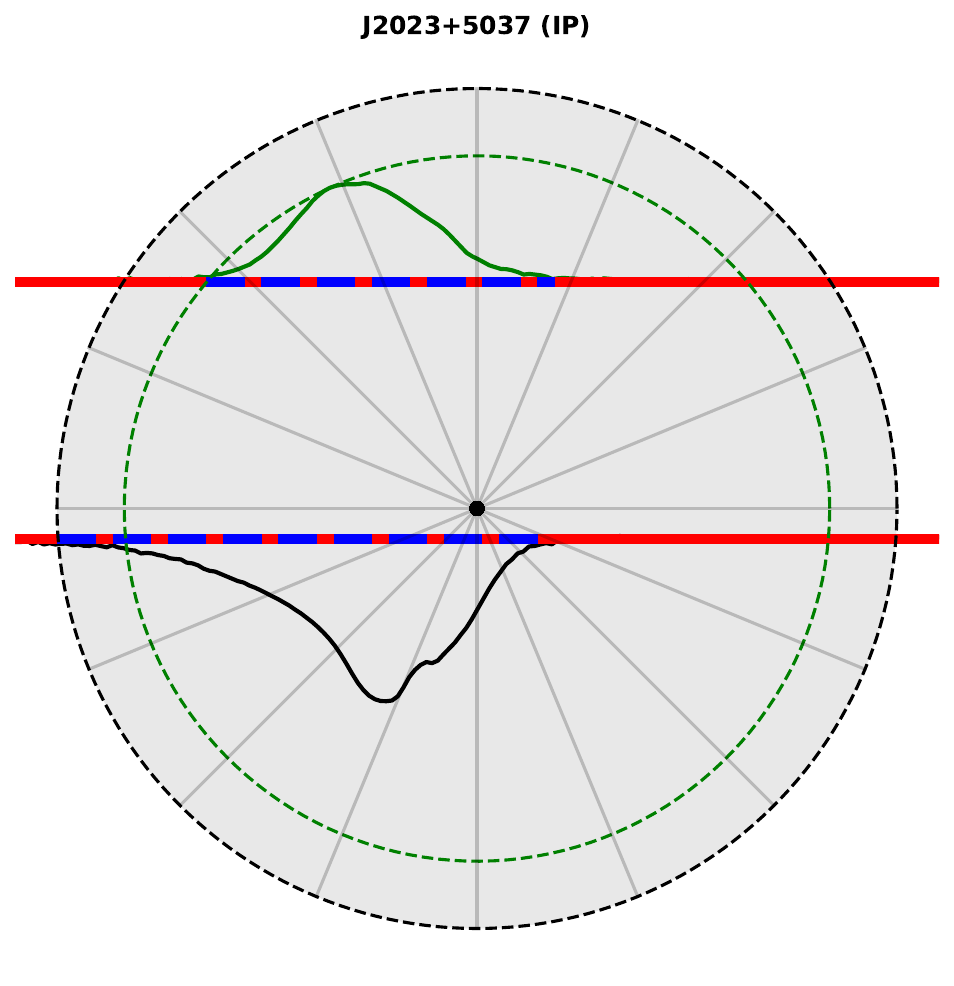}
\includegraphics[width=0.25\textwidth]{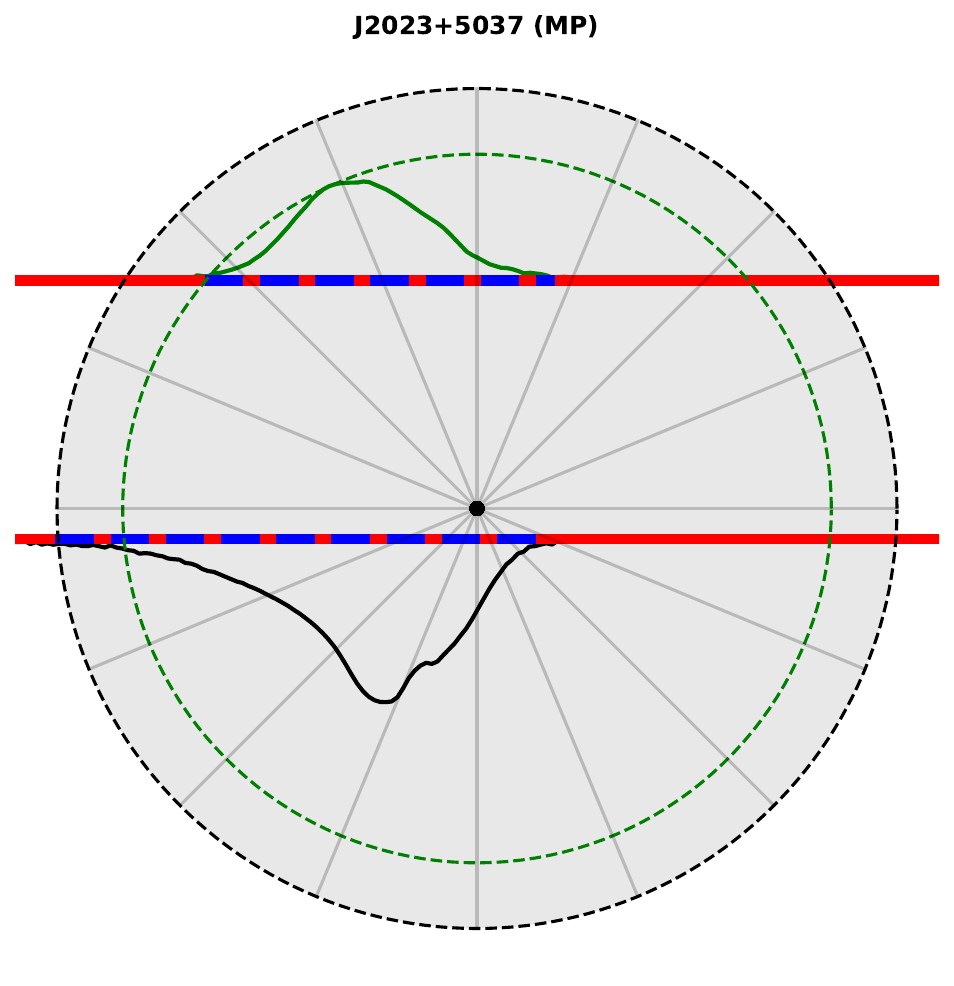}
\includegraphics[width=0.25\textwidth]{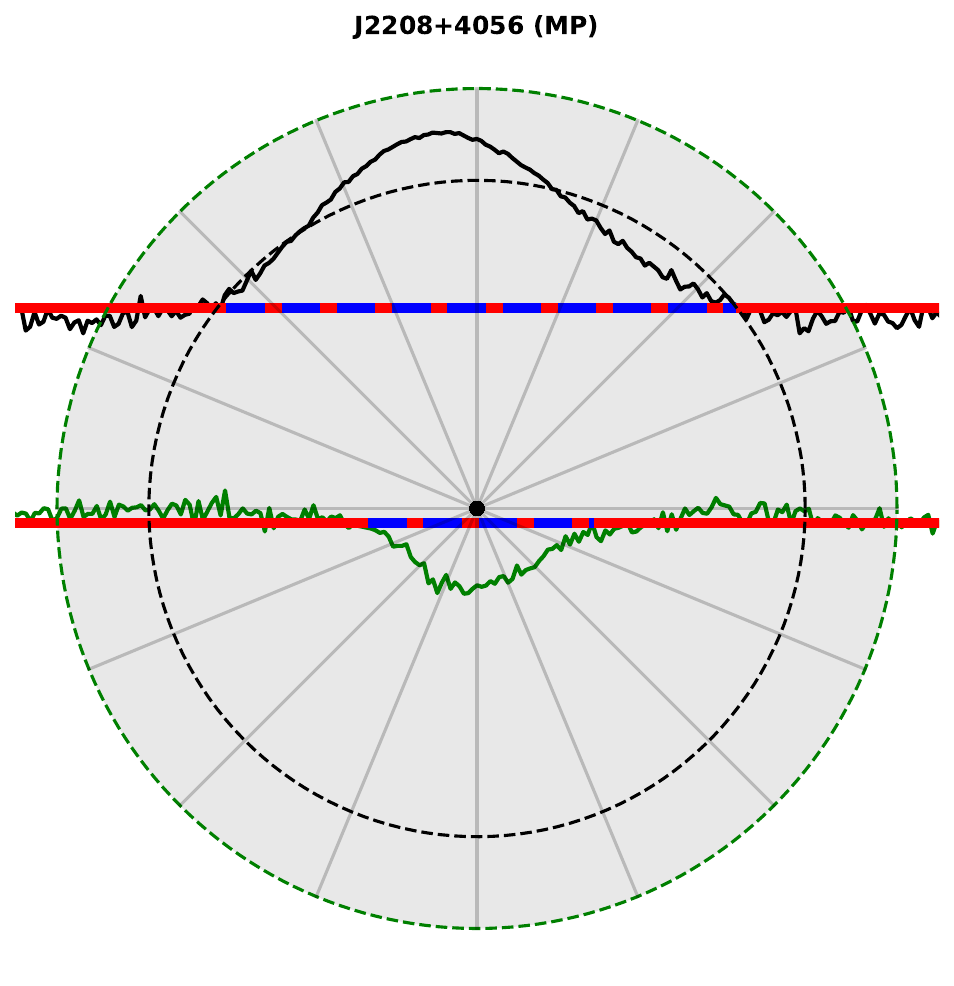}
}
\caption{The schematic configuration for the beams of 8 pulsars with opposite pole IP emissions. The red solid lines represent the LOS, while the blue dashed lines, overlapping with the red lines, indicate the on-pulse windows. The black and green dashed lines represent the emission beams of the MP and IP, respectively. The black and green solid lines within the beams represent the average profiles of the MP and IP, respectively. The baselines of the profiles are set to 1 and are shown on a logarithmic scale. The MP and IP in the titles refer to the beam configurations with the beams of the MP and IP taken as the reference beams, respectively..} 
\label{fig:2beam}
\end{figure*}

\begin{figure}
\centering
\includegraphics[width=0.45\textwidth]{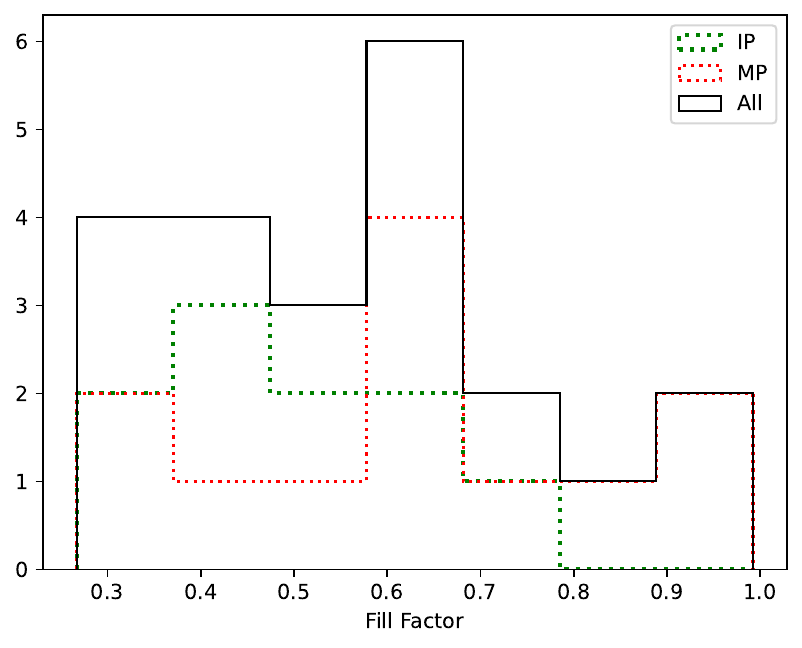}
\caption{The distribution of filling factor for 8 pulsars with opposite pole IP emissions. The red doted, green doted and black solid lines represent the distributions of MPs, IPs, and combination of MPs and IPs, respectively. }
\label{fig:factor}
\end{figure}

\begin{figure}
\centering
\includegraphics[width=0.45\textwidth]{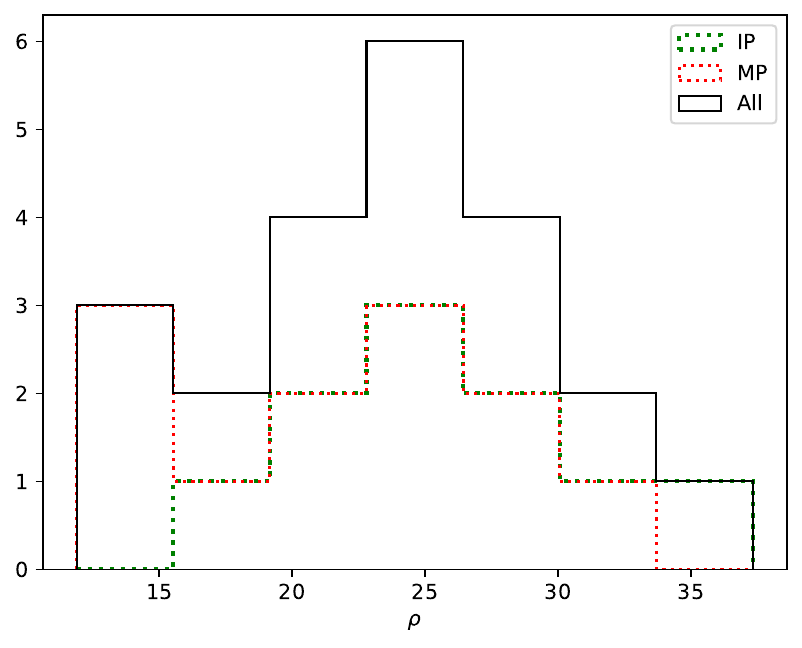}
\caption{The distribution of estimated beam radius for 8 pulsars with opposite pole IP emissions. The red doted, green doted and black solid lines represent the distributions of MPs, IPs, and combination of MPs and IPs, respectively.}
\label{fig:half_angle}
\end{figure}

\begin{figure}
\centering
\includegraphics[width=0.45\textwidth]{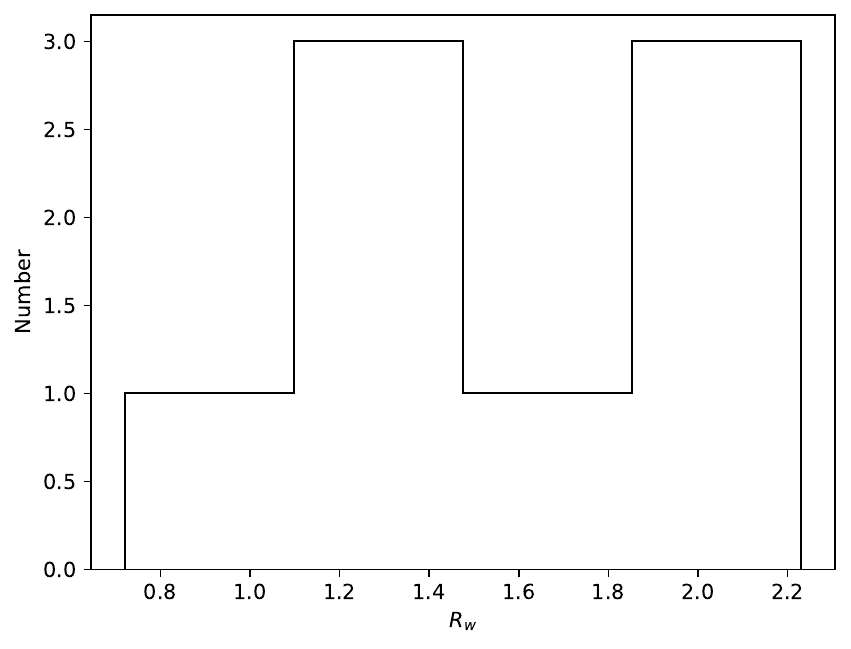}
\caption{The distribution of the ratio of pulse widths at larger $\beta$ compared to that at smaller  ($R_{\rm W}=W_{\rm \beta,large}/W_{\rm \beta,small}$) for 8 pulsars with the opposite pole IP emissions.}
\label{fig:width_beta}
\end{figure}

\begin{figure}
\centering
\includegraphics[width=0.45\textwidth]{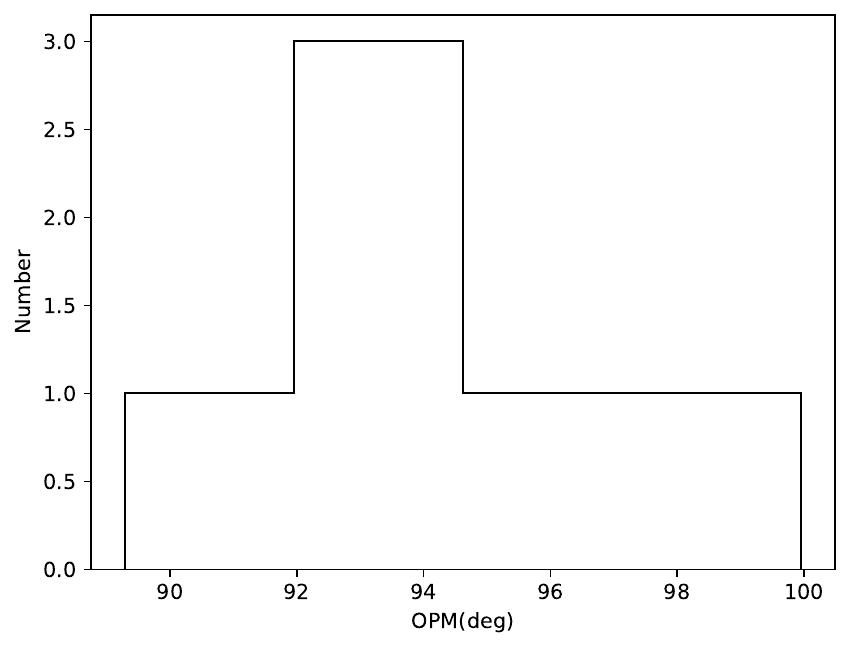}
\caption{The distribution of PA jump degrees for pulsars.}
\label{fig:opm}
\end{figure}

The average polarization profiles with PPAs for 23 sources in our sample, superposed on single-pulse PPAs, are shown in Figure~\ref{fig:prof}. 
Among these 23 sources, 16 exhibit smooth PPAs that can be fitted using the RVM, with the determined geometries presented in Table~\ref{fit}. 
The PPAs, along with the fitted RVM curves, are shown in Figure~\ref{fig:pafit}.
We note that RVM fitting for 14 of these 16 pulsars has been conducted in previous studies (e.g., \citealt{bcw1991,ew2001,2023MNRAS.520.4562S, 2023RAA....23j4002W,jk2019,sjk+2021,yzk+2023}). 
Given that the S/N in our observations is significantly higher than in previous studies, our results may provide new insights into pulsar emission geometry and location. 
For comparison, the RVM fitting results from previous studies are also listed in Table~\ref{fit}. For the remaining 7 pulsars, we are unable to determine the geometries because of their PPAs do not conform to the RVM.
Note that for simplicity, we define profiles with the larger peak intensities as MPs.

The distributions of $\alpha$ and $\zeta$ for these 16 pulsars are shown in Figure~\ref{fig:dist}.
For a pulsar exhibiting IP emission, if $\alpha$ is close to $0^{\circ}$ or $180^{\circ}$ and/or if there is bridge emission between the MP and IP, the IP emission is initially assumed to originate from the same magnetic pole as the MP.
Conversely, if $\alpha$ is close to $90^{\circ}$, the IP emission is assumed to originate from the opposite magnetic pole. 
In our sample of 16 pulsars, we found that the IP emissions for 5 pulsars originate from the same pole as the MP (red squares in Figure~\ref{fig:dist}), while the IP emissions for 11 pulsars (black circles in Figure~\ref{fig:dist}) originate from the opposite poles.  
For pulsars with opposite pole IP emissions, only the $\alpha$ and $\zeta$ values for the MPs are shown in Figure~\ref{fig:dist}. 
We found that $\alpha$ for pulsars with same pole IP emissions is generally less than $\sim 31^{\circ}$ or greater than $\sim 141^{\circ}$, whereas $\alpha$ for pulsars with opposite pole IP emissions exhibits a wider distribution, ranging from $53^{\circ}$ to $99^{\circ}$. 
A detailed analysis of the emission geometry for each pulsar is provided in Section~\ref{sec31}.

For the 5 pulsars with same pole IP emissions, we did not measure the polarization fractions for MPs or IPs separately due to the presence of bridge emissions. Instead, we measured the fractional linear polarization ($F_{\rm L}$), net circular polarization ($F_{\rm V}$), and absolute circular polarization ($F_{\rm |C|}$) for the entire on-pulse window and listed the values in Table~\ref{1per}. 
For the 11  pulsars with opposite pole IP emissions, $F_{\rm L}$, $F_{\rm V}$, and $F_{\rm |C|}$ for both MPs and IPs were measured and are shown in Table~\ref{2per}. For the 7 pulsars with unsmooth PPAs, we do not detected any bridge emissions from them. Therefore, we also measured the $F_{\rm L}$, $F_{\rm V}$, and $F_{\rm |C|}$ for both MPs and IPs, and the results are shown in Table~\ref{2per}, labeled by asterisk.

For pulsars with opposite pole IP emissions, the emission heights of IP and MP are different. 
Using Equations~\ref{eq50} and~\ref{eq5}, we measured the value of $\Delta$, and the $h_{\rm MI}$ for each pulsar (Table~\ref{fit} and Figure~\ref{fig:height_diff}). 
For most pulsars, the $\Delta$ is less than $30^{\circ}$, and the $h_{\rm MI}$ are approximately several hundred kilometers. However, for PSRs J1913+0832 and J2208+4056, the $h_{\rm MI}$ can beyond 1000\,km. 
It is unclear why the difference in emission height between MP and IP varies over a large range.

The emission regions within a pulsar beam are generally not fully active (e.g., \citealt{lm1988}), making it difficult to define the beam boundary. 
To estimate the beam size, we assumed that the emission beam is circular and neglected the effects of aberration and retardation. 
The beam boundary was then defined as the critical size at which the beam just encompasses MP/IP on-pulse windows. 
The filling factor was calculated as the ratio of the observed pulse width to the expected maximum pulse width for a fully active beam.
Note that this method provides a lower limit for the beam size, as the actual beam could be larger.

Figures~\ref{fig:1beam} and \ref{fig:2beam} present schematic configurations of beam structures for pulsars based on RVM fitting results. 
Due to the high sensitivity of our observations, the on-pulse window is defined as the region where the intensity exceeds $3\sigma$ of the off-pulse window, corresponding to the pulse width ($W_{3\sigma}$), rather than the conventional 10\% peak intensity threshold. 
To better visualize weak IPs, we set the baseline level to 1 and present the profiles using a logarithmic scale. 
For some pulsars, $\alpha$ or $\zeta$ values could not be determined precisely, leading to significant uncertainties in the beam configurations. 
Therefore, beam configurations for these pulsars (e.g., PSRs J1852$-$0118, J1913+0832, J1952+3252, and J2032+4127) are not presented.

For pulsars with same pole IP emissions, the LOS sweeps within the beam, providing information on emission regions in both longitude and latitude. 
We found that the emission regions are not fully active in either longitude or latitude within the beam (Figure~\ref{fig:1beam}) and exhibit complex variations. 
However, due to the LOS trajectory is within the beam, we could not measure the filling factor for this type of pulsars.  
For pulsars with opposite pole IP emissions, we took the IP beam as the reference beam, assuming it just encompasses the IP on-pulse window. 
The beam radius (or half-opening angle) was then obtained using Equation~\ref{eq4}. 
The emission height for the IP was determined by substituting the half-opening angle into Equation~\ref{eq3}. 
The emission height for the MP was subsequently estimated by considering the height difference $h_{\rm MI}$. 
The half-opening angle for the MP was then obtained using Equations~\ref{eq3} and \ref{eq4}. 
Finally, the filling factors for both the IP and MP were determined.  
Alternatively, if we used the MP beam as the reference beam, where it just encompassed the MP on-pulse window, we obtained another set of filling factors and half-opening angles using the same method. 
We then averaged the values obtained from both approaches to derive the final filling factors and half-opening angles for the IP and MP.  

The distribution of filling factors for these 8 pulsars with opposite pole IP emissions is shown in Figure~\ref{fig:factor}, ranging from 0.27 to 0.99, with a mean value of $\sim 0.58$. 
The filling factors for MPs and IPs exhibit similar distributions.
The distribution of the estimated beam radius is presented in Figure~\ref{fig:half_angle}, which varies widely from $12^{\circ}$ to $37^{\circ}$, with a peak at $\sim 25^{\circ}$. 
For most pulsars, the beam radius is less than $30^{\circ}$~\citep{lm1988}, but for PSRs J1755$-$0903 and J2208+4056 the beam radii are larger, measuring $37^{\circ}$ and $32^{\circ}$, respectively.

We also examined the relationship between the observed pulse width $W_{3\sigma}$ and $\beta$ for pulsars with opposite pole IP emissions.  
We define $R_{\rm W}=W_{\rm \beta,large}/W_{\rm \beta,small}$ as the ratio of pulse widths at larger $\beta$ compared to those at smaller $\beta$.  
The distribution of $R_{\rm W}$ is shown in Figure~\ref{fig:width_beta}. 
For 7 pulsars, $W_{3\sigma}$ increases with increasing $\beta$, while for the remaining 1 pulsar, $W_{3\sigma}$ decreases, suggesting the observed pulse width maybe increase with increasing $\beta$.

The phenomenon of OPMs has been detected in some pulsars, as evidenced by the PPAs of single-pulses, such as PSRs J0953+0755 and J1952+3252. 
However, for many pulsars, their single-pulses are too weak to detect PPAs, making it difficult to confirm whether a jump in PPA is due to OPMs. 
We simply assume that jumps in the PPAs of average profiles correspond to OPMs and determine the degrees of these jumps using RVM fitting. 
The degrees of OPM jumps for these pulsars are listed in Tables~\ref{fit}, with their distribution shown in Figure~\ref{fig:opm}. 
For most pulsars, the OPM jumps are approximately $90^{\circ}$. Some pulsars exhibit arbitrary PA jumps in their PPAs that deviate from $90^{\circ}$. For example, PSR J1755$-$0903 shows a PA jump of approximately $55^{\circ}$ in the MP. Such non-90$^{\circ}$ jumps have been observed in many pulsars, but their physical origin remains unclear (e.g., \citealt{2001A&A...378..883P,jk2018}).

\subsection{Pulsar geometry}
\label{sec31}

\subsubsection{The same pole}

\textbf{J0304+1932:} This pulsar exhibits weak IP emission, with a peak intensity of about 0.07\% of the MP (see Figure~\ref{fig:prof} and Table~\ref{1per}). The separation between the MP and IP is $101.6^{\circ}$, significantly smaller than $180^{\circ}$, with no detectable bridge emission.  
The PPAs of the MP follow an S-shape, whereas those of the IP are flat. The OPM phenomenon is detected in the PPAs of single-pulses but not in the PPAs of the average profile (Figure~\ref{fig:prof}).  
\citet{yzk+2023} performed RVM fitting for the MP alone and suggested that the IP originates from the same magnetic pole as the MP but at a higher altitude. We fit the PPAs of the entire profile using Equation~\ref{eq60}, and the results, presented in Table~\ref{fit}, indicate a large $\alpha$, supporting the same pole origin hypothesis.  
Our results are consistent with \citet{yzk+2023}. 
Note that our results are limited by the low S/N of IP, and follow up observation with longer durations are needed. 
The beam configuration is shown in Figure~\ref{fig:1beam}.
The estimated beam radius is approximately $37^{\circ}$, assuming it encompasses the on-pulse window (blue dashed line in Figure~\ref{fig:1beam}).

\textbf{J0953+0755:} The IP emission of this pulsar has a peak intensity of 1.5\% relative to the MP (see Figure~\ref{fig:prof} and Table~\ref{1per}). The MP and IP are separated by $152.6^{\circ}$, which is less than $180^{\circ}$, with bridge emission detected between them. The average profile has a pulse width of $W_{3\sigma} = 300.9^{\circ}$.  
Figure~\ref{fig:prof} shows the average polarization profile superposed on the PPAs of single pulses, revealing the OPM phenomenon. In the PPAs of the average profile, a bump appears at the profile peak, and a gradual OPM jump occurs at a pulse phase of $65^{\circ}$.  
For RVM fitting, we excluded PPAs at the bump and gradual OPM jump (black dots in Figure~\ref{fig:pafit}) and treated the OPM jump degree as a free parameter using Equation~\ref{eq6} (Figure~\ref{fig:pafit}). The results, listed in Table~\ref{fit}, indicate that the MP and IP originate from the same pole. The OPM jump degree is measured as $99.96^{\circ}$$ ^{+0.48} _{-0.49}$.  
Our findings are consistent with \citet{bcw1991} but have significantly smaller uncertainties. The beam configuration is shown in Figure~\ref{fig:1beam}. The estimated beam radius is about $2^{\circ}$, assuming it encompasses the entire on-pulse window (blue dashed line in Figure~\ref{fig:1beam}).  

\textbf{J1852$-$0118:} This pulsar's IP emission has a peak intensity of 38.1\% relative to the MP (see Figure~\ref{fig:prof} and Table~\ref{1per}). The MP and IP are separated by $133.9^{\circ}$, with weak bridge emission detected. The average profile pulse width is $W_{3\sigma} = 152.6^{\circ}$.  
Figure~\ref{fig:prof} presents the average polarization profile superposed on PPAs of single pulses, where the OPM phenomenon is evident. Two OPM jumps are detected in the PPAs of the average profile at pulse phases $-70^{\circ}$ and $-20^{\circ}$.  
For RVM fitting, we treated the OPM jump degrees as free parameters using Equation~\ref{eq6} (Figure~\ref{fig:pafit}), and the results, listed in Table~\ref{fit}, suggest that the MP and IP originate from the same pole. The OPM jump degree is measured as $94.09^{\circ}$$ ^{+1.44} _{-1.20}$.  
Compared to \citet{2023RAA....23j4002W}, our results have a constant $\beta$ but a different $\alpha$. Given the larger uncertainties in \citet{2023RAA....23j4002W}, we believe our results are reasonable. However, due to significant uncertainties, we do not analyze the beam configuration for this pulsar.  

\textbf{J1932+1059:} This pulsar has a weak IP, with a peak intensity of only 0.1\% of the MP (see Figure~\ref{fig:prof} and Table~\ref{1per}). Weak bridge emission is detected, and the MP and IP are separated by $186.3^{\circ}$. The average profile pulse width is $W_{3\sigma} = 185.6^{\circ}$.  
Figure~\ref{fig:prof} shows the average polarization profile superposed on the PPAs of single pulses, where the OPM phenomenon is observed. An OPM jump is detected in the PPAs of the average profile at a pulse phase of $-90^{\circ}$.  
For RVM fitting, we treated the OPM jump degree as a free parameter using Equation~\ref{eq6} (Figure~\ref{fig:pafit}), and the results, presented in Table~\ref{fit}, indicate that the MP and IP originate from the same pole. The OPM jump degree is measured as $89.29^{\circ}$$ ^{+1.57} _{-1.59}$.  
Compared to \citet{ew2001}, our results maintain a constant $\beta$ but suggest a slightly smaller $\alpha$. 
The beam configuration is shown in Figure~\ref{fig:1beam}. The estimated beam radius is about $54^{\circ}$, assuming it encompasses the entire on-pulse window (blue dashed line in Figure~\ref{fig:1beam}).

\textbf{J1946+1805:} This pulsar's IP emission has a peak intensity of about 0.4\% of the MP (see Figure~\ref{fig:prof} and Table~\ref{1per}). The MP and IP are separated by $183.9^{\circ}$, with no detectable bridge emission.  
Figure~\ref{fig:prof} shows the average polarization profile superposed on the PPAs of single pulses, where the OPM phenomenon is evident. Two OPM jumps are detected in the PPAs of the average profile at pulse phases $71^{\circ}$ and $110^{\circ}$.  
For RVM fitting, we treated the OPM jump degree as a free parameter using Equation~\ref{eq6} (Figure~\ref{fig:pafit}), and the results, shown in Table~\ref{fit}, suggest that the MP and IP originate from the same pole. The OPM jump degree is measured as $94.49^{\circ}$$ ^{+0.76} _{-0.76}$.  
Our findings agree with \citet{2023RAA....23j4002W} but with significantly smaller uncertainties. The beam configuration is shown in Figure~\ref{fig:1beam}. The estimated beam radius is about $10^{\circ}$, assuming it encompasses the entire on-pulse window (blue dashed line in Figure~\ref{fig:1beam}).

%\iffalse

%\fi

\begin{table*}[ht]
\setlength{\tabcolsep}{18pt}
\renewcommand{\arraystretch}{1.2}
\centering
\caption{Parameters for 5 pulsars with same pole IP emissions. $W_{\rm 3\sigma}$ denotes the pulse width, while $F_{\rm L}$, $F_{\rm V}$, and $F_{\rm |C|}$ refer to the fractions of linear polarization, net circular polarization, and absolute circular polarization, respectively. The separation between the peaks of the MP and IP, as well as the ratio of peak intensities of the IP and MP, are shown in the sixth and seventh columns, respectively.}
\label{1per}
\begin{tabular}{ccccccc}
\hline
Name & $W_{3\sigma}$ & $F_{\rm L}$ & $F_{\rm V}$ & $F_{\rm | C |}$ & Sep & IP/MP$_{\rm peak}$
\\
 & $(^{\circ})$ & \% & \% & \% & $(^{\circ})$ &
\\
\hline
J0304+1932 & 57.0  & 35.14(1)  & 12.21(2)   & 12.27(2) & 101.6  & 0.0007
\\
J0953+0755 & 300.9 & $13.99(1)$ & $-4.80(1)$ & $4.92(1)$ &  152.6 & 0.015
\\
J1852$-$0118 & 152.6 & $63.4(8)$ & $-6.38(3)$ & $12.31(3)$ &  133.9 & 0.381
\\
J1932+1059 & 185.6 & $76.53(5)$ & $-5.16(1)$ & $6.12(1)$ &  186.3 & 0.001
\\
J1946+1805 & 139.9 & $37.08(1)$ & $-5.99(1)$ & $6.14(1)$ & 183.9 & 0.004
\\
\hline
 \end{tabular}
\end{table*}

\begin{table*}
\setlength{\tabcolsep}{6pt}
\renewcommand{\arraystretch}{1}
\centering
\caption{ Polarization parameters for pulsars with the opposite pole IP emissions and unclassified IP emissions (denoted by an asterisk superscript). See Table~\ref{1per} for details.}
\label{2per}
\begin{tabular}{ccccccccccc}
\hline
Name & \multicolumn{4}{c}{MP} & \multicolumn{4}{c}{IP}  & Sep & IP/MP$_{\rm peak}$
\\
& $W_{3\sigma}$ & $F_{\rm L}$ & $F_{\rm V}$ & $F_{\rm | C |}$ & $W_{3\sigma}$ & $F_{\rm L}$  & $F_{\rm V}$ & $F_{\rm | C |}$ & 
\\
 & $(^{\circ})$ & \% & \% & \%  & $(^{\circ})$ & \%  & \% & \%  & $(^{\circ})$ 
\\
\hline
J0627+0706$^{*}$ & 23.2 & $23.82(1)$ & $-3.73(1)$ & $5.89(1)$ & 25.3 & $19.24(1)$ & $-4.66(1)$ & $6.37(1)$ & 183.2 & 0.266
\\
J0826+2637$^{*}$ & 66.4 & $23.74(1)$ & $1.29(1)$ & $2.81(1)$ & 32.7 & $26.1(7)$ & $0.0(2)$ & $3.2(2)$ & 179.3 & 0.002
\\
J1755$-$0903 & 29.9  & $32.4(2)$ & $5.4(1)$ & $13.9(1)$ & 33.4 & $16(1)$ & $0.4(8)$ & $6.9(8)$ & 156.4 & 0.112
\\
J1816$-$0755$^{*}$ & 26.7 & $18.65(1)$ & $-4.04(1)$ &$5.40(1)$ & 33.0 & $15.44(4)$ & $3.35(1)$ & $6.85(1)$ & 175.4 & 0.220
\\
J1825$-$0935$^{*}$ & 44.6  & $32.02(1)$ & $3.52(1)$ & $4.20(1)$ & 25.0 & $15.5(1)$ & $-3.3(1)$ & $3.6(1)$ & 187.0 & 0.054
\\
J1842+0358 & 21.4 & $36.82(2)$ & $-3.40(4)$ & $4.64(4)$ & 14.4 & $2.98(8)$ & $-0.1(1)$ & $4.4(1)$ & 177.5 & 0.201
\\
J1843$-$0702 & 19.3  & $15.5(2)$ & 0.0(1) & $3.1(1)$ & 21.8 & $31.9(4)$ & $-3.5(2)$ & $4.6(2)$ & 172.3 & 0.373
\\
J1849+0409$^{*}$ & 11.6 & $54.64(2)$ & $22.26(9)$ & $22.32(9)$ & 18.6 & $26.83(4)$ & $1.8(1)$ & $2.2(1)$ & 185.6 & 0.840
\\
J1909+0749 & 16.2 & $52.1(1)$ & $-9.51(8)$ & $10.62(8)$ & 21.1 & $90.1(1)$ & $6.84(9)$ & $7.48(9)$ & 186.3 & 0.837
\\
J1913+0832 & 45.7 & $55.0(2)$ & $-0.3(4)$ & $1.7(4)$ & 37.3 & $41.0(4)$ & $-14.8(7)$ & $15.4(7)$ & 192.7 & 0.590
\\ 
J1918+1541 & 30.9 & $70.5(1)$ & $18.8(6)$ & $22.7(6)$ & 14.4 & $33.9(2)$ & $0(1)$ & $11(1)$ & 163.5 & 0.227
\\
J1926+0737$^{*}$ & 16.9 & $11.2(2)$ & $0.0(6)$ & $4.4(6)$ & 10.2 & $14.4(4)$ & $2(1)$ & $4(1)$  & 166.3 & 0.655
\\
J1935+2025 & 32.3 & $96.36(4)$ & $-10.87(7)$ & $10.92(7)$ & 15.1 &  $87.0(1)$ & $-35.4(1)$ & $35.5(1)$ & 189.5 & 0.362
\\ 
J1952+3252 & 78.1 & $44.29(9)$ & $-6.69(2)$ & $6.77(2)$ & 21.1 & $18(2)$ & $-1.6(5)$ & $7.6(5)$ & 184.9 & 0.026
\\ 
J2023+5037 & 30.2 & $18.89(1)$ & $-3.12(8)$ & $4.94(8)$ & 21.8 & $52.02(2)$ & $-2.0(2)$ & $2.6(2)$ & 178.9 & 0.131
\\
J2032+4127 & 49.2 & $82.50(7)$ & $-1.9(1)$ & $2.5(1)$ & 16.2 & $78.5(5)$ & $-0(1)$ & $10(1)$ &  165.9 & 0.093
\\
J2047+5029$^{*}$ & 16.2 & $29.04(2)$ & $-1.50(1)$ & $1.71(1)$ & 16.2 & $49.94(4)$ & $34.56(2)$ & $34.79(2)$ & 179.6 & 0.619
\\
J2208+4056 & 40.8 & $76.15(7)$ & $-5(2)$ & $7(2)$ & 18.3 & $88.8(3)$ & $-3(10)$ & $11(10)$ & 181.4 & 0.197
\\
\hline
 \end{tabular}
\end{table*}

\subsubsection{The opposite poles}

\textbf{J1755$-$0903:} This pulsar exhibits IP emission with a peak intensity of approximately 11.2\% of the MP (Figure~\ref{fig:prof} and Table~\ref{2per}). The $W_{3\sigma}$ values of the MP and IP are 29.9$^\circ$ and 33.4$^\circ$, respectively, with a separation of $156.4^{\circ}$. 
The average polarization profile, superimposed on the PPAs of single pulses, is shown in Figure~\ref{fig:prof}, where a jump of $\sim 55^{\circ}$ is observed at the pulse phase of 2.5$^{\circ}$. Considering the jump significantly deviates from $90^{\circ}$, we exclude the PPAs before 2.5$^{\circ}$ in the RVM fitting. 
RVM fitting was performed using Equation~\ref{eq50}, and the results presented in Table~\ref{fit} suggest that the IP emission originates from the pole opposite to the MP. 
The negative value of $\Delta =-9.90^{\circ}$$ ^{+3.21} _{-4.78}$ indicates that the emission height of the IP is lower than that of the MP, with an estimated height difference $h_{\rm IM} = 393.34_{-189.92}^{+127.54}$\,km.
Compared to the results of~\citet{sjk+2021}, the values of $\zeta$ and $\Delta$ are consistent, though our results yield a slightly smaller $\alpha_{\rm M}$. 
We note that the uncertainties in some parameters in our results are larger than those in~\citet{sjk+2021}. This discrepancy may be attributed to differences S/Ns. 
Using the method described in Section~\ref{sec:res}, we take the IP beam as the inference beam, assuming that it fully encompasses the on-pulse window of the IP. The measured beam radii for the IP and MP are $32.78^{\circ}$ and $37.33^{\circ}$, respectively, corresponding to filling factors of 0.55 and 0.42. 
Conversely, assuming the MP beam fully encompasses its on-pulse window, we find that IP emission extends beyond the expected IP beam boundary, rendering this assumption unreasonable. Thus, we present the schematic beam configuration for this pulsar by adopting the IP beam as the inference beam (Figure~\ref{fig:2beam}).

\textbf{J1842+0358:} This pulsar exhibits IP emission with a peak intensity of approximately 20.1\% of the MP (Figure~\ref{fig:prof} and Table~\ref{2per}). The $W_{3\sigma}$ values for the MP and IP are 21.4$^\circ$ and 14.4$^\circ$, respectively, with a separation of $177.5^{\circ}$. 
The average polarization profile, superimposed on the PPAs of single pulses, is shown in Figure~\ref{fig:prof}. OPM is observed in the MP in the PPAs of single-pulses but is not evident in the PPAs of the average profile. 
RVM fitting was performed using Equation~\ref{eq50}, and the results, presented in Table~\ref{fit}, suggest that the IP emission originates from the pole opposite to the MP. 
The negative value of $\Delta = -2.18^{\circ}$$ ^{+0.80} _{-0.79} $ indicates a lower emission height for the IP compared to the MP, with $h_{\rm IM}= $ $105.97_{-38.40}^{+38.89}$\,km.
Notably, only two PPAs are available for the IP, which may limit the reliability of our results. Follow-up observations with longer durations are necessary to confirm these findings. 
Using the MP beam as the reference beam, we determine that the beam radii for the IP and MP are $12.49^{\circ}$ and $15.04^{\circ}$, respectively, corresponding to filling factors of 0.66 and 0.87. 
If the IP beam is taken as the reference beam, the MP emission extends beyond the expected MP beam boundary. Therefore, we illustrate the schematic configuration of the beam for this pulsar using the MP beam as the reference (Figure~\ref{fig:2beam}).

\textbf{J1843$-$0702:} This pulsar exhibits IP emission with a peak intensity of approximately 37.3\% of the MP (Figure~\ref{fig:prof} and Table~\ref{2per}). The $W_{3\sigma}$ values for the MP and IP are 19.3$^\circ$ and 21.8$^\circ$, respectively, with a separation of $172.3^{\circ}$. 
The average polarization profile, superimposed on the PPAs of single pulses, is shown in Figure~\ref{fig:prof}. The PPAs of the IP exhibit smooth variations, whereas a reversal is observed in the leading component of the MP. 
RVM fitting was performed using Equation~\ref{eq50}, excluding the PPAs of the leading component of the MP. The results, presented in Table~\ref{fit}, suggest that the IP emission originates from the pole opposite to the MP. 
The negative value of $\Delta =  -5.11^{\circ}$$ ^{+0.32} _{-0.34} $ indicates a lower emission height for the IP compared to the MP, with $h_{\rm IM}= $ $203.99_{-13.57}^{+12.77}$\,km. 
Compared to previous results by~\citet{sjk+2021}, the value of $\zeta$ remains constant, while the values of $\alpha_{\rm M}$ and $\Delta$ differ slightly.  
Using the IP beam as the reference beam, the beam radii for the IP and MP are determined to be $23.31^{\circ}$ and $26.61^{\circ}$, respectively, corresponding to filling factors of 0.53 and 0.37. 
If the MP beam is taken as the reference beam, the emission height of MP is about $150$\,km, which is smaller than $h_{\rm MI}$, rendering this assumption unreasonable. Therefore, we illustrate the schematic configuration of the beam for this pulsar using the IP beam as the reference (Figure~\ref{fig:2beam}).

\textbf{J1909+0749:} This pulsar has an IP emission with a peak intensity of about 83.7\% of the MP (Figure~\ref{fig:prof} and Table~\ref{2per}). The $W_{3\sigma}$ values of the MP and IP are 16.2$^\circ$ and 21.1$^\circ$, respectively, with a separation of $186.3^{\circ}$. 
The average polarization profile, superposed on PPAs of single pulses, is shown in Figure~\ref{fig:prof}, in which the PPAs of both MP and IP show smooth variations. We carried out RVM fitting using Equation~\ref{eq50}, and the results are shown in Table~\ref{fit}, which suggests that the IP emission originates from the opposite pole of the MP. 
The positive value of $\Delta = 8.99^{\circ}$$ ^{+1.23} _{-0.86} $ indicates a higher emission height of IP compared to MP, with $h_{\rm IM}= $ $444.18_{-42.49}^{+60.77}$\,km. Compared to previous studies by~\citet{sjk+2021} and \citet{2023RAA....23j4002W}, our results agree with those of \citet{2023RAA....23j4002W}, but do not coincide with those of~\citet{sjk+2021}, which may be due to the effect of S/N. 
We take the beam of IP as the inference beam and obtain that the beam radii for IP and MP are measured to be $27.79^{\circ}$ and $21.97^{\circ}$, respectively, corresponding to filling factors of 0.44 and 0.37. If we take the beam of MP as the inference beam, we obtain that the beam radii for IP and MP are measured to be $24.20^{\circ}$ and $17.21^{\circ}$, respectively, corresponding to filling factors of 0.54 and 0.47.
The schematic configurations of the beams for this pulsar, by taking the beams of IP and MP as the inference beams, are shown in Figure~\ref{fig:2beam}. Taking the mean value of these two approaches, we find that the beam radii for IP and MP are $26.00^{\circ}$ and $19.59^{\circ}$, respectively, with corresponding filling factors of 0.49 and 0.42.

\textbf{J1913+0832:} This pulsar exhibits IP emission with a peak intensity of approximately 59.0\% of the MP (Figure~\ref{fig:prof} and Table~\ref{2per}). The $W_{3\sigma}$ values of the MP and IP are 45.7$^\circ$ and 37.3$^\circ$, respectively, with a separation of $192.7^{\circ}$. 
The average polarization profile, superposed on the PPAs of single pulses, is shown in Figure~\ref{fig:prof}, where the PPAs of both MP and IP show smooth variations.
We performed RVM fitting using Equation~\ref{eq50}, and the results, shown in Table~\ref{fit}, suggest that the IP emission originates from the opposite pole of the MP. 
The negative value of $\Delta = -43.25^{\circ}$$ ^{+8.72} _{-8.60} $ indicates a lower emission height for the IP compared to the MP, with $h_{\rm IM}= $$1211.08_{-240.82}^{+244.18}$\,km, which is significantly large. 
Compared to the previous study by~\citet{2023RAA....23j4002W}, the value of $\beta$ remains constant, but the value of $\alpha$ differs, possibly because \citet{2023RAA....23j4002W} did not consider the difference in emission height between the MP and IP.
Due to the large uncertainties in our fitting results, we do not present the schematic configuration of the beam.

\textbf{J1918+1541:} This pulsar exhibits IP emission with a peak intensity of approximately 22.7\% of the MP (Figure~\ref{fig:prof} and Table~\ref{2per}). The $W_{3\sigma}$ values of the MP and IP are 30.9$^\circ$ and 14.4$^\circ$, respectively, with a separation of $163.5^{\circ}$. 
The average polarization profile, superposed on the PPAs of single pulses, is shown in Figure~\ref{fig:prof}, where the PPAs of both MP and IP display smooth variations, along with a PA jump at the first component of the MP. 
Given the large uncertainties and the limited number of PPAs in the first component of the MP, we exclude these PPAs from the RVM fitting. 
We performed RVM fitting using Equation~\ref{eq50}, and the results, shown in Table~\ref{fit}, suggest that the IP emission originates from the opposite pole of the MP. 
The negative value of $\Delta =  -9.38^{\circ}$$ ^{+0.56} _{-0.80}$ indicates a lower emission height for the IP compared to the MP, with $h_{\rm IM}= $$724.77_{-61.81}^{+43.27}$\,km.
Our results are consistent with the previous studies by~\citet{2023RAA....23j4002W} and~\citet{sjk+2021}.
We take the beam of the IP as the inference beam and find that the beam radii for the IP and MP are $17.20^{\circ}$ and $24.45^{\circ}$, respectively, corresponding to filling factors of 0.45 and 0.74, respectively.
Alternatively, if we take the beam of the MP as the inference beam, we obtain beam radii of $11.89^{\circ}$ and $21.06^{\circ}$, respectively, corresponding to filling factors of 0.72 and 0.92, respectively.
The schematic configurations of the beams for this pulsar, taking the beams of the IP and MP as inference beams, are shown in Figure~\ref{fig:2beam}.
By averaging the values obtained from these two approaches, we determine that the beam radii for the IP and MP are $14.54^{\circ}$ and $22.76^{\circ}$, respectively, corresponding to filling factors of 0.58 and 0.83, respectively.

\textbf{J1935+2025:} This pulsar exhibits IP emission with a peak intensity of approximately 36.2\% of the MP (Figure~\ref{fig:prof} and Table~\ref{2per}). The $W_{3\sigma}$ values of the MP and IP are 32.3$^\circ$ and 15.1$^\circ$, respectively, with a separation of $189.5^{\circ}$.
The average polarization profile, superposed on the PPAs of single pulses, is shown in Figure~\ref{fig:prof}, where the PPAs of both MP and IP display smooth variations. 
We performed RVM fitting using Equation~\ref{eq50}, and the results, shown in Table~\ref{fit}, suggest that the IP emission originates from the opposite pole of the MP. 
The positive value of $\Delta = 1.06^{\circ}$$ ^{+0.17} _{-0.17} $ indicates a higher emission height for the IP compared to the MP, with $h_{\rm IM}= $$17.69_{-2.84}^{+2.84}$\,km.
Our results are consistent with previous studies by~\citet{jk2019} and~\citet{2023RAA....23j4002W}. 
We take the beam of the IP as the inference beam and find that the MP emission extends beyond the expected MP beam boundary, which is not reasonable.
However, if we take the beam of the MP as the inference beam, we obtain beam radii for the IP and MP of $28.63^{\circ}$ and $28.03^{\circ}$, respectively, corresponding to filling factors of 0.27 and 0.64, respectively.
The schematic configuration of the beams for this pulsar, using the MP as the inference beam, is shown in Figure~\ref{fig:2beam}.

\textbf{J1952+3252:} This pulsar exhibits weak IP emission, with a peak intensity of approximately 2.6\% of the MP (Figure~\ref{fig:prof} and Table~\ref{2per}). The $W_{3\sigma}$ values of the MP and IP are 78.8$^\circ$ and 23.2$^\circ$, respectively, with a separation of $184.9^{\circ}$. 
The average polarization profile, superposed on the PPAs of single pulses, is shown in Figure~\ref{fig:prof}, where the PPAs of the IP display smooth variations, while an OPM jump occurs in the leading component of the MP at a pulse phase of $12.5^{\circ}$.
We performed RVM fitting using Equation~\ref{eq5}, taking $\phi_{\rm OPM} = 12.5^{\circ}$ and fitting the degree of the OPM jump as a free parameter. The results, shown in Table~\ref{fit}, suggest that the IP emission originates from the opposite pole of the MP.
The measured degree of the OPM jump is $93.95^{\circ}$$ ^{+1.23} _{-1.37}$.
The positive value of $\Delta = 15.90^{\circ}$$ ^{+14.90} _{-6.99} $ indicates a higher emission height for the IP compared to the MP, with $h_{\rm IM}= $$130.95_{-57.57}^{+122.71}$\,km.
Our results are consistent with the previous study by~\citet{2023RAA....23j4002W}.
Due to the large uncertainties in the RVM fitting results, we do not present the schematic configuration of the beams.

\textbf{J2023+5037:} The peak intensity of IP emission for this pulsar is approximately 13.1\% of the MP (Figure~\ref{fig:prof} and Table~\ref{2per}). The $W_{3\sigma}$ values for the MP and IP are 30.2$^\circ$ and 21.8$^\circ$, respectively, with a separation of $178.9^{\circ}$. The average polarization profile, superposed on the PPAs of single pulses, is shown in Figure~\ref{fig:prof}, where an OPM jumps is detected in MP at pulse phases of $1.8^{\circ}$.
In the IP, there is an PA jump of $\sim41^{\circ}$ at pulse phases of $186.7^{\circ}$, and we exclude the PPAs beyond $186.7^{\circ}$ in the RVM fitting. 
We use Equation~\ref{eq5}, incorporating a PA jump with $\phi_{\rm OPM} = 1.8^{\circ}$, and treat the degree of this OPM jump as a free parameter. The results, presented in Table~\ref{fit}, suggest that the IP emission originates from the opposite pole of the MP. The measured degree of the OPM jump is $95.64^{\circ}$$^{+1.93} _{-1.90}$. The negative value of $\Delta = -5.33^{\circ}$$^{+0.65} _{-0.65}$ indicates a lower emission height for the IP compared to the MP, with $h_{\rm IM}= 413.76_{-50.46}^{+50.46}$\,km. 
We consider the beam of the IP as the inference beam, and find that the beam radii for the IP and MP are $21.69^{\circ}$ and $25.34^{\circ}$, respectively, corresponding to filling factors of 0.65 and 0.59. If we instead take the beam of the MP as the inference beam, we obtain similar results, with beam radii for the IP and MP of $22.07^{\circ}$ and $25.67^{\circ}$, respectively, and filling factors of 0.64 and 0.59, respectively. The schematic configurations of the beams for this pulsar, using either the IP or MP beam as the inference beam, are shown in Figure~\ref{fig:2beam}. Taking the mean values from these two approaches, we obtain beam radii of 21.88$^{\circ}$ and 25.51$^{\circ}$ for the IP and MP, respectively, corresponding to filling factors of 0.65 and 0.59.

\textbf{J2032+4127:} The peak intensity of IP emission for this pulsar is about 9.3\% of the MP (Figure~\ref{fig:prof} and Table~\ref{2per}). The $W_{3\sigma}$ values for the MP and IP are 49.2$^\circ$ and 19.7$^\circ$, respectively, with a separation of $165.9^{\circ}$. The average polarization profile, superposed on the PPAs of single pulses, is shown in Figure~\ref{fig:prof}, where both the MP and IP exhibit smooth variations. 
We performed RVM fitting using Equation~\ref{eq50}, and the results, presented in Table~\ref{fit}, suggest that the IP emission originates from the opposite pole of the MP. The positive value of $\Delta = 11.05^{\circ}$$^{+9.29} _{-6.88}$ indicates a higher emission height for the IP compared to the MP, with $h_{\rm IM}= 329.77_{-205.32}^{+277.24}$\,km. Compared to the results of \citet{2023RAA....23j4002W}, the value of $\alpha$ remains constant, but $\beta$ differs, likely because \citet{2023RAA....23j4002W} did not account for the emission height difference between the MP and IP. 
Due to large uncertainties in the RVM fitting results, we do not show the schematic configurations of the beams.

\textbf{J2208+4056:} This pulsar exhibits IP emission, with a peak intensity of about 19.7\% of the MP (Figure~\ref{fig:prof} and Table~\ref{2per}). The $W_{3\sigma}$ values for the MP and IP are 39.0$^\circ$ and 17.2$^\circ$, respectively, with a separation of $181.4^{\circ}$. The average polarization profile, superposed on the PPAs of single pulses, is shown in Figure~\ref{fig:prof}, where the PPAs of both the MP and IP show smooth variations.
We performed RVM fitting using Equation~\ref{eq50}, and the results presented in Table~\ref{fit}, suggest that the IP emission originates from the opposite pole of the MP. The positive value of $\Delta = 10.69^{\circ}$$^{+0.52} _{-0.42}$ indicates a higher emission height for the IP compared to the MP, with $h_{\rm IM}= 1418.56_{-55.73}^{+69.00}$\,km. Compared to the previous study by \citet{2023RAA....23j4002W}, both values of $\alpha$ and $\beta$ differ slightly, likely because \citet{2023RAA....23j4002W} did not account for the emission height difference between the MP and IP.
We consider the beam of the IP as the inference beam and found that the emission height from the MP is smaller than $h_{\rm IM}$, which is unreasonable. However, if we take the beam of the MP as the inference beam, the beam radii for the IP and MP are measured to be $31.88^{\circ}$ and $25.92^{\circ}$, respectively, corresponding to filling factors of 0.28 and 0.99. The schematic configurations of the beams for this pulsar, using the MP beam as the inference beam, are shown in Figure~\ref{fig:2beam}.

\subsubsection{\bf Unclassified}

\textbf{J0627+0706:} This pulsar exhibits an IP emission with a peak intensity of approximately 26.6\% of the MP (Figure~\ref{fig:prof} and Table~\ref{2per}). The $W_{3\sigma}$ values of the MP and IP are 23.2$^\circ$ and 25.3$^\circ$, respectively, with a separation of $183.2^{\circ}$. 
The average polarization profile, superimposed on the PPAs of single pulses, is shown in Figure~\ref{fig:prof}. The PPAs of the MP display smooth variations, whereas those of the IP exhibit complex variations.
Therefore, RVM fitting was not performed for this pulsar.

\textbf{J0826+2637:} This pulsar exhibits weak IP emission with a peak intensity of approximately 0.2\% of the MP (Figure~\ref{fig:prof} and Table~\ref{2per}). The $W_{3\sigma}$ values of the MP and IP are 66.4$^\circ$ and 32.7$^\circ$, respectively, with a separation of $179.3^{\circ}$. 
The average polarization profile, superimposed on the PPAs of single pulses, is shown in Figure~\ref{fig:prof}. The PPAs of the IP display smooth variations with a PA jump, whereas those of the MP exhibit complex variations.
Based on the PPAs of single pulses, an OPM phenomenon is observed in the center of the MP. 
Due to the unsmooth PPAs, RVM fitting was not performed for this pulsar.

\textbf{J1816$-$0755:} This pulsar exhibits IP emission with a peak intensity of approximately 22.0\% of the MP (Figure~\ref{fig:prof} and Table~\ref{2per}). The $W_{3\sigma}$ values of the MP and IP are 26.8$^\circ$ and 32.3$^\circ$, respectively, with a separation of $175.4^{\circ}$. 
The average polarization profile, superimposed on the PPAs of single pulses, is shown in Figure~\ref{fig:prof}. The PPAs of the MP display smooth variations with a PA jump at the trailing component, whereas the IP does not follow an S-shape.
Consequently, RVM fitting was not performed for this pulsar.

\textbf{J1825$-$0935:} This pulsar exhibits IP emission with a peak intensity of approximately 5.4\% of the MP (Figure~\ref{fig:prof} and Table~\ref{2per}). The $W_{3\sigma}$ values of the MP and IP are 44.6$^\circ$ and 25.0$^\circ$, respectively, with a separation of $187.0^{\circ}$. 
The average polarization profile, superimposed on the PPAs of single pulses, is shown in Figure~\ref{fig:prof}, in which neither the MP nor the IP follow an S-shape. 
Based on the PPAs of single pulses, OPMs are present in both the MP and IP. The profile properties of PSR J1825$-$0935 have been extensively studied in the literature (e.g., ~\citealt{1994A&A...282...45G}). However, it remains unclear whether the MP and IP emissions originate from the same or opposite magnetic poles. Due to the unsmooth PPAs in our observation, RVM fitting was not performed for this pulsar.

\textbf{J1849+0409:} This pulsar exhibits IP emission with a peak intensity of approximately 84.0\% of the MP (Figure~\ref{fig:prof} and Table~\ref{2per}). The $W_{3\sigma}$ values for the MP and IP are 11.6$^\circ$ and 18.6$^\circ$, respectively, with a separation of $185.6^{\circ}$. 
The average polarization profile, superimposed on the PPAs of single pulses, is shown in Figure~\ref{fig:prof}. The PPAs of the MP exhibit smooth variations, whereas those of the IP do not follow an S-shape. 
Based on the PPAs of single pulses, OPMs are observed in both the MP and IP, although they are not evident in the PPAs of the average profile.  
RVM fitting was not performed for this pulsar due to the unsmooth PPAs.

\textbf{J1926+0737:} This pulsar exhibits IP emission with a peak intensity of approximately 65.5\% of the MP (Figure~\ref{fig:prof} and Table~\ref{2per}). The $W_{3\sigma}$ values of the MP and IP are 17.0$^\circ$ and 10.2$^\circ$, respectively, with a separation of $166.3^{\circ}$. 
The average polarization profile, superposed on the PPAs of single pulses, is shown in Figure~\ref{fig:prof}, where the PPAs of the IP show smooth variations, whereas those of the MP exhibit a V-shape. 
We did not perform RVM fitting for this pulsar because the PPAs do not follow an S-shape.

\textbf{J2047+5029:} This pulsar exhibits weak IP emission, with a peak intensity of about 61.9\% of the MP (Figure~\ref{fig:prof} and Table~\ref{2per}). The $W_{3\sigma}$ values for the MP and IP are both 16.2$^\circ$, with a separation of $179.6^{\circ}$. The average polarization profile, superposed on the PPAs of single pulses, is shown in Figure~\ref{fig:prof}, where the PPAs of the IP show smooth variations, while those for the MP do not follow an S-shape.
The OPM phenomenon in the center of the MP is detected by analyzing the PPAs of single pulses. However, we did not carry out RVM fitting for this pulsar because the PPAs do not follow an S-shape.

\section{DISCUSSION and CONCLUSIONS}\label{sec:discussion}

\subsection{Beams Structures}

The structure of pulsar radio beams remains unclear. Three main beam structure models have been proposed: the core-cone beam model~\citep{r1983}, the patch beam model~\citep{ml1977}, and the fan beam model~\citep{wpz+2014}. According to these models, the pulse width exhibits different behaviors with respect to the impact angle ($\beta$): it decreases with increasing $\beta$ in the core-cone beam model, increases in the fan beam model, and follows no obvious trend in the patch beam model. Pulsars with opposite pole IP emissions provide a unique opportunity to test these models.
By analyzing 8 such pulsars, we found that in 7 cases, the pulse width increases with increasing $\beta$, whereas in 1 case, it decreases (Figure~\ref{fig:width_beta}). This trend appears to favor the fan beam model. However, considering the underfilled nature of the emission beam and the limited sample size, we cannot definitively exclude any of the beam models.

For pulsars with same-pole IP emissions, the LOS sweeps within the beam, providing constraints on the beam structure (Figure~\ref{fig:1beam}). In these cases, we found that the MP and IP emissions for some pulsars roughly correspond to regions associated with the smallest and largest angles between the magnetic axis and the LOS. This observation may support the core-cone model, in which MP and IP emissions are related to two distinct cones, respectively~\citep{ml1977,g1985}, with weak bridge emission potentially arising from regions between these two cones.
The complex profiles of MP/IP emissions suggest that active emission regions may be patchy within the conal ring~\citep{1993ApJ...405..285R,2007MNRAS.380.1678K}. By analyzing an extensive dataset for PSR J1906+0746, which exhibits relativistic spin precession, \cite{dkl+2019} presented a two-dimensional beam map and found that the observed beam of PSR J1906+0746 is elongated and smaller than the expected full open field-line region. Therefore, the shape of pulsar beams is likely more complex than predicted by the conal-core or patch beam models.

\subsection{Emission Heights}

\begin{figure}
\centering
\includegraphics[width=0.45\textwidth]{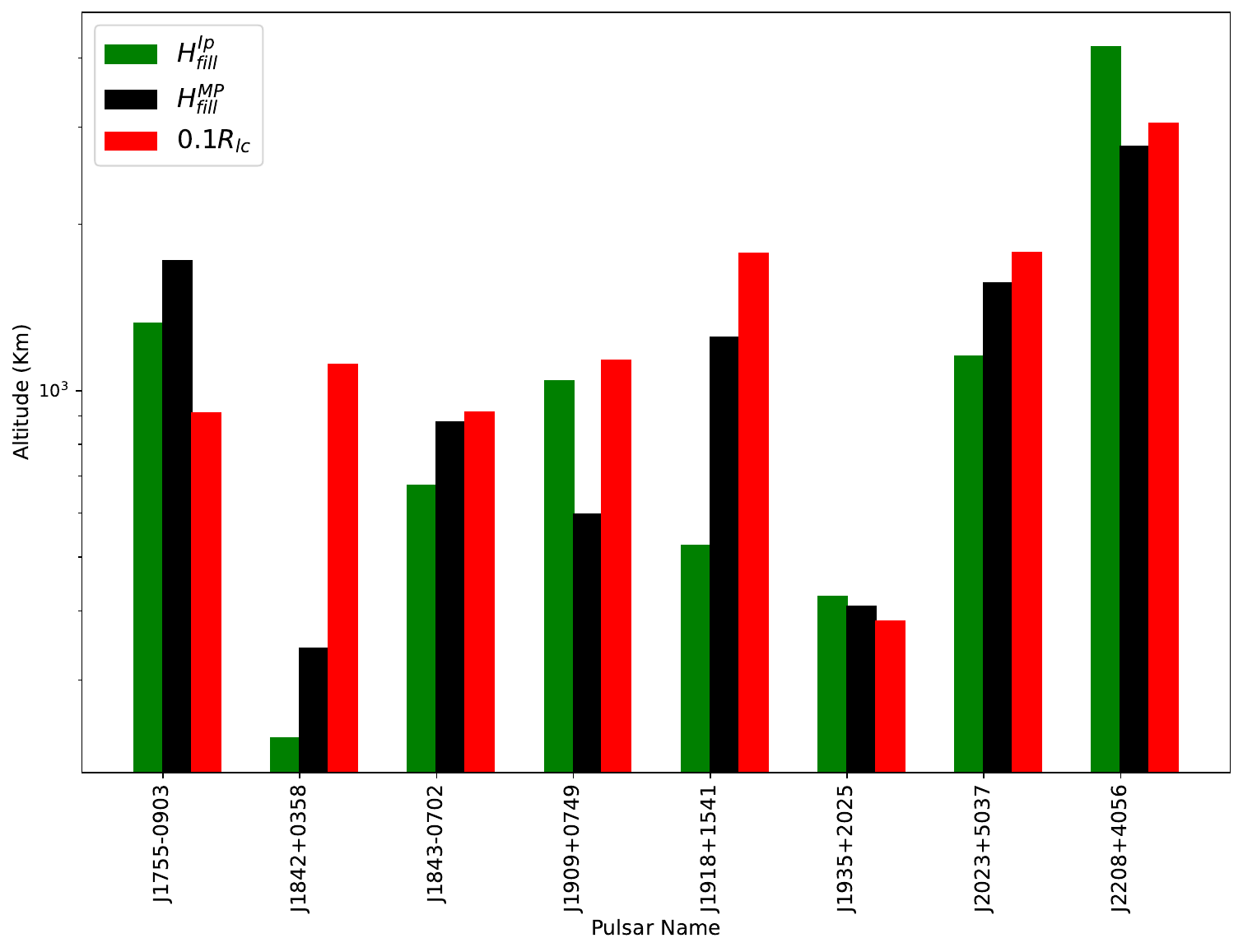}
\caption{The emission heights for 8 pulsars with opposite pole IP emissions. The green and black histograms represent the emission heights estimated using the expected maximum profile width according to the geometrical method. The red histogram indicates the value of $0.1 R_{\rm lc}$.}
\label{fig:altitude}
\end{figure}

For pulsars, it is generally accepted that radio emission occurs within \(0.1 R_{\rm lc}\) (e.g., \citealt{2003A&A...397..969K}). The height of radio emission is typically estimated using the delay-radius and geometrical methods. The delay-radius method has been applied to many pulsars, with measured emission heights ranging from approximately \(200\) to \(500\) km (e.g., \citealt{bcw1991,1997A&A...324..981V,2004A&A...421..215M,2008MNRAS.391.1210W}). However, this method relies on several assumptions, such as the emission being fully filled (see the review by \citealt{2017JApA...38...52M}), and it may only be valid at low emission heights~\citep{2008MNRAS.391..859D}.
By analyzing the polarization properties of pulsars with IP emissions, we found that the radio beam is underfilled, with filling factors ranging from 0.27 to 0.99. Given these underfilled beams, the emission heights derived from the delay-radius method are likely to have large uncertainties, and those obtained from geometrical methods may be underestimated. For pulsars with IP emissions, the difference in emission height between the MP and IP can be measured by RVM fitting~\citep{jk2019}, which does not depend on the shape of the profile. We find that \(h_{\rm MI}\) in our sample has a wide distribution, ranging from \(17.69_{-2.84}^{+2.84}\) to \(1418.56_{-55.73}^{+69.00}\) km, which further supports the idea that radio emissions from pulsars generally occur at heights of several hundred kilometers.

For pulsars with IP emissions, \(\alpha\) and \(\beta\) can be determined precisely, potentially providing an approach to measure emission height. We assume that the emission beam is circular and neglect the effects of aberration and retardation. The boundary of the beam is defined by a critical size that encompasses the on-pulse windows of the MPs/IPs. It should be noted that the actual beam may be larger. The expected maximum profile width can then be calculated based on the fraction of the LOS within the beam. We use this expected maximum profile width to estimate the emission height via the geometrical method, rather than relying on the observed profile width.
The emission heights obtained using the expected maximum profile width of IPs/MPs for each pulsar are shown in Figure~\ref{fig:altitude}. The emission heights for these pulsars range from a few hundred kilometers to a few thousand kilometers. The emission heights for some of them are notably high, even exceeding \(0.1 R_{\rm lc}\), such as PSRs J1935+2025 and J2208+4056. PSR J1935+2025 is relatively young, with a characteristic age of about \(10^{4}\) years, which agrees with the conclusion that emissions can occur at high altitudes for young pulsars~\citep{2006MNRAS.368.1856J,2007MNRAS.380.1678K}. However, not all pulsars with high emission heights are young.
PSRs J1755$-$0903 and J2208+4056 are much older, with characteristic ages of about \(10^{6}\) years, and their beam radii are relatively large, measuring \(37^{\circ}\) and \(32^{\circ}\), respectively. \citet{2009MNRAS.395.2117W} presented an RVM fitting for PSR B1055$-$52, which has a characteristic age of \(5\times 10^{5}\) years and exhibits IP emission. They concluded that the emission beam of PSR B1055$-$52 is wide and suggested that the emission occurs beyond the polar cap boundary and outside the null-charge surface. It is possible that radio emissions for some pulsars originate outside the polar cap, such as in the slot gap~\citep{2003ApJ...588..430M} or the outer gap~\citep{1995ApJ...438..314R}. The cases of PSRs J1755$-$0903 and J2208+4056 may be similar to that of PSR B1055$-$52. However, it remains unclear why emissions for some pulsars can extend beyond the polar cap boundary.

\subsection{OPMs}

The OPM is a common phenomenon in pulsars, though its origin remains unclear. It is believed that the OPM phenomenon may be related to propagation effects in the birefringent pulsar plasma. The observed $90^\circ$ jumps may result from the separation of the X-mode and O-mode beams caused by refraction~\citep{1986ApJ...302..138B,2001A&A...378..883P}. However, in many pulsars, the OPM jumps significantly deviate from $90^\circ$, a discrepancy that cannot be explained solely by the existence of two completely orthogonal modes.
These deviations from $90^\circ$ may be attributed to the superposition of two non-orthogonal modes originating from different magnetic field lines (e.g., \citealt{scw+1984}), wave mode coupling, and cyclotron absorption~\citep{Petrova2006}, or interstellar scattering~\citep{Karastergiou2009}. Observations across a wide bandwidth have revealed that the OPM jump is frequency-dependent. For example, for PSR J0953+0755, the OPM jump is approximately $90^\circ$ at 1404 MHz~\citep{scr+1984,ryw2022}, while it decreases to $65^\circ$ at 430 MHz~\citep{br1980}. Given that plasma waves at different frequencies traverse distinct paths through the pulsar magnetosphere, it is unsurprising that the OPM jumps vary with frequency~\citep{2001A&A...378..883P}.
For a given pulsar, the PA jumps at different pulse phases can also vary. This phenomenon may further indicate that emissions from different components occur at varying heights within the pulsar magnetosphere~\citep{2007MNRAS.380.1678K}. Further observations using wideband capabilities, such as the ultra-wide bandwidth receiver on the Parkes radio telescope~\citep{2020PASA...37...12H}, will provide more insight into the physical process of the escape of X and O mode in pulsar magnetosphere~\citep{2023MNRAS.525..840O}.

\section*{Acknowledgments}

This work is sponsored by the National Natural Science Foundation of China (NSFC) project (No.12288102, 12041303, 12041304, 12403060, 12273100, 12203092),  the Major Science and Technology Program of Xinjiang Uygur Autonomous Region (No. 2022A03013-1), the National Key R\&D Program of China (No. 2022YFC2205201, 2022YFC2205202, 2020SKA0120200), the Natural Science Foundation of Xinjiang Uygur Autonomous Region (No. 2022D01B218, 2022D01B71, 2022D01D85), the West Light Foundation of Chinese Academy of Sciences (No. WLFC 2021-XBQNXZ-027), the 2023 project Xinjiang uygur autonomous region of China for Tianchi talents, the Tianshan Talent Training Program for Young Elite Scientists (No. 2023TSYC-551QNTJ0024), and the Tianshan Talent Program of Xinjiang Uygur autonomous region (No. 2023TSYCTD0013).
The research is supported by the Scientific Instrument Developing Project of the Chinese Academy of Sciences, Grant No. PTYQ2022YZZD01, and partly supported by the Operation, Maintenance and Upgrading Fund for Astronomical Telescopes and Facility Instruments, budgeted from the Ministry of Finance of China (MOF) and administrated by the Chinese Academy of Sciences (CAS). We thank the investigators of PT2023$\_$0108 for their useful data. 
This work made use of the data from FAST (Five-hundred-meter Aperture Spherical radio Telescope) (https://cstr.cn/31116.02.FAST).  
FAST is a Chinese national mega-science facility, operated by National Astronomical Observatories, Chinese Academy of Sciences. 
We thank the referee for the valuable comments, which improved our manuscript.

\software{DSPSR \citep{vb2011}, PSRCHIVE \citep{hvm2004}}

\bibliography{sample63}{}

\begin{thebibliography}{}
\expandafter\ifx\csname natexlab\endcsname\relax\def\natexlab#1{#1}\fi
\providecommand{\url}[1]{\href{#1}{#1}}
\providecommand{\dodoi}[1]{doi:~\href{http://doi.org/#1}{\nolinkurl{#1}}}
\providecommand{\doeprint}[1]{\href{http://ascl.net/#1}{\nolinkurl{http://ascl.net/#1}}}
\providecommand{\doarXiv}[1]{\href{https://arxiv.org/abs/#1}{\nolinkurl{https://arxiv.org/abs/#1}}}

\bibitem[{{Backer} \& {Rankin}(1980)}]{br1980}
{Backer}, D.~C., \& {Rankin}, J.~M. 1980, \apjs, 42, 143,
  \dodoi{10.1086/190647}

\bibitem[{{Barnard} \& {Arons}(1986)}]{1986ApJ...302..138B}
{Barnard}, J.~J., \& {Arons}, J. 1986, \apj, 302, 138, \dodoi{10.1086/163979}

\bibitem[{{Blaskiewicz} {et~al.}(1991){Blaskiewicz}, {Cordes}, \&
  {Wasserman}}]{bcw1991}
{Blaskiewicz}, M., {Cordes}, J.~M., \& {Wasserman}, I. 1991, \apj, 370, 643,
  \dodoi{10.1086/169850}

\bibitem[{{Desvignes} {et~al.}(2019){Desvignes}, {Kramer}, {Lee}, {van
  Leeuwen}, {Stairs}, {Jessner}, {Cognard}, {Kasian}, {Lyne}, \&
  {Stappers}}]{dkl+2019}
{Desvignes}, G., {Kramer}, M., {Lee}, K., {et~al.} 2019, Science, 365, 1013,
  \dodoi{10.1126/science.aav7272}

\bibitem[{{Dyks}(2008)}]{2008MNRAS.391..859D}
{Dyks}, J. 2008, \mnras, 391, 859, \dodoi{10.1111/j.1365-2966.2008.13923.x}

\bibitem[{{Everett} \& {Weisberg}(2001)}]{ew2001}
{Everett}, J.~E., \& {Weisberg}, J.~M. 2001, \apj, 553, 341,
  \dodoi{10.1086/320652}

\bibitem[{{Foreman-Mackey} {et~al.}(2013){Foreman-Mackey}, {Hogg}, {Lang}, \&
  {Goodman}}]{fhl+2013}
{Foreman-Mackey}, D., {Hogg}, D.~W., {Lang}, D., \& {Goodman}, J. 2013, \pasp,
  125, 306, \dodoi{10.1086/670067}

\bibitem[{{Gil}(1985)}]{g1985}
{Gil}, J. 1985, \apj, 299, 154, \dodoi{10.1086/163688}

\bibitem[{{Gil} \& {Kijak}(1993)}]{gk1993}
{Gil}, J.~A., \& {Kijak}, J. 1993, \aap, 273, 563

\bibitem[{{Gil} {et~al.}(1994){Gil}, {Jessner}, {Kijak}, {Kramer}, {Malofeev},
  {Malov}, {Seiradakis}, {Sieber}, \& {Wielebinski}}]{1994A&A...282...45G}
{Gil}, J.~A., {Jessner}, A., {Kijak}, J., {et~al.} 1994, \aap, 282, 45

\bibitem[{{Hankins} \& {Cordes}(1981)}]{hc1981}
{Hankins}, T.~H., \& {Cordes}, J.~M. 1981, \apj, 249, 241,
  \dodoi{10.1086/159281}

\bibitem[{{Hobbs} {et~al.}(2020){Hobbs}, {Manchester}, {Dunning}, {Jameson},
  {Roberts}, {George}, {Green}, {Tuthill}, {Toomey}, {Kaczmarek}, {Mader},
  {Marquarding}, {Ahmed}, {Amy}, {Bailes}, {Beresford}, {Bhat}, {Bock},
  {Bourne}, {Bowen}, {Brothers}, {Cameron}, {Carretti}, {Carter}, {Castillo},
  {Chekkala}, {Cheng}, {Chung}, {Craig}, {Dai}, {Dawson}, {Dempsey}, {Doherty},
  {Dong}, {Edwards}, {Ergesh}, {Gao}, {Han}, {Hayman}, {Indermuehle},
  {Jeganathan}, {Johnston}, {Kanoniuk}, {Kesteven}, {Kramer}, {Leach},
  {Mcintyre}, {Moss}, {Os{\l}owski}, {Phillips}, {Pope}, {Preisig}, {Price},
  {Reeves}, {Reilly}, {Reynolds}, {Robishaw}, {Roush}, {Ruckley}, {Sadler},
  {Sarkissian}, {Severs}, {Shannon}, {Smart}, {Smith}, {Smith}, {Sobey},
  {Staveley-Smith}, {Tzioumis}, {van Straten}, {Wang}, {Wen}, \&
  {Whiting}}]{2020PASA...37...12H}
{Hobbs}, G., {Manchester}, R.~N., {Dunning}, A., {et~al.} 2020, \pasa, 37,
  e012, \dodoi{10.1017/pasa.2020.2}

\bibitem[{{Hotan} {et~al.}(2004){Hotan}, {van Straten}, \&
  {Manchester}}]{hvm2004}
{Hotan}, A.~W., {van Straten}, W., \& {Manchester}, R.~N. 2004, \pasa, 21, 302,
  \dodoi{10.1071/AS04022}

\bibitem[{{Jiang} {et~al.}(2019){Jiang}, {Yue}, {Gan}, {Yao}, {Li}, {Pan},
  {Sun}, {Yu}, {Liu}, {Tang}, {Qian}, {Lu}, {Yan}, {Peng}, {Zhang}, {Wang},
  {Li}, \& {Li}}]{jyg+2019}
{Jiang}, P., {Yue}, Y., {Gan}, H., {et~al.} 2019, Science China Physics,
  Mechanics, and Astronomy, 62, 959502, \dodoi{10.1007/s11433-018-9376-1}

\bibitem[{{Johnston} \& {Kerr}(2018)}]{jk2018}
{Johnston}, S., \& {Kerr}, M. 2018, \mnras, 474, 4629,
  \dodoi{10.1093/mnras/stx3095}

\bibitem[{{Johnston} \& {Kramer}(2019)}]{jk2019}
{Johnston}, S., \& {Kramer}, M. 2019, \mnras, 490, 4565,
  \dodoi{10.1093/mnras/stz2865}

\bibitem[{{Johnston} {et~al.}(2023){Johnston}, {Kramer}, {Karastergiou},
  {Keith}, {Oswald}, {Parthasarathy}, \& {Weltevrede}}]{jkk+2023}
{Johnston}, S., {Kramer}, M., {Karastergiou}, A., {et~al.} 2023, \mnras, 520,
  4801, \dodoi{10.1093/mnras/stac3636}

\bibitem[{{Johnston} \& {Weisberg}(2006)}]{2006MNRAS.368.1856J}
{Johnston}, S., \& {Weisberg}, J.~M. 2006, \mnras, 368, 1856,
  \dodoi{10.1111/j.1365-2966.2006.10263.x}

\bibitem[{{Karastergiou}(2009)}]{Karastergiou2009}
{Karastergiou}, A. 2009, \mnras, 392, L60,
  \dodoi{10.1111/j.1745-3933.2008.00585.x}

\bibitem[{{Karastergiou} \& {Johnston}(2007)}]{2007MNRAS.380.1678K}
{Karastergiou}, A., \& {Johnston}, S. 2007, \mnras, 380, 1678,
  \dodoi{10.1111/j.1365-2966.2007.12237.x}

\bibitem[{{Kijak} \& {Gil}(2003)}]{2003A&A...397..969K}
{Kijak}, J., \& {Gil}, J. 2003, \aap, 397, 969,
  \dodoi{10.1051/0004-6361:20021583}

\bibitem[{{Kou} {et~al.}(2021){Kou}, {Yan}, {Peng}, {Lu}, {Liu}, {Zhang},
  {Strom}, {Wang}, {Yuan}, {Yuen}, {Yu}, {Yao}, {Liu}, {Yan}, {Jiang}, {Jin},
  {Li}, {Qian}, {Yue}, {Zhu}, \& {FAST Collaboration}}]{kyp+2021}
{Kou}, F.~F., {Yan}, W.~M., {Peng}, B., {et~al.} 2021, \apj, 909, 170,
  \dodoi{10.3847/1538-4357/abd545}

\bibitem[{{Li} {et~al.}(2018){Li}, {Wang}, {Qian}, {Krco}, {Jiang}, {Yue},
  {Jin}, {Zhu}, {Pan}, {Nan}, \& {Dunning}}]{lwq+2018}
{Li}, D., {Wang}, P., {Qian}, L., {et~al.} 2018, IEEE Microwave Magazine, 19,
  112, \dodoi{10.1109/MMM.2018.2802178}

\bibitem[{{Lyne} \& {Manchester}(1988)}]{lm1988}
{Lyne}, A.~G., \& {Manchester}, R.~N. 1988, \mnras, 234, 477,
  \dodoi{10.1093/mnras/234.3.477}

\bibitem[{{Manchester} \& {Lyne}(1977)}]{ml1977}
{Manchester}, R.~N., \& {Lyne}, A.~G. 1977, \mnras, 181, 761,
  \dodoi{10.1093/mnras/181.4.761}

\bibitem[{{Mitra}(2017)}]{2017JApA...38...52M}
{Mitra}, D. 2017, Journal of Astrophysics and Astronomy, 38, 52,
  \dodoi{10.1007/s12036-017-9457-6}

\bibitem[{{Mitra} \& {Li}(2004)}]{2004A&A...421..215M}
{Mitra}, D., \& {Li}, X.~H. 2004, \aap, 421, 215,
  \dodoi{10.1051/0004-6361:20034094}

\bibitem[{{Muslimov} \& {Harding}(2003)}]{2003ApJ...588..430M}
{Muslimov}, A.~G., \& {Harding}, A.~K. 2003, \apj, 588, 430,
  \dodoi{10.1086/368162}

\bibitem[{{Nan} {et~al.}(2011){Nan}, {Li}, {Jin}, {Wang}, {Zhu}, {Zhu},
  {Zhang}, {Yue}, \& {Qian}}]{2011IJMPD..20..989N}
{Nan}, R., {Li}, D., {Jin}, C., {et~al.} 2011, International Journal of Modern
  Physics D, 20, 989, \dodoi{10.1142/S0218271811019335}

\bibitem[{{Oswald} {et~al.}(2023){Oswald}, {Karastergiou}, \&
  {Johnston}}]{2023MNRAS.525..840O}
{Oswald}, L.~S., {Karastergiou}, A., \& {Johnston}, S. 2023, \mnras, 525, 840,
  \dodoi{10.1093/mnras/stad2271}

\bibitem[{{Petrova}(2001)}]{2001A&A...378..883P}
{Petrova}, S.~A. 2001, \aap, 378, 883, \dodoi{10.1051/0004-6361:20011297}

\bibitem[{{Petrova}(2006)}]{Petrova2006}
---. 2006, \mnras, 366, 1539, \dodoi{10.1111/j.1365-2966.2005.09941.x}

\bibitem[{{Posselt} {et~al.}(2023){Posselt}, {Karastergiou}, {Johnston},
  {Parthasarathy}, {Oswald}, {Main}, {Basu}, {Keith}, {Song}, {Weltevrede},
  {Tiburzi}, {Bailes}, {Buchner}, {Geyer}, {Kramer}, {Spiewak}, \&
  {Krishnan}}]{pkj+2023}
{Posselt}, B., {Karastergiou}, A., {Johnston}, S., {et~al.} 2023, \mnras, 520,
  4582, \dodoi{10.1093/mnras/stac3383}

\bibitem[{{Radhakrishnan} \& {Cooke}(1969)}]{rc1969}
{Radhakrishnan}, V., \& {Cooke}, D.~J. 1969, \aplett, 3, 225

\bibitem[{{Rankin}(1983)}]{r1983}
{Rankin}, J.~M. 1983, \apj, 274, 333, \dodoi{10.1086/161450}

\bibitem[{{Rankin}(1990)}]{r1990}
---. 1990, \apj, 352, 247, \dodoi{10.1086/168530}

\bibitem[{{Rankin}(1993)}]{1993ApJ...405..285R}
---. 1993, \apj, 405, 285, \dodoi{10.1086/172361}

\bibitem[{{Rejep} {et~al.}(2022){Rejep}, {Yan}, \& {Wang}}]{ryw2022}
{Rejep}, R., {Yan}, W.-M., \& {Wang}, N. 2022, Research in Astronomy and
  Astrophysics, 22, 065005, \dodoi{10.1088/1674-4527/ac6735}

\bibitem[{{Romani} \& {Yadigaroglu}(1995)}]{1995ApJ...438..314R}
{Romani}, R.~W., \& {Yadigaroglu}, I.~A. 1995, \apj, 438, 314,
  \dodoi{10.1086/175076}

\bibitem[{{Serylak} {et~al.}(2021){Serylak}, {Johnston}, {Kramer}, {Buchner},
  {Karastergiou}, {Keith}, {Parthasarathy}, {Weltevrede}, {Bailes}, {Barr},
  {Camilo}, {Geyer}, {Hugo}, {Jameson}, {Reardon}, {Shannon}, {Spiewak}, {van
  Straten}, \& {Venkatraman Krishnan}}]{sjk+2021}
{Serylak}, M., {Johnston}, S., {Kramer}, M., {et~al.} 2021, \mnras, 505, 4483,
  \dodoi{10.1093/mnras/staa2811}

\bibitem[{{Song} {et~al.}(2023){Song}, {Weltevrede}, {Szary}, {Wright},
  {Keith}, {Basu}, {Johnston}, {Karastergiou}, {Main}, {Oswald},
  {Parthasarathy}, {Posselt}, {Bailes}, {Buchner}, {Hugo}, \&
  {Serylak}}]{2023MNRAS.520.4562S}
{Song}, X., {Weltevrede}, P., {Szary}, A., {et~al.} 2023, \mnras, 520, 4562,
  \dodoi{10.1093/mnras/stad135}

\bibitem[{{Stinebring} {et~al.}(1984{\natexlab{a}}){Stinebring}, {Cordes},
  {Rankin}, {Weisberg}, \& {Boriakoff}}]{scr+1984}
{Stinebring}, D.~R., {Cordes}, J.~M., {Rankin}, J.~M., {Weisberg}, J.~M., \&
  {Boriakoff}, V. 1984{\natexlab{a}}, \apjs, 55, 247, \dodoi{10.1086/190954}

\bibitem[{{Stinebring} {et~al.}(1984{\natexlab{b}}){Stinebring}, {Cordes},
  {Weisberg}, {Rankin}, \& {Boriakoff}}]{scw+1984}
{Stinebring}, D.~R., {Cordes}, J.~M., {Weisberg}, J.~M., {Rankin}, J.~M., \&
  {Boriakoff}, V. 1984{\natexlab{b}}, \apjs, 55, 279, \dodoi{10.1086/190955}

\bibitem[{{Sun} {et~al.}(2022){Sun}, {Yan}, {Wang}, {Wang}, {Wang}, \&
  {Dang}}]{2022ApJ...934...57S}
{Sun}, S.~N., {Yan}, W.~M., {Wang}, N., {et~al.} 2022, \apj, 934, 57,
  \dodoi{10.3847/1538-4357/ac7c15}

\bibitem[{{van Straten} \& {Bailes}(2011)}]{vb2011}
{van Straten}, W., \& {Bailes}, M. 2011, \pasa, 28, 1, \dodoi{10.1071/AS10021}

\bibitem[{{von Hoensbroech} \& {Xilouris}(1997)}]{1997A&A...324..981V}
{von Hoensbroech}, A., \& {Xilouris}, K.~M. 1997, \aap, 324, 981

\bibitem[{{Wang} {et~al.}(2014){Wang}, {Pi}, {Zheng}, {Deng}, {Wen}, {Ye},
  {Guan}, {Liu}, \& {Xu}}]{wpz+2014}
{Wang}, H.~G., {Pi}, F.~P., {Zheng}, X.~P., {et~al.} 2014, \apj, 789, 73,
  \dodoi{10.1088/0004-637X/789/1/73}

\bibitem[{{Wang} {et~al.}(2023){Wang}, {Han}, {Xu}, {Wang}, {Yan}, {Jing},
  {Su}, {Zhou}, \& {Wang}}]{2023RAA....23j4002W}
{Wang}, P.~F., {Han}, J.~L., {Xu}, J., {et~al.} 2023, Research in Astronomy and
  Astrophysics, 23, 104002, \dodoi{10.1088/1674-4527/acea1f}

\bibitem[{{Wang} {et~al.}(2024){Wang}, {Wang}, {Wang}, {Hobbs}, {Xu}, {Wang},
  {Dai}, {Dang}, {Li}, {Feng}, \& {Zhang}}]{2024ApJ...964....6W}
{Wang}, S.~Q., {Wang}, N., {Wang}, J.~B., {et~al.} 2024, \apj, 964, 6,
  \dodoi{10.3847/1538-4357/ad217b}

\bibitem[{{Weltevrede} \& {Johnston}(2008)}]{2008MNRAS.391.1210W}
{Weltevrede}, P., \& {Johnston}, S. 2008, \mnras, 391, 1210,
  \dodoi{10.1111/j.1365-2966.2008.13950.x}

\bibitem[{{Weltevrede} \& {Wright}(2009)}]{2009MNRAS.395.2117W}
{Weltevrede}, P., \& {Wright}, G. 2009, \mnras, 395, 2117,
  \dodoi{10.1111/j.1365-2966.2009.14643.x}

\bibitem[{{Weltevrede} {et~al.}(2012){Weltevrede}, {Wright}, \&
  {Johnston}}]{wws2012}
{Weltevrede}, P., {Wright}, G., \& {Johnston}, S. 2012, \mnras, 424, 843,
  \dodoi{10.1111/j.1365-2966.2012.21207.x}

\bibitem[{{Weltevrede} {et~al.}(2007){Weltevrede}, {Wright}, \&
  {Stappers}}]{wws2007}
{Weltevrede}, P., {Wright}, G.~A.~E., \& {Stappers}, B.~W. 2007, \aap, 467,
  1163, \dodoi{10.1051/0004-6361:20066957}

\bibitem[{{Yan} {et~al.}(2019){Yan}, {Manchester}, {Wang}, {Yuan}, {Wen}, \&
  {Lee}}]{ymw+2019}
{Yan}, W.~M., {Manchester}, R.~N., {Wang}, N., {et~al.} 2019, \mnras, 485,
  3241, \dodoi{10.1093/mnras/stz650}

\bibitem[{{Yuan} {et~al.}(2023){Yuan}, {Zhu}, {Kramer}, {Peng}, {Lu}, {Xu},
  {Shao}, {Wang}, {Meng}, {Niu}, {Zhao}, {Miao}, {Miao}, {Xue}, {Feng}, {Wang},
  {Li}, {Zhang}, {Champion}, {Fonseca}, {Hu}, {Yao}, {Freire}, \&
  {Guo}}]{yzk+2023}
{Yuan}, M., {Zhu}, W., {Kramer}, M., {et~al.} 2023, \apj, 949, 115,
  \dodoi{10.3847/1538-4357/accb9a}

\end{thebibliography}
\bibliographystyle{aasjournal}

\end{document}